\documentclass[12pt]{article}
\usepackage{amssymb,amsthm,latexsym,authblk,amsmath,pstricks,hyperref,multirow}
\usepackage{graphicx}
\usepackage{psfrag}
\usepackage{color} 
\usepackage{subfig}
\usepackage{epsfig}
\usepackage{epstopdf} 
\newcommand{\re}{\mathrm{Re}\, }

\newcommand{\I}{{\rm{i}}}
\newcommand{\D}{{\rm{d}}}
\newcommand{\E}{e}
\newcommand{\tr}{\operatorname{tr}}

\newcommand{\spec}{{\rm{sp}}}
\newcommand{\dss}{\displaystyle}
\newcommand{\nn}{\nonumber}

\newcommand{\be}{\begin{equation}}
\newcommand{\ee}{\end{equation}}
\newcommand{\bea}{\begin{eqnarray}}
\newcommand{\eea}{\end{eqnarray}}
\newcommand{\range}{\operatorname{ran}}

\newcommand{\rank}{\operatorname{rank}}
\newcommand{\ket}[1]{| #1 \rangle}
\newcommand{\bra}[1]{\langle #1 |}
\newcommand{\braket}[2]{\langle #1 | #2 \rangle}
\newcommand{\ketbra}[2]{| #1 \rangle \langle #2 |}

\def\real{{\mathbb{R}}}

\def\complex{{\mathbb{C}}}

\def\proba{{\rm I\kern -.18em P}}

\newcommand{\Proof}{\noindent {\bf Proof. }}
\newcommand{\Proofof}[1]{\noindent {\bf Proof of #1. }}
\newcommand{\finpro}{\hfill $\Box$}

\newcommand{\ie}{i.e.}
\newcommand{\ifif}{if and only if\;\,}
\newcommand{\RHS}{r.h.s.\;\,}

\newcommand{\onehalf}{\frac{1}{2}}

\newcommand{\mmax}{{\rm max}}

\newcommand{\ONB}{orthonormal basis\;\,}
\newcommand{\ONBs}{orthonormal bases\;\,}
\newcommand{\QO}{quantum operation\;\,}
\newcommand{\QOs}{quantum operations\;\,}
\newcommand{\QSD}{quantum state discrimination\;\,}
\newcommand{\QCs}{quantum correlations\;\,}
\newcommand{\GD}{geometric discord\;\,}
\newcommand{\meas}{measurement\;\,}
\newcommand{\meass}{measurements\;\,}
\newcommand{\clas}{{\rm clas}}
\newcommand{\sep}{{\rm sep}}
\newcommand{\product}{{\rm prod}}
\newcommand{\CCQ}{closest $\AAA$-classical\;}
\newcommand{\Aclass}{{A\rm{-cl}}}

\newcommand{\class}{{\rm clas}}
\newcommand{\ent}{{\,\rm ent}}
\newcommand{\opt}{{\rm{opt}}}
\newcommand{\vN}{{\rm{v.N.}}}
\newcommand{\observables}{{\cal{B}} ({\cal{H}} ) }
\newcommand{\saobservables}{ {\cal{B}} ({\cal{H}} )_{\rm s.a.} }
\newcommand{\states}{{\cal E}}
\newcommand{\EoF}{{\rm EoF}}
\newcommand{\Bu}{{\rm Bu}}
\newcommand{\Hel}{{\rm He}}
\newcommand{\HS}{{\rm HS}}

\newtheorem{theorem}{Theorem}
\newtheorem{definition}{Definition}
\newtheorem{proposition}{Proposition}
\newtheorem{lemma}{Lemma}
\newtheorem{corollary}{Corollary}
\newtheorem{remark}{Remark}

\newcommand{\pv}{{\bf{p}}}
\newcommand{\qv}{{\bf{q}}}

\newcommand{\uv}{{\vec{u}}}

\newcommand{\sigmav}{{\vec{\sigma}}}

\newcommand{\AAA}{A}

\newcommand{\AB}{{AB}}

\newcommand{\BB}{B}

\newcommand{\EE}{{E}}

\newcommand{\SSS}{{S}}

\newcommand{\SE}{{SE}}

\newcommand{\Bb}{{\cal B}}
\newcommand{\Cc}{{\cal C}}

\newcommand{\Ee}{{\cal E}}
\newcommand{\Ff}{{\cal F}}

\newcommand{\Hh}{{\cal H}}
\newcommand{\Ii}{{\cal I}}

\newcommand{\Kk}{{\cal K}}

\newcommand{\Mm}{{\cal M}}
\newcommand{\Nn}{{\cal N}}
\newcommand{\Oo}{{\cal O}}
\newcommand{\Pp}{{\cal P}}

\newcommand{\Rr}{{\cal R}}
\newcommand{\Ss}{{\cal S}}

\begin{document}

\title{Geometric measures of quantum correlations with Bures and Hellinger distances}
\author[1,2]{D. Spehner}
\affil[1]{\normalsize Universit\'e Grenoble Alpes, Institut Fourier,  F-38000 Grenoble, France}
\affil[2]{\normalsize CNRS, Laboratoire de Physique et Mod\'elisation des
Milieux Condens\'es, F-38000 Grenoble, France}
\author[3,4]{F. Illuminati}
\affil[3]{\normalsize Universit\`a degli Studi di Salerno, Dipartimento di Ingegneria Industriale, Via Giovanni Paolo II 132, I-84084 Fisciano (SA), Italy}
\affil[4]{\normalsize INFN, Sezione di Napoli, Gruppo collegato di Salerno, I-84084 Fisciano (SA), Italy}
\author[5]{M. Orszag}
\affil[5]{\normalsize Pontificia Universidad Cat\'olica, Instituto de F\'{\i}sica, Casilla 306, Santiago 22, Chile} 
\author[6]{W. Roga}
\affil[6]{\normalsize University of Strathclydeb, Department of Physics, John Anderson Building, 107 Rottenrow, Glasgow, G4 0NG, UK} 


\date{\today}


\maketitle

\begin{abstract}
This article contains a survey of the geometric approach to quantum correlations, 
to be published in the book ``{\it Lectures on General Quantum Correlations and their Applications}''
edited by F.~Fanchini, D.~Soares-Pinto, and G.~Adesso (Springer, 2017).
We focus mainly on the geometric measures of quantum correlations 
based on the Bures and quantum Hellinger distances. 
\end{abstract}

\tableofcontents

\section{Introduction}

Quantum correlations in composite quantum systems are at the origin of the
most peculiar features of quantum mechanics such as the violation of Bell's inequalities and non-locality.
In quantum information theory, they are viewed as quantum resources used by quantum algorithms 
and communication protocols to outperform their classical analogs.
If the  composite system is in a mixed state, 
classical correlations between the parties -- arising e.g. from a random state preparation --
may be present at the same time as quantum correlations. 
In two seminal papers,
Ollivier and Zurek~\cite{Ollivier01} and Henderson and Vedral~\cite{Henderson01}
proposed a way to separate   in bipartite systems classical from quantum correlations
and introduced the quantum discord as a quantifier of the latter.
For pure states, this quantifier 
coincides with the entanglement of formation, in agreement with the fact that
\QCs in pure states are synonymous to entanglement. 
For mixed states, however,
the states with a vanishing discord, i.e. those states which possess only classical correlations, 
form a small (zero-measure) subset of the set of separable states. 
It has been argued that a non-zero discord could
be responsible for the quantum speed-up of the DQC1 algorithm~\cite{Datta05,Datta08}.
Furthermore, the discord can be interpreted as the cost of quantum communication 
in certain protocols such as quantum state merging~\cite{Madhok11,Cavalcanti11,Modi_review}
and can be related to the distillable entanglement between one subsystem and a \meas apparatus~\cite{Streltsov11a,Piani11}. 
On the other hand, the evaluation of the quantum discord remains a difficult challenge, even in the
simplest case of two qubits (see~\cite{Girolami2011,Modi_review} and references therein).
 
In this chapter, we study alternative measures of \QCs  which share many of the properties of 
the quantum discord while being easier to compute and enabling for 
operational interpretations in terms of  state distinguishability.
Such measures are related to the geometry  of the set of quantum states $\states ( \Hh_\AB)$
of the bipartite system $\AB$. Actually, they are defined in terms of a  distance on $\states ( \Hh_\AB)$. 
Apart from easier computability and  operational interpretations,
a notable advantage of the geometric approach is that it provides additional tools 
going beyond  the quantification of correlations. In particular, one can determine 
the closest separable and closest classically-correlated states to a given state $\rho$, 
as well as the geodesics linking $\rho$ to those states.   
These tools may be useful 
when studying dissipative dynamical evolutions. For instance, one can gain some insight
on the efficiency of a dynamical process in changing the amount of entanglement or quantum
correlations by comparing the physical trajectory $t \mapsto \rho_t$
in  $\states ( \Hh_{\AB} )$  with the geodesics connecting
$\rho_t$ to its closest separable or classically-correlated state(s).

The aim of what follows is to introduce and review the main properties
of a few geometric measures of quantum correlations depending on the choice of a distance
on $\states ( \Hh_\AB)$. Instead of discussing the (huge amount of) different measures present in the
literature, we shall restrict  our attention to three quantities.
We will mainly focus on (1) the 
{\it geometric discord}~\cite{Dakic10}, defined as the minimal distance between 
the bipartite state $\rho$ and a classically-correlated state.
We compare this discord with two other measures characterizing the
 sensitivity of the state to local measurements and  unitary perturbations on one subsystem, namely
(2) the {\it measurement-induced geometric discord}~\cite{Luo_Fu10},  defined as  the minimal 
distance between $\rho$ and the corresponding post-measurement state
after an arbitrary local measurement on subsystem $\AAA$, and
(3) the {\it discord of response}~\cite{Gharibian2012,Giampaolo2013}, defined as
 the minimal distance between $\rho$ and its time-evolved version after 
an arbitrary local unitary evolution  on $\AAA$ implemented by a unitary operator  with a fixed non-degenerate spectrum.
As indicated in the title of the chapter, we will only consider two distinguished distances
on the set of quantum states, namely the  Bures and  Hellinger distances. 
The 
discord of response for these two distances corresponds (in a sense that will become clear below)
to well known measures of quantum correlations  having clear operational interpretations, called the 
{\it interferometric power}~\cite{Girolami2014} and 
{\it Local Quantum Uncertainty (LQU)}~\cite{Girolami2013}.
We will show that the geometric discord with Bures and Hellinger distances are  
related to a quantum  state discrimination task, thereby establishing an explicit link between 
\QCs and state distinguishability.
We will also demonstrate that the geometric discord and discord of response with the Hellinger distance 
are almost as easy to compute as their analogs for the Hilbert-Schmidt distance
(for instance, an explicit formula valid for arbitrary qubit-qudit states, which
involves the coefficients of the expansion of the square root of the state
in terms of generalized Pauli matrices, will be derived in Sec.~\ref{sec-computability_Hellinger_GD}).
We point out that for the Bures and Hellinger distances, the measures (1)-(3) 
obey all the basic axioms of {\it bona fide} measures of quantum correlations, 
in contrast to what happens for the Hilbert-Schmidt distance~\cite{Piani14}.
Hence, the geometric discord and discord of response with the Hellinger distance offer the 
advantage of easy computability while being physically reliable.

The material of this chapter is to a large extend self-contained. The proofs
of most results save for basic theorems related in textbooks (e.g. in Ref.~\cite{Nielsen})
are included. A few technical proofs are, however, omitted. 
We apologize to the authors of many papers related to geometric measures of \QCs for not
 citing their works, either because they are not directly related to the results 
presented here or because we are not aware of them.

The remaining of the chapter is organized as follows. We recall in Sec.~\ref{sec-Q_class_correlations} the definitions of the 
entropic quantum discord and classically correlated states and formulate the basic postulates on 
measures of quantum correlations.  
The three measures outlined above are defined properly in Sec.~\ref{sec-geo_meas_QCs}. Sufficient conditions on the 
distance insuring that they obey the basic postulates are given in this section.
A detailed review on the Bures and 
Hellinger distances and their metrics is provided in Sec.~\ref{sec-Bures_and_Hellinger_dist}. 
Sections~\ref{sec-Bures_GD} and~\ref{sec-Hellinger_GD} are devoted  to the geometric discord with
the Bures and Hellinger distances, respectively.  
We present without proofs in Sec.~\ref{sec_meas_ind_geo_disc_and_disc_resp}
 some results on the other two measures (2)-(3), in particular
some bounds involving these measures and the geometric discord.
The last section~\ref{sec-conclusion} contains a few concluding remarks.

\section{Quantum vs classical correlations} \label{sec-Q_class_correlations}
\subsection{Entropic quantum discord} \label{sec-def_entropic_discord}

In all what follows, we consider a bipartite quantum system $\AB$,
 formed by putting together two systems $\AAA$ and $\BB$, with
Hilbert space $\Hh_\AB = \Hh_\AAA \otimes \Hh_\BB$, $\Hh_\AAA$ and $\Hh_\BB$ being the Hilbert spaces
of the two subsystems. In the whole chapter, we only consider systems with finite dimensional Hilbert spaces,
$n_\AAA = \dim \Hh_\AAA < \infty$ and $n_\BB = \dim \Hh_\BB < \infty$. 
Let us recall that a state of $\AB$ is given by a density matrix $\rho$, that is, a non-negative operator on $\Hh_\AB$ with 
unit trace $\tr \rho = 1$. We write $\states ( \Hh )$ the convex set formed by all density matrices on 
the Hilbert  space $\Hh$.
The  extreme points of this convex set 
are the pure states $\rho_\psi = \ketbra{\psi}{\psi}$, with $\ket{\psi} \in \Hh$, $\| \psi\|=1$. 
We often abusively write $\ket{\psi}$ 
instead of $\rho_\psi$.  
Given a state $\rho \in \states (\Hh_\AB)$ of the bipartite system $\AB$, 
the reduced states of  $\AAA$ and $\BB$ are defined by partial tracing over the other subsystem, that is,
 $\rho_A = \tr_{B} (\rho) \in \states ( \Hh_\AAA)$ and $\rho_B=\tr_A(\rho) \in \states ( \Hh_\BB)$.
They correspond to the marginals of a joint probability in classical probability theory.

The quantum discord of Ollivier and Zurek~\cite{Ollivier01} and Henderson and Vedral~\cite{Henderson01} 
is defined as follows.
The total correlations between the two parties are
characterized by the mutual information
\begin{equation} \label{mutual1}
I_{\AAA : \BB } ( \rho) =  S(\rho_B) + S(\rho_A) - S(\rho ) \; ,
\end{equation}
where the information (ignorance) about the state of $\AB$
is given by the von Neumann entropy $S(\rho)= -\tr \rho \ln \rho $, and similarly for subsystems $\AAA$ and $\BB$. In
classical information theory, the mutual information is equal to the
difference between the Shannon entropy of $\BB$ and the conditional
entropy of $\BB$ conditioned  on $\AAA$. In the quantum setting, the corresponding difference is
the Holevo quantity\footnote{
We recall that $\chi ( \{ \rho_{B|i} , \eta_i\})$ gives an upper
bound  on the classical mutual information between $ \{ \eta_i\}$
and the outcome probabilities when performing a \meas to discriminate the states $\rho_{\BB |i}$.
}
%
\begin{equation} \label{eq-Holevo_quantity_discord}
\chi ( \{ \rho_{B|i} , \eta_i\}) =  S(\rho_B)-\sum_i \eta_i S(\rho_{B|i} ) \; ,
\end{equation}
where $\eta_i$  and $\rho_{B|i}$ are respectively the probability of outcome $i$ and the corresponding
conditional state of $\BB$ after a local von Neumann measurement  on $\AAA$,
\begin{equation} \label{eq-post_meas_cond_states_B}
\eta_i=\tr \rho \,\Pi_{i}^{A}\otimes 1 
\quad , \quad \rho_{B |i}= \eta_i^{-1}
\tr_A( \rho \, \Pi_{i}^{A} \otimes 1 )\;.
\end{equation}
Here, the \meas is given by a family  $\{ \Pi_{i}^{A}\}$ 
of projectors satisfying $\Pi_i^{A} \Pi_j^{A} = \delta_{ij} \Pi_i^{A}$ and
$\sum_i \Pi_i^{A} = 1$
(hereafter, $1$ stands for the identity operator on
$\Hh_\AAA$, $\Hh_\BB$, or another space).

It turns out that, unlike in the classical case,
$I_{\AAA: \BB}(\rho)$ and $\chi ( \{ \rho_{B|i} , \eta_i\})$ are
not equal for general quantum states $\rho$, whatever the \meas on $\AAA$.
One defines the {\em quantum discord} as the difference~\cite{Ollivier01}
\begin{equation} \label{def:discord}
D_{\AAA}^\ent ( \rho ) = I_{\AAA : \BB} ( \rho)  -  J_{\BB | \AAA} (\rho) 
\quad , \quad J_{\BB | \AAA} (\rho) = \max_{\{\Pi_{i}^{A} \}} \chi ( \{ \rho_{B|i} , \eta_i\})  \; ,
\end{equation}
where the maximum is over all projective measurements\footnote{
By using the concavity of the entropy $S$, one can show that the maximum is achieved for 
projectors $\Pi_{i}^{A}$ of rank one.
}
$\{ \Pi_{i}^{A}\} $ on $\AAA$.
Alternatively, one can maximize over all POVMs\footnote{
Let us recall that a POVM associated to a (generalized) \meas
is a family $\{ M_i\}$ of operators $M_i \geq 0$ such that 
$\sum_i M_i = 1$. The probability of outcome $i$ is  
$\eta_i = \tr M_i \rho$ and the corresponding post-\meas conditional state is
$\eta_i^{-1} A_i \rho A_i^\dagger$, where the Kraus operators $A_i$ satisfy $A_i^\dagger A_i = M_i$. 
}
$\{ M_i^{A} \}$ on $\AAA$~\cite{Henderson01}. 
The quantum discord $D_{\AAA}^{\rm \,ent} $ is interpreted as a quantifier of the non-classical
correlations in the bipartite system. 
Note that it is not
symmetric under the exchange of the two parties. One defines the
discord $D_{\BB}^\ent$  analogously, by considering local measurements on subsystem $\BB$. 

The two discords $D_\AAA^\ent$ and
$D_\BB^\ent$ give the amount of mutual
information   that cannot be retrieved by measurements on one of the
subsystems. Actually, it is not difficult to show that:

\begin{proposition}
{\rm ~\cite{Ollivier01}} For any state $\rho \in \states ( \Hh_\AB)$,
\begin{equation} \label{def:discord_bis}
D_{\AAA}^\ent ( \rho ) =  I_{\AAA : \BB} ( \rho) - \max_{\{\Pi^{A}_i \}} I_{\AAA: \BB} \big( 
\Mm_{A}^\Pi \otimes 1 ( \rho)  \big)  \; ,
\end{equation}
where the maximum is over all projective measurements  on $\AAA$  with
rank-one projectors $\Pi_{i}^{A}$ (or with rank-one  operators $M_i^{A}$ if the maximization is taken 
over all POVMs in (\ref{def:discord})) and
\begin{equation} \label{eq-QO_meas}
\Mm_{A}^\Pi \otimes 1 ( \rho) = \sum_{i=1}^{n_A} \Pi_{i}^{A} \otimes 1 \,\rho\,\Pi_{i}^{A} \otimes 1
\end{equation}
is the post-measurement state in the absence of readout.
\end{proposition}

By using (\ref{def:discord_bis}) and the contractivity of the mutual information 
under local quantum operations (data processing inequality), 
one finds that $D_{\AAA}^\ent ( \rho ) \geq 0$ for any state $\rho$.
Furthermore,   $J_{\BB | \AAA} (\rho)=I_{\AAA : \BB} ( \rho) -  D_{\AAA}^\ent ( \rho )$
is equal to the maximum in the \RHS of (\ref{def:discord_bis}) and
 can thus be interpreted as the amount of {\it classical correlations}
between the two parties (in fact, local measurements on $\AAA$
remove all quantum correlations between $\AAA$ and $\BB$). 
One can show that $J_{\BB | \AAA } (\rho) = 0$ \ifif $\rho = \rho_\AAA \otimes \rho_\BB$ is a product state.

It is not difficult to show that $D_{\AAA}^\ent$ and $D_{\BB}^\ent$ coincide  for pure states 
with the von Neumann entropy of the reduced states, \ie, with
the entanglement of formation $E_\EoF$~\cite{Bennett96a,Bennett96},
\begin{equation} \label{eq-Qdiscord_pure_states}
D_{\AAA}^{\rm \,ent} ( \ket{\Psi} ) = D_{\BB}^{\rm \,ent} ( \ket{\Psi} ) = E_\EoF ( \ket{\Psi} )
= S ( [ \rho_\Psi]_A ) = S ( [ \rho_\Psi]_B ) \;.
\end{equation}
In contrast, for mixed
states $\rho$,  $D_{\AAA}^{\rm \,ent} (\rho)$ and $D_{\BB}^{\rm \,ent}(\rho)$  capture quantum correlations different
from entanglement. 
In fact, mixed states can have a non-zero discord even if they are separable.
Such states are obtained by preparing  locally mixtures  
of non-orthogonal states, which cannot be perfectly discriminated by local measurements. 
An example of an $\AAA$- and $\BB$-discordant two-qubit state with no entanglement is
\begin{equation} \label{eq-ex_separable_state_non_zero_discord}
\rho = \frac{1}{4} \bigl( \ketbra{+}{+} \otimes \ketbra{0}{0} + \ketbra{-}{-} \otimes \ketbra{1}{1} + \ketbra{0}{0} \otimes \ketbra{-}{-} 
+ \ketbra{1}{1} \otimes \ketbra{+}{+} \bigr)
\end{equation} 
 with $\ket{\pm} = (\ket{0} \pm \ket{1} )/\sqrt{2}$.
The separable state (\ref{eq-ex_separable_state_non_zero_discord}) cannot be classified as ``classical'' and actually 
contains quantum correlations that are not detected by any entanglement measure.  

It turns out that the 
evaluation of the discord $D_{\AAA}^{\rm \,ent} (\rho)$ for mixed states $\rho$ is a 
challenging task, even for two-qubits~\cite{Girolami2011,Modi_review}. 
For the latter system, an analytical expression has been 
found so far for Bell-diagonal states  only~\cite{Luo08}, while the formula proposed in~\cite{Ali10} for the larger family
of $X$-states happen to be only approximate~\cite{Huang13,Modi_review}. For a
large number of qubits, the computation of the discord is an NP-complete problem~\cite{Huang14}.

\subsection{Classical-quantum and classical states} \label{sec-classical_Q_states}

 States of a
bipartite system $\AB$ with a vanishing quantum discord with respect to $\AAA$ possess only classical correlations. 
They  are usually called classical-quantum states, but we shall prefer here the terminology ``$\AAA$-classical states''. 
One can show that a state $\sigma_\Aclass$ is $\AAA$-classical \ifif it
is left unchanged by a local von Neumann \meas  on $\AAA$ with rank-one projectors
$\Pi_i^A$, \ie, $\sigma_\Aclass= \Mm_{A}^\Pi \otimes 1 ( \sigma_\Aclass )$, where  $\Mm_{A}^\Pi$ is defined
by (\ref{eq-QO_meas})\footnote{
This can be justified by using the identity~(\ref{def:discord_bis}) and 
a theorem due to Petz, which gives a necessary and sufficient condition on $\rho$ such  that
$I_{\AAA : \BB } ( \rho )= I_{\AAA : \BB } ( \Mm_{A} \otimes 1 ( \rho) )$
for a fixed quantum operation $\Mm_{A}$ on $\AAA$
(saturation of the data processing inequality)~\cite{Petz03,Hayden04}. We refer the reader 
to~\cite{my_review_JMP} for more detail. Note that the proof originally given in 
Ref.~\cite{Ollivier01} is not correct.
}.
Therefore, all $\AAA$-classical states are of the form
\begin{equation} \label{eq-A-classical_states}
\sigma_\Aclass = \sum_{i=1}^{n_A}  q_{i}\ket{\alpha_i}\bra{\alpha_i} \otimes \rho_{B | i} \;,
\end{equation}
where $\{ \ket{\alpha_i }\}_{i=1}^{n_A}$ is an orthonormal basis of $\Hh_A$, 
$\{ q_i\}$ is a probability distribution (\ie, $q_i\geq 0$ and
$\sum_i q_i=1$), and $\rho_{B| i}$ are arbitrary states of $\BB$.
Equation  (\ref{eq-A-classical_states}) means that the zero-discord states are mixtures of locally discernable states, 
that is, of states which can be perfectly discriminated by local measurements on $\AAA$.

The $\AAA$-classical states form a non-convex  set
$\Cc_\AAA$, the convex hull of which is the set of  all separable
states $\Ss_\AB$ of the bipartite system. It is clear from (\ref{eq-A-classical_states}) that a pure state 
$\ket{\Phi}$ is $\AAA$-classical \ifif it is a product state $\ket{\Phi} = \ket{\alpha} \ket{\beta}$.
Thus, for pure states classicality is
equivalent to separability, as already evidenced by the relation (\ref{eq-Qdiscord_pure_states}). 
In contrast, most separable mixed states are not
$\AAA$-classical. 

The $\BB$-classical states are defined  analogously as the states with a vanishing discord
with respect to subsystem $\BB$. They are of the form  (\ref{eq-A-classical_states}) with 
$\{ \ket{\alpha_i}\}$ replaced by  an \ONB $\{ \ket{\beta_i }\}$ of $\Hh_\BB$ and
$\rho_{B| i}$ by arbitrary states $\rho_{A| i}$ of $\AAA$.
The states which are both $\AAA$- and $\BB$-classical are called {\it classical states}. They
are of the form
\begin{equation}\label{eq-classical_states}
\sigma_\class = \sum_{i,j=1}^{n_\AAA, n_\BB} q_{ij}\ketbra{\alpha_i}{\alpha_i} \otimes \ketbra{\beta_j}{\beta_j} \; .
\end{equation}
We denote by $\Cc_\BB$ and  $\Cc_\AB = \Cc_\AAA \cap \Cc_\BB$  the sets 
of $\BB$-classical states and of classical states, respectively. 
An illustrative picture of these subsets of the set of quantum states is displayed in Fig.~\ref{fig1}.
Note that this  picture does not reflect all geometrical aspects
  (in particular, $\Cc_\AAA$, $\Cc_\BB$, and $\Cc_\AB$ typically have a lower dimensionality than 
$\states ( \Hh_\AB)$ and $\Ss_\AB$).

\begin{figure}
\begin{center}
\includegraphics[width=8.7cm]{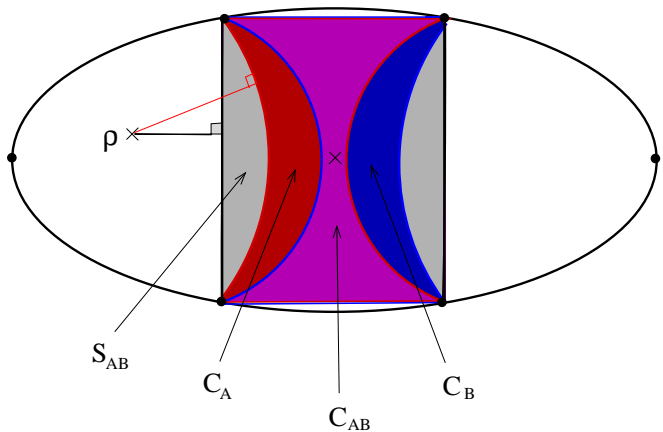}
\end{center}
\caption{
Schematic view of the set of quantum states $\states (\Hh_\AB)$ of a bipartite system $\AB$. The subset $\Cc_\AB$  of classical states (in magenta) 
is the intersection of the subsets $\Cc_\AAA$ and $\Cc_\BB$ of $\AAA$- and $\BB$-classical states  (in red and blue). The convex hull of 
$\Cc_\AAA$ (or $\Cc_\BB$) is the subset $\Ss_\AB$ of separable states  (gray rectangle). All  these subsets intersect the border of 
$\states (\Hh_\AB)$ (which contains all pure states of $\AB$) at the pure product states, represented by the four vertices of the rectangle.
The maximally mixed state $\rho=1/(n_\AAA n_\BB)$ lies at the center (cross).  
The two points at the left and right extremities of the ellipse represent the maximally entangled  pure states, which are the most distant states 
from $\Ss_\AB$ (and also from $\Cc_\AAA$, $\Cc_\BB$, and $\Cc_\AB$).
The square distances between a given state $\rho$ and $\Ss_\AB$ (black line) and  between $\rho$ and $\Cc_\AAA$ (red line)  
define the geometric entanglement $E_\AB^{\rm G} ( \rho)$  and the geometric discord 
$D_{\AAA}^{\rm G} ( \rho)$, respectively.
}
\label{fig1}
\end{figure}

\subsection{Axioms on   measures of quantum correlations} \label{sec-axioms_measure_QC}

Before proceeding further, let us briefly recall the definition of a quantum operation
(or quantum channel).
We denote by $\Bb (\Hh )$ the $C^\ast$-algebra  of bounded linear operators from 
$\Hh $ into itself, that is, $n \times n$ complex matrices with
 $\dim \Hh = n$ in our finite dimensional setting.
 Mathematically, a \QO is
a completely positive (CP)\footnote{
A linear map
$\Mm : \Bb ( \Hh ) \rightarrow \Bb ( \Hh ' )$ is positive if it transforms a non-negative operator $\rho \geq 0$
into a non-negative operator $\Mm ( \rho) \geq 0$. It  is CP if the map
\begin{equation}
\Mm \otimes 1 : X \in \Bb ( \Hh   \otimes \complex^m)\; \mapsto \; \sum_{k,l=1}^m \Mm ( X_{kl} ) \otimes \ketbra{k}{l} \in \Bb ( \Hh ' \otimes \complex^m)
\end{equation}
is positive for any integer $m \geq 1$.
}
trace-preserving linear map $\Mm : \Bb ( \Hh ) \rightarrow \Bb ( \Hh ' )$.
Physically, \QOs represent either quantum evolutions or changes in the system state
due to \meass without readout like in (\ref{eq-QO_meas}).
More precisely,  let a quantum system $\SSS$
initially in state $\rho_\SSS$ be coupled at time $t=0$ to its environment $\EE$, with 
which it has never interacted at prior times. If the joint state $\rho_ \SE (t)$ of $\SE$
either evolves unitarily according to the Schr\"odinger equation or 
is modified by a measurement process, then the reduced state of $\SSS$ at time $t>0$ is 
given by  $\rho_\SSS (t)= \Mm_t ( \rho_\SSS )$ where $\Mm_t$ is a quantum operation.

By studying the properties of the quantum discord, 
one is led to define the following axioms for a 
{\it bona fide} measure of quantum correlations~\cite{Girolami2013,Ciccarello14,Girolami2014,Roga2014,my_review_JMP}.

\begin{definition} \label{def-measure_of_QC}
A measure of quantum correlations on the bipartite system $\AB$  is a
non-negative function $D_A$ on the set of quantum states $\states ( \Hh_\AB)$
such that :
\begin{itemize}
\item[{\rm (i)}] $D_A (\rho)=0$ \ifif $\rho$ is $\AAA$-classical;
\item[{\rm (ii)}] $D_A$ is invariant under local unitary transformations, \ie,  
$D_A ( U_A \otimes U_B \rho\, U_A^\dagger \otimes U_B^\dagger) = D_A ( \rho)$ for any
 unitaries  $U_A$ and $U_B$ acting on $\Hh_\AAA$ and $\Hh_\BB$;
\item[{\rm (iii)}] $D_A$ is monotonically non-increasing under quantum operations on $\BB$, 
\ie, $D_A ( 1 \otimes \Mm_\BB ( \rho) ) \leq D_A (\rho)$ for any \QO 
$\Mm_\BB : \Bb ( \Hh_\BB) \rightarrow \Bb ( \Hh_\BB')$;
\item[{\rm (iv)}] $D_A$ reduces to an entanglement monotone on pure states\footnote{
Recall that an entanglement monotone $E$ on pure states 
is a function which does not increase under Local 
Operations and Classical Communication (LOCC), \ie, $E ( \ket{\Phi} ) \leq E ( \ket{\Psi})$ 
whenever $\ket{\Psi}$ can be transformed into $\ket{\Phi}$ by a LOCC operation~\cite{Nielsen,Horedecki_review}. 
}.
\end{itemize}
\end{definition}

These axioms are satisfied in particular 
by the entropic discord\footnote{  
Actually, $D_A^\ent$ obeys axiom (i) by definition. Axiom (ii) follows
from the unitary invariance of the entropy $S$. Axiom (iv) is a consequence of  
(\ref{eq-Qdiscord_pure_states}) and the entanglement monotonicity of the entanglement of formation. 
The proof of axiom (iii) is given e.g. in~\cite{my_review_JMP}.
}.

It can be shown\footnote{
This follows from the facts that a function $D_A$ on $\states ( \Hh_{AB})$ satisfying (iii) 
is maximal on pure states if $n_A\leq n_B$~\cite{Streltsov12} and that any pure state can be obtained from a maximally
entangled pure state via a LOCC.
}
that axioms (iii) and (iv) imply that, if  
the space dimensions of $\Hh_\AAA$ and $\Hh_\BB$ are such that $n_A \leq n_B$, 
$D_A$ is maximum on maximally entangled pure states $\ket{\Psi_{\rm me}}$, \ie, if  $\rho= \rho_{\Psi_{\rm me}}$
then  $D_A(\rho)=D_{\mmax}$~\cite{Piani14}.
It is argued in Ref.~\cite{Roga_Spehner_Illuminatti2016} that 
a proper measure of \QCs $D_A$ must actually be such that the maximally entangled
states are the {\it only} states satisfying $D_A(\rho)=D_{\mmax}$. 
We will  thus consider the following additional axiom,  
fulfilled in particular by the entropic discord $D_A^\ent$~\cite{my_review_JMP}:

{\it
\begin{itemize}
\item[{\rm (v)}] if  $n_A \leq n_B$
then $D_A (\rho)$ is maximum if and only if $\rho$ is maximally entangled, that is,
$\rho$ has maximal entanglement of formation $E_\EoF ( \rho) = \ln n_A$.
\end{itemize}
}

Many authors have looked for  functions $D_A \not= D^\ent_A$ on $\states ( \Hh_\AB)$ fulfilling axioms (i-iv), 
which  can be used as $D^\ent_A$ to quantify quantum correlations in bipartite
systems while being easier to compute and having operational interpretations.
Among such measures, the distance-based measures studied in this chapter
 are especially appealing since
they provide a geometric understanding of quantum correlations not limited to their quantification,
as stressed in the Introduction.

\section{Geometric measures of quantum correlations} \label{sec-geo_meas_QCs}
\subsection{Contractive distances on the set of quantum states} \label{sec-distances__Q_states}

 A fundamental issue in quantum information theory is the problem of distinguishing quantum states, that is,
quantifying how ``different'' or how ``far from each other'' are   two given  states $\rho$ and $\sigma$.
A natural  way to deal with this problem is to endow  the set of quantum states $\Ee ( \Hh )$ with a distance $d$.  
One has {\it a priori} the choice between many distances.  The most common ones are
the $L^p$-distances
\begin{equation} \label{eq-Lp_norm}
d_p ( \rho, \sigma) = \| \rho-\sigma\|_p = \bigl[ \tr ( |\rho-\sigma|^p ) \bigr]^{\frac{1}{p}}
\end{equation}
with $p \geq 1$ (here $|X|$ denotes the non-negative operator $|X|= \sqrt{X^\dagger X }$).

In quantum information theory, it is important to consider distances $d$ having the following property: 
if two identical systems in states $\rho$ and $\sigma$ undergo the same quantum  evolution
or are subject to the same measurement described by a quantum operation $\Mm$, then the time-evolved or post-measurement states 
$\Mm (\rho)$ and $\Mm (\sigma)$ cannot be farther from each other than the initial states  $\rho$ and $\sigma$. 
In other words, the two states are less distinguishable after the evolution or the measurement, because some information has been lost
in the environment or in the measurement apparatus. 
Distances $d$ on the sets of quantum states satisfying this property are said to be 
{\it contractive under quantum operations} (or ``contractive'' for short). 
More precisely, $d$ is contractive if for any finite-dimensional Hilbert spaces $\Hh$ and $\Hh'$, any quantum operation 
$\Mm : \Bb(\Hh) \rightarrow \Bb(\Hh' )$, and any $\rho$, $\sigma \in \Ee (\Hh)$,
it holds
\begin{equation} \label{eq-def_contractivity}
d ( \Mm (\rho), \Mm (\sigma) ) \leq d (\rho,\sigma)\;.
\end{equation}
Note that a contractive distance is in particular unitary invariant, \ie,
\begin{equation} \label{eq-unitary_invariance}
d \bigl( U \rho\, U^\dagger, U \sigma\, U^\dagger \bigr) = d(\rho,\sigma)
\quad \text{if $U$ is unitary on $\Hh$}
\end{equation}
(in fact, $\rho \mapsto U \rho \, U^\dagger$ is an invertible \QO on $\Bb(\Hh)$).

The relative von Neumann entropy $S ( \rho || \sigma ) =
 \tr [ \rho ( \ln \rho - \ln \sigma ) ]$ is a prominent example of contractive 
function on $\states ( \Hh) \times \states (\Hh)$ and
 has a fundamental interpretation in terms of information.
However, $S ( \rho || \sigma )$ is not a distance 
(it is not symmetric under the exchange of $\rho$ and $\sigma$).
It is desirable
 to work with contractive functions $d$ on $\states ( \Hh) \times \states (\Hh)$  
which can be interpreted  like $S$ in terms of information while being true distances.  
It turns out that the $L^p$-distances $d_p$ are {\it not} contractive, with the notable exception of  
the trace distance $d_1$~\cite{Ruskai94,Ozawa00,Perez-Garcia2006}. Hence $d_p$, $p>1$ (and in particular
the Hilbert-Schmidt distance $d_2$) cannot be reliably used to distinguish
quantum states. 
We will focus in what follows on two particular distances, called the Bures and Hellinger distances, defined by
\begin{eqnarray} 
\label{eq-def_Bures_dist_2}
d_\Bu ( \rho,\sigma) 
& = & \bigl( 2 - 2 \sqrt{F(\rho,\sigma )} \bigr)^{\frac{1}{2}}
\\
\label{eq-Q_Hellinger_distance}
d_{\Hel}(\rho ,\sigma )
& = &  
\big( 2 -2 \tr\sqrt{\rho}\sqrt{\sigma} \big)^\onehalf  \; ,
\end{eqnarray}
where the Uhlmann fidelity  is given by
\be \label{eq-fidelity} 
F(\rho,\sigma) = \| \sqrt{\rho} \sqrt{\sigma} \|_1^2 = \Bigl( \tr \bigl[ (  \sqrt{\sigma} \rho \sqrt{\sigma} )^{\onehalf} \bigr] \Bigr)^2 \;.  
\ee
These distances will be studied in Sec.~\ref{sec-Bures_and_Hellinger_dist}. We will show that they are
contractive, enjoy a number of other nice properties, and are related to the R\'enyi relative entropies.

\subsection{Distances to separable, classical-quantum, and product states} \label{sec-geo_discord} 

From a geometrical viewpoint, it is quite natural to quantify the amount of quantum correlations in a state $\rho$ of a bipartite system $\AB$ by
the distance $d( \rho, \Cc_\AAA)$ of $\rho$ to the subset $\Cc_\AAA$ of  $\AAA$-classical states, \ie, by
the minimal distance between $\rho$ and an arbitrary $\AAA$-classical state (see Fig~\ref{fig1}). 
This idea goes back to Vedral and Plenio~\cite{Vedral97,Vedral98}, who
proposed to define the entanglement in $\AB$ by the (square) distance from $\rho$ 
to the set of separable states $\Ss_\AB$, 
\begin{equation} \label{eq-def_geometric_entanglement}
E^{\rm G}_\AB (\rho) = d ( \rho, \Ss_\AB )^2 = \min_{\sigma_\sep \in \Ss_\AB} d ( \rho,\sigma_\sep )^2 
\;.
\end{equation}
These authors have shown that $E^{\rm G}_\AB$
is an entanglement monotone if the distance $d$ is contractive.
By analogy,  Daki\'c, Vedral, and Brukner~\cite{Dakic10} introduced the geometric discord
\begin{equation} \label{eq-def_geo_discord}
D_\AAA^{\rm G} (\rho) = d ( \rho, \Cc_\AAA )^2 = \min_{\sigma_\Aclass \in \Cc_\AAA} d ( \rho,\sigma_\Aclass )^2 
\; .
\end{equation} 
Unfortunately, the distance $d$ was chosen  in Ref.~\cite{Dakic10} to be the Hilbert-Schmidt distance $d_2$, which
 is not a good choice because $d_2$ is not contractive (Sec.~\ref{sec-distances__Q_states}). 
Further works have studied  the geometric discords based on the more physically reliable
Bures distance (see~\cite{Abad2012,Streltsov2011c,moi_NJP,moi_JPA,Aaronson13}), Hellinger
distance (see~\cite{Marian15,Roga_Spehner_Illuminatti2016}), and  trace distance
(see~\cite{Nakano13,Paula2013,Ciccarello14} and references therein).

The discord $D^{\rm G}_\BB$ relative to subsystem $\BB$ is defined by replacing $\Cc_\AAA$ by $\Cc_\BB$
in (\ref{eq-def_geo_discord}). Like for the entropic discord,  one has in general $D_\AAA^{\rm G} \not= D_\BB^{\rm G}$.
Symmetric measures of quantum correlations are obtained by considering the square distance
$D_\AB^{\rm G} ( \rho) =  d ( \rho, \Cc_\AB )^2$ to the set of classical states $\Cc_\AB = \Cc_\AAA \cap \Cc_\BB$.

We emphasize that since 
$\Cc_\AB \subset \Cc_\AAA \subset \Ss_\AB$ (see Fig.~\ref{fig1}), the geometric measures are ordered as 
\begin{equation} \label{eq-order_geometrical_meas}
E^{\rm G}_\AB  (\rho) \leq D_\AAA^{\rm G} (\rho)  \leq D_\AB^{\rm G} (\rho)\;.
\end{equation} 
This ordering  is a nice feature of the geometrical approach. In contrast, depending on $\rho$,
the 
entanglement of formation $E_\EoF (\rho)$ 
can be larger or smaller than the entropic discord $D_\AAA^{\rm ent} (\rho)$.

It is easy  to show that $E_\AB^{\rm G}$ is
an entanglement monotone  if the distance $d$ is contractive
(this follows from the invariance of $\Ss_\AB$ under 
LOCC operations, see~\cite{Vedral98,my_review_JMP}) and that
$E^{\rm G}_\AB (\rho)=0$ \ifif $\rho$ is separable (since a distance $d$ separates points). Hence 
$E^{\rm G}_\AB$ qualifies as a reliable measure of entanglement\footnote{
Furthermore, $E^{\rm G}_\AB$ is convex if $d$ is the Bures or the Hellinger distance 
since then $d^2$ is jointly convex, see Sec.~\ref{sec-contractivity}. Convexity is sometimes
considered as another axiom for entanglement measures, apart from entanglement monotonicity and vanishing
for separable states and only for those states.
}.
Similarly, one may ask whether
the geometric discord $D_A^{\rm G}$ satisfies axioms (i-iv) 
of Definition~\ref{def-measure_of_QC}.
If $d$ is contractive, one easily shows\footnote{
Actually, 
$D_A^{\rm G}$ clearly obeys axiom (i), irrespective of the choice of the distance. It satisfies 
axiom (ii) for any unitary-invariant distance, thus in particular for contractive distances.
One shows that it fulfills axiom (iii) by using the contractivity of $d$ 
and the fact that the set of $\AAA$-classical states $\Cc_\AAA$ is invariant under quantum operations acting
on $\BB$, as is evident from (\ref{eq-A-classical_states}).
}
 that $D_A^{\rm G}$
obeys the first three axioms (i-iii). Finding general conditions on $d$ insuring the validity
of the last axiom (iv) is still an open question. 
We will  show below that (see Sec.~\ref{sec_pure_states},~\ref{sec-link_QSD}, and~\ref{sec-geo_disc_Hell_mixed_states})

\begin{proposition} \label{prop_geometric_discord_bona_fide_meas_QCs}
$D_A^{\rm G}$ is a {\emph{bona fide}} measure of \QCs when $d$ is the Bures or the Hellinger distance.
Furthermore, if $d=d_\Bu$ then $D_A^{\rm G}$ satisfies the additional axiom (v).
\end{proposition}

It can be proven
that  $D_A^{\rm G}$ obeys axiom (v) also for the Hellinger distance $d_\Hel$ when 
 $\AAA$ is a qubit or a qutrit ($n_\AAA=2$ or $n_\AAA=3$)~\cite{Roga_Spehner_Illuminatti2016}, 
and we believe that this is still true for higher dimensional spaces $\Hh_\AAA$. 
It is conjectured by several authors that $D^{\rm G}_{\AAA}$ 
is a {\it bona fide} measure of quantum correlations for the trace distance $d=d_1$, but as far as we are aware
the justification of  axiom (iv) is still open 
(however, this axiom  holds for $n_\AAA=2$, see e.g.~\cite{Roga_Spehner_Illuminatti2016}). 
In contrast, $D_A^{\rm G}$ is {\it not} a  measure of \QCs for the Hilbert-Schmidt distance $d=d_2$. Indeed, as one
might expect from the fact  
that $d_2$ is not contractive, $D_A^{\rm G}$ does not fulfill axiom (iii)
(an explicit counter-example is given in Ref.~\cite{Piani12}).

One can replace the square distance by the relative entropy $S$
in formulas (\ref{eq-def_geometric_entanglement})-(\ref{eq-def_geo_discord}).
Since $S$ is contractive under \QOs and satisfies  
 $S(\rho || \sigma)=0$ \ifif $\rho=\sigma$, one shows in the same way as for 
contractive distances that  
the corresponding entanglement measure $E^{\rm S}_\AB$ is entanglement monotone~\cite{Vedral97} and that 
the discord $D_A^{\rm S}$ obeys axioms (i-iii). Furthermore, one finds 
that the closest separable state to a pure state $\ket{\Psi}$ 
for the relative entropy is a classical state and that $E^{\rm S}_\AB (\ket{\Psi})$ is equal to 
the entanglement of formation $E_\EoF ( \ket{\Psi})$~\cite{Vedral98}. 
Hence  
$D_A^{\rm S} ( \ket{\Psi})= E^{\rm S}_\AB (\ket{\Psi})=E_\EoF ( \ket{\Psi})$ for any pure state $\ket{\Psi}\in \Hh_\AB$.
As a result, $D_A^{\rm S}$ is a {\it bona fide} measure of quantum correlations.

The mutual information  (\ref{mutual1}) quantifying the total amount of correlations between 
$\AAA$ and $\BB$ is equal to\footnote{
This identity follows from the relations 
$I_{\AAA : \BB} (\rho)= S( \rho || \rho_\AAA \otimes \rho_\BB)$
and $S( \rho || \sigma_\AAA \otimes \sigma_\BB)- S( \rho || \rho_\AAA \otimes \rho_\BB) =
S( \rho_\AAA || \sigma_\AAA) +S( \rho_\BB || \sigma_\BB) \geq 0$.
It means in particular that the ``closest'' product state
to $\rho$ for the relative entropy is the product 
$\rho_\AAA \otimes \rho_\BB$ of the marginals of $\rho$~\cite{Modi10}.
}
%
\begin{equation} \label{eq-mutual_info_and_relative_ent}
I_{\AAA : \BB} ( \rho) 
=  \min_{\sigma_\product \in \Pp_\AB} S ( \rho || \sigma_\product )\;,
\end{equation}
where $\Pp_\AB = \{ \sigma_\AAA \otimes \sigma_\BB ; \sigma_S \in \states (\Hh_S), S = \AAA, \BB \}$ is the set of 
product (\ie, uncorrelated) states.
In analogy with (\ref{eq-mutual_info_and_relative_ent}), 
one can define a geometrical mutual information $I_\AB^{\rm G}$ and a measure $C_\AAA^{\rm G}$ of classical correlations
by~\cite{Modi10,Adesso2014} 
\begin{equation} \label{eq-geo_total_and_classical_correlations}
I_\AB^{\rm G} ( \rho) = d( \rho , \Pp_\AB)^2
\quad , \quad 
C_\AAA^{\rm G} (\rho ) = \;\min_{\sigma_\rho \in \Cc_\AAA} d ( \sigma_\rho , \Pp_\AB)^2\;,
\end{equation}
where the minimum is over all\footnote{
As we shall see below, $\rho$ may have an infinite family of closest $\AAA$-classical states.
}
 closest $\AAA$-classical states $\sigma_\rho$ to $\rho$. Unlike in the case of the entropic discord,
the total correlations $I_\AB^{\rm G} ( \rho) $ is not the sum of the quantum and classical correlations
$D_\AAA^{\rm G} ( \rho)$ and $C_\AAA^{\rm G} ( \rho)$~\cite{Modi10}.
However, the triangle inequality yields
$I_\AB^{\rm G} ( \rho) \leq ( \sqrt{D_\AAA^{\rm G} ( \rho)} + \sqrt{C_\AAA^{\rm G} ( \rho)} )^2$.

\subsection{Response to local measurements and unitary perturbations} \label{sec-disc_response}

An alternative way to quantify \QCs with the help of a distance $d$ is 
to consider the sensitivity of the state $\rho \in \states ( \Hh_\AB)$ to local measurements 
or local unitary perturbations.

\begin{itemize}

\item[(1)] The  distinguishability of $\rho$ with the corresponding post-measurement state
after a local projective  measurement on subsystem $\AAA$ 
is characterized by the {\it measurement-induced geometric discord}, defined by~\cite{Luo_Fu10}
\begin{equation} \label{def:geomdisc2}
D_{A}^{\rm M}  (\rho ) = \min_{\{ \Pi_{i}^A \} } d \bigl(\rho, \Mm^\Pi_A \otimes 1 (\rho) \bigr)^2 \;,
\end{equation}
where the minimum is over all measurements on $\AAA$ with rank-one projectors $\Pi_{i}^{A}$  and 
 $\Mm^\Pi_A$ is the associated \QO (\ref{eq-QO_meas}).
Since the outputs of such measurements are always 
$\AAA$-classical, one has 
$D^{\rm G}_{A} ( \rho ) \leq D^{\rm M}_{A} (\rho)$ for any state $\rho$. This inequality is an equality
if $d=d_2$ is the Hilbert-Schmidt distance~\cite{Luo_Fu10}.
For the Bures and Hellinger distances, $D_{A}^{\rm G} $ and
$D_{A}^{\rm M}$ are in general different, even if $\AAA$ is a qubit~\cite{Roga_Spehner_Illuminatti2016}.
For the trace distance, $D_{A}^{\rm G} =D_{A}^{\rm M}$ when $\AAA$ is a qubit but this is no longer true
for $n_\AAA>2$~\cite{Nakano13}. 

\item[(2)] 
The distinguishability of $\rho$ with the corresponding state 
after a local unitary evolution on subsystem $\AAA$ is characterized by the
 {\it  discord of response}~\cite{Gharibian2012,Giampaolo2013,Roga2014}
\begin{equation} \label{def:quantumn}
D_{A}^{\rm R}  (\rho )  = \frac{1}{\Nn} \min_{U_A , \spec ( U_A) = \E^{\I \Lambda}} d \bigl( \rho,U_A \otimes 1 \,\rho \,U_A^{\dagger}\otimes 
1 \bigr)^2 \, ,
\end{equation}
where the minimum is over all  unitary operators 
$U_A$ on $\Hh_A$ separated from the identity by the condition of having a fixed
spectrum $\spec (U_A) = e^{\I \Lambda} = \{ e^{2 \I \pi j/n_A} ; j=0 ,
\ldots , n_A -1\} $ given by the roots  of unity\footnote{
See~\cite{Roga2014} for a discussion on the choice of the non-degenerate spectrum $\E^{\I \Lambda}$.
}.
  The normalization factor $\Nn$ in (\ref{def:quantumn}) depends on the distance and is chosen such that
$D_{A}^{\rm R} (\rho)$ has maximal value equal to unity.
For instance, $\Nn=2$ for the Bures, Hellinger, and Hilbert-Schmidt
distances and $\Nn=4$ for the trace distance~\cite{Roga_Spehner_Illuminatti2016}. 

\end{itemize}

The measurement-induced \GD and discord of response are special instances of measures of quantumness given by
\begin{equation} \label{eq-measure_quantumness_theo_Piani}
Q_{\delta, \Ff_A} (\rho) = \inf_{\Mm_A \in \Ff_A} \delta ( \rho, \Mm_A \otimes 1 ( \rho))\;,
\end{equation}
where $\Ff_A$ is a family of  \QOs  on $\AAA$ and $\delta$ is a (square) distance or a relative entropy.
The following result  of Piani, Narasimhachar, and Calsamiglia~\cite{Piani14}
is useful to check that such measures are {\it bona fide} measures of quantum correlations.

\begin{theorem} {\rm~\cite{Piani14}} \label{prop_Piani_et_al}
For all spaces $\Hh$ with $\dim \Hh < \infty$, let $\delta (\rho,\sigma)$ 
be non-negative functions on $\states ( \Hh) \times \states ( \Hh)$ which are contractive under \QOs and  
satisfy the `flags' condition 
\begin{equation} \label{eq-flag_condition}
\delta \Big( \sum_i \eta_i \rho_i \otimes \ketbra{i}{i}, \sum_i \eta_i \sigma_i \otimes \ketbra{i}{i} \Big)
=
\sum_i \eta_i \delta ( \rho_i, \sigma_i )\;,  
\end{equation}
where $\{ \ket{i}\}$ is an \ONB of an ancilla Hilbert space $\Hh_\EE$. Assume that $n_\AAA \leq n_\BB$.
Let the family $\Ff_A$ of \QOs  on $\Bb ( \Hh_\AAA)$ be closed 
under unitary conjugations. Then the measure of quantumness (\ref{eq-measure_quantumness_theo_Piani}) 
satisfies axioms (ii-iv) of Definition~\ref{def-measure_of_QC}.
\end{theorem}

{\small

\Proof We first show that $Q_{\delta, \Ff_A}$ is an entanglement monotone when restricted to pure states.
It is known (see e.g.~\cite{Nielsen}) that when $n_\AAA \leq n_\BB$,
a LOCC acting on  a pure state $\ket{\Psi}$ may always be
simulated by a one-way communication protocol involving only three steps: (1) Bob first performs a
\meas with Kraus operators $\{ B_i \}$ on subsystem $\BB$; (2) 
he sends his \meas result to Alice; (3) Alice performs a unitary evolution $U_i$ on subsystem $\AAA$ conditional to Bob's result.
Therefore, it is enough to show that for any pure state $\ket{\Psi} \in \Hh_\AB$,
any family $\{ B_i \}$  of Kraus operators on  $\Hh_\BB$ (satisfying $\sum_i B_i^\dagger B_i = 1$), and any 
family $\{ U_i \}$ of unitaries on $\Hh_\AAA$, it holds
\begin{equation} \label{eq-proof_theorem1}
\sum_i \eta_i Q_{\delta, \Ff_A} ( \ket{\Phi_i}) \leq Q_{\delta, \Ff_A} ( \ket{\Psi} ) 
\;,
\end{equation}
where $\eta_i = \| 1 \otimes B_i \ket{\Psi}\|^2$ is the probability that Bob's outcome is $i$ and
$\ket{\Phi_i}= \eta_i^{-\onehalf} U_i \otimes B_i \ket{\Psi}$ is the corresponding conditional post-\meas  state after
Alice's unitary evolution.
The inequality (\ref{eq-proof_theorem1}) is proven by considering the following \QO 
$\Mm : \Bb ( \Hh_\AB) \rightarrow \Bb ( \Hh_{\AB\EE} )$
\begin{equation}
\Mm ( \rho ) = \sum_{i} U_i \otimes B_i\, \rho \, U_i^\dagger \otimes B_i^\dagger \otimes \ketbra{i}{i}
\;.
\end{equation}
From the contractivity of $\delta$ and the flags condition, one gets
\begin{eqnarray}
\nonumber
 Q_{\delta, \Ff_A} ( \ket{\Psi} ) 
& \geq &
\inf_{\Mm_A \in \Ff_A}  \delta \bigl( \Mm ( \ketbra{\Psi}{\Psi} ) \, , \, \Mm \circ \Mm_\AAA \otimes 1 ( \ketbra{\Psi}{\Psi} ) \bigr)
\\
\nonumber
& = & 
\inf_{\Mm_A \in \Ff_A}  \delta \biggl(  
 \sum_i \eta_i \ketbra{\Phi_i}{\Phi_i} \otimes \ketbra{i}{i} \, , \,
\sum_i \eta_i \Mm_\AAA^{(i)} \otimes 1 (  \ketbra{\Phi_i}{\Phi_i} )  \otimes \ketbra{i}{i}\biggr)
\\
& = & 
\inf_{\Mm_A \in \Ff_A} \sum_i \eta_i \,\delta \Bigl(  \ketbra{\Phi_i}{\Phi_i} \, , \, 
 \Mm_\AAA^{(i)} \otimes 1 (  \ketbra{\Phi_i}{\Phi_i} ) \Bigr)
\end{eqnarray}
with $\Mm_\AAA^{(i)} ( \cdot) = U_i \Mm_\AAA ( U_i^\dagger \cdot U_i ) U_i^\dagger$.
Bounding the infimum of the sum by the sum of the infima and using the 
assumption $U_i \Ff_\AAA U_i^\dagger = \Ff_\AAA$, one is led to the desired result
\begin{equation}
 Q_{\delta, \Ff_A} ( \ket{\Psi} ) 
 \geq 
  \sum_i \eta_i \inf_{\Mm_\AAA^{(i)} \in \Ff_\AAA} \delta \Bigl(  \ketbra{\Phi_i}{\Phi_i}  , 
 \Mm_\AAA^{(i)} \otimes 1 (  \ketbra{\Phi_i}{\Phi_i} ) \Bigr)
= \sum_i \eta_i  Q_{\delta, \Ff_A} ( \ket{\Phi_i}) \;.
\end{equation}
In particular, if the pure state $\ket{\Psi}$ can be transformed by a LOCC operation into the pure state $\ket{\Phi}$,
which means that $\ket{\Phi_i} = \ket{\Phi}$ for all outcomes $i$, 
then $ Q_{\delta, \Ff_A} ( \ket{\Psi} ) \geq  Q_{\delta, \Ff_A} ( \ket{\Phi} )$.
Axiom (ii) follows from a similar argument and the unitary invariance of $\delta$ (which is a consequence of the contractivity
assumption, see Sec.~\ref{sec-distances__Q_states}). Finally, one easily
verifies that $ Q_{\delta, \Ff_A}$ fulfills axiom (iii) by exploiting the contractivity of $\delta$.
\finpro

}

\vspace{1mm}

\begin{proposition} \label{prop-meas_ind_disc_and_disc_resp_are_meas_QCs}
$D_A^{\rm M}$ and $D_A^{\rm R}$ 
are  {\emph{bona fide}} measures of \QCs if the distance $d$ is contractive and 
$d^2$ satisfies the flag condition (\ref{def:quantumn}).
\end{proposition}

It is easy to show that the square Bures and Hellinger distances
$ d_\Bu^2$ and $d_\Hel^2$ satisfy the flags condition, so that
Proposition~\ref{prop-meas_ind_disc_and_disc_resp_are_meas_QCs} applies in particular
to these distances. The result applies to the trace distance $d_1$ as well, see~\cite{Piani14}.

\vspace{2mm}

{\small

\Proof Let us first discuss axiom (i). For $D_A^{\rm M}$, its validity comes from the fact that
a state is $\AAA$-classical if and only if it is invariant under a
von Neumann measurement on $A$ with rank-one projectors (Sec.~\ref{sec-classical_Q_states}).
Note that this axiom would not hold if the minimization in (\ref{def:geomdisc2}) was performed over projectors $\Pi_i^A$
with ranks larger than  one.
For $D_A^{\rm R}$, one uses an equivalent characterization of $\AAA$-classical states
as the states $\rho$ which are left invariant by  a local unitary
transformation on $A$ for some unitary $U_A$ on $\Hh_\AAA$ having a non-degenerate
spectrum~\cite{Roga2014}. Actually, $U_A \otimes 1 \rho U_A^\dagger \otimes 1 = \rho$ \ifif 
$\rho$ commutes with $U_A \otimes 1$, or, equivalently, with all its spectral projectors $\Pi_i^\AAA$.
This means that $\Mm_A^\Pi \otimes 1 (\rho)= \rho$ with $\Mm_A^\Pi$ defined by 
(\ref{eq-QO_meas}).
Since  $\spec(U_A) $ is not degenerate, the spectral projectors $\Pi_i^\AAA$ have rank one. Consequently, the above condition
on unitary transformations is equivalent to the  invariance of $\rho$ under a \meas on $A$ with rank-one projectors and thus to $\rho$ being $\AAA$-classical.
This proves that  $D_A^{\rm R}$ satisfies axiom (i).
The fact that $D_A^{\rm M}$ and $D_A^{\rm R}$ obey the other axioms (ii-iv) is a consequence of Theorem~\ref{prop_Piani_et_al}.
\finpro
}

\vspace{2mm}

As for the geometric discord, we do not have a general argument implying that 
$D_A^{\rm M}$ and  $D_A^{\rm R}$  fulfill the additional axiom (v) under appropriate assumptions on the distance.
However, one can show that $D^{\rm R}_\AAA$ obeys axiom (v) for the Bures, Hellinger, and trace distances,
and  this is also true for $D^{\rm M}_\AAA$ for the Bures distance~\cite{Roga_Spehner_Illuminatti2016}.

\subsection{Speed of response to local unitary perturbations}

All distance-based measures of \QCs defined above are global geometric quantities, in the sense that they
depend on the distance between $\rho$ and states that are not in the neighborhood
of $\rho$ (excepted of course when the measure vanishes).
It is natural to look for quantifiers of \QCs involving  local geometric quantities\footnote{
The word ``local'' refers here to the geometry on $\states ( \Hh_\AB )$ and should not be confused with
the usual notion of locality in quantum mechanics.
}
depending only on the properties of $\states ( \Hh_\AB)$ in the vicinity of $\rho$.
The idea of the sensitivity to local unitary perturbations sustaining the definition of the discord of response
is well suited for this purpose. Indeed, instead of considering the minimal distance between
$\rho$ and its perturbation by a local unitary with a fixed spectrum, one may
consider the minimal speed at time $t=0$ of the time-evolved states 
\begin{equation}
\label{eq-ouput_states}
\rho_{\rm out} (t)= \E^{-\I t H_A \otimes 1} \rho\, \E^{\I t H_A \otimes 1}\;.
\end{equation}
This leads to the definition of the {\it discord of speed of response}
\begin{equation} \label{eq-disc_inf_response}
D_A^{\rm SR} (\rho) = \min_{H_A , \spec ( H_A) = \Lambda} \lim_{t \rightarrow 0} t^{-2} d \big( \rho, \rho_{\rm out} (t) \big)^2
\;,
\end{equation}
where the minimum is over all self-adjoint operators $H_A$ on $\Hh_\AAA$ with a fixed 
non-degenerate spectrum
$\Lambda =  \{ 2 \pi j/n_A ; j=0 , \ldots , n_A -1 \}$.
This local geometric version of the discord of response has apparently not been defined
before in the literature. The results of Propositions~\ref{prop_dic_inf_repsonse_bona_fide_meas_QCs} 
and~\ref{prop_dic_inf_repsonse_Bures_Hellinger} below have up to our knowledge not been published elsewhere.

\begin{proposition} \label{prop_dic_inf_repsonse_bona_fide_meas_QCs}
If the distance $d$ is contractive and Riemannian, and if $d^2$ satisfies the 
flags condition (\ref{eq-flag_condition}), then $D_A^{\rm SR} $ is a {\emph{bona fide}} measure of quantum correlations.
Furthermore, one has
\begin{equation} \label{eq-bound_discord_of_inf_resp}
\Nn D_A^{\rm R} ( \rho) \leq D_A^{\rm SR} (\rho)\;.
\end{equation} 
\end{proposition}

{\small

\Proof
If $d$ is a Riemannian distance then the limit in  (\ref{eq-disc_inf_response}) exists and is equal to
$g_\rho ( - i [ H_A \otimes 1 , \rho], - i [ H_A \otimes 1, \rho])$ where $g$ is the metric associated to $d$
(see Sec.~\ref{sec-charac_contractive_dist}). Since $g_\rho$ is a scalar product, 
$D_A^{\rm SR} (\rho)=0$ \ifif $ [ H_A \otimes 1 , \rho]=0$ for some observable $H_A$ of $\AAA$ with a non-degenerate
spectrum $\spec ( H_A) = \Lambda$. As in the proof of Proposition~\ref{prop-meas_ind_disc_and_disc_resp_are_meas_QCs},
this is equivalent to  $\rho$ being $\AAA$-classical. Hence axiom (i) holds true.
One easily convinces oneself that $D_A^{\rm SR}$ fulfills
axioms (ii) and (iii) by using the unitary invariance and the contractivity of $d$, respectively. 
One deduces from the flags condition  (\ref{eq-flag_condition}) for $\delta=d^2$ that
the metric $g$ satisfies
\begin{equation}
g_{\rho_{\AB \EE}} \bigg( \sum_i \eta_i \dot{\rho_i} \otimes \ketbra{i}{i}
\, ,\,  \sum_i \eta_i \dot{\rho_i} \otimes \ketbra{i}{i} \bigg)
= \sum_i \eta_i g_{\rho_i} ( \dot{\rho_i} , \dot{\rho_i} )
\quad \text{ for } 
\rho_{\AB \EE}= \sum_i \eta_i \rho_i \otimes \ketbra{i}{i}\;.
\end{equation}
Similarly, one infers from the contractivity of $d$ that the metric $g$ satisfies the inequality (\ref{eq-metric_contractive}) below.
By repeating the arguments in the proof of Theorem~\ref{prop_Piani_et_al}, one finds
  that $D_A^{\rm SR}$ is an entanglement monotone for pure states, \ie, it obeys axiom (iv).

The bound (\ref{eq-bound_discord_of_inf_resp}) is a consequence of the triangle inequality and the unitary invariance of $d$.
Actually, one has
\begin{eqnarray}
\nonumber
d \big( \rho, U_A \otimes 1 \rho U_A^\dagger \otimes 1 \big)^2 
& \leq  & 
\lim_{N \rightarrow \infty} 
\biggl\{ \sum_{n=1}^{N} d \Bigl( U_A^{\frac{n-1}{N}} \otimes 1 \rho  \big( U_A^{\frac{n-1}{N}}  \big)^\dagger \otimes 1 , 
 U_A^{\frac{n}{N}} \otimes 1 \rho  (U_A^{\frac{n}{N}})^\dagger \otimes 1 \Bigr) \biggr\}^2
\\
& = & \lim_{N \rightarrow \infty} N^2 d \Bigl( \rho,  U_A^{\frac{1}{N}} \otimes 1 \rho  (U_A^{\frac{1}{N}})^\dagger \otimes 1 \Bigr)^2
= \lim_{t \rightarrow 0} t^{-2} d \big( \rho, \rho_{\rm out} (t) \big)^2 \; ,
\end{eqnarray}
where $\rho_{\rm out} (t)$ is given by (\ref{eq-ouput_states}) and $U_A=e^{\I H_A}$.
\finpro
}

\vspace{2mm}

Note that when subsystem $\AAA$ is a qubit ($n_A=2$),
 the dependence  of $D_A^{\rm SR}$ on the choice of the spectrum $\Lambda$ reduces to a multiplication
factor\footnote{
This is a consequence of the following observations~\cite{Girolami2013}: (a) any $H \in \Bb ( \complex^2)_{\rm sa}$
with spectrum $\{ \lambda_-, \lambda_+\}$ has the form $(\lambda_+-\lambda_-) \sigma_\uv /2+ (\lambda_++\lambda-)/2$,
where  $\sigma_\uv = \sum_{m=1}^3 u_m \sigma_m$, $\uv$ is a unit vector in $\real^3$, and $\sigma_1$, $\sigma_2$, and $\sigma_3$  
are the three Pauli matrices;
(b) as noted in the proof of Proposition~\ref{prop_dic_inf_repsonse_bona_fide_meas_QCs},  the limit in the \RHS of (\ref{eq-disc_inf_response}) is equal to
$g_\rho (- \I [ H_A \otimes 1 , \rho], -\I  [ H_A \otimes 1, \rho])$ where $g_\rho$ is a scalar product.
Hence changing the spectrum $\Lambda$ from $\{ 0, \pi\}$ to $\{ \lambda_-, \lambda_+\}$ amounts to
multiply $D_A^{\rm SR}$ by the constant factor $[( \lambda_{+} - \lambda_-)/\pi ]^2$.
}.
It follows from the physical interpretations  of the Bures and Hellinger metrics (see Sec.~\ref{sec-interpretation_metrics} below) that

\begin{proposition}  \label{prop_dic_inf_repsonse_Bures_Hellinger}
 For invertible density matrices $\rho>0$, the discord of speed of response (\ref{eq-disc_inf_response}) coincides with
\begin{itemize}
\item the {\emph{interferometric power}~\cite{Girolami2014}}  if $d=d_\Bu$ is the Bures distance:
\begin{equation}
D_A^{\rm SR} (\rho) = {\cal P}_\Lambda(\rho) = \frac{1}{4}  \min_{H_A , \spec ( H_A) = \Lambda} \Ff_Q ( \rho, H_A \otimes 1)\;,
\end{equation}
where 
$\Ff_Q ( \rho, H_A \otimes 1)$ is the quantum Fisher information giving the best precision in the estimation
of the unknown parameter $t$  from arbitrary measurements on the output states (\ref{eq-ouput_states}).
\item (twice the) {\emph{Local Quantum Uncertainty (LQU)}~\cite{Girolami2013}} if $d=d_\Hel$ is the Hellinger distance:  
\begin{equation}
D_A^{\rm SR} (\rho) = 2 LQU_\Lambda ( \rho) = 2 \min_{H_A , \spec ( H_A) = \Lambda}  \Ii_{\rm skew} ( \rho, H_A \otimes 1)\;,
\end{equation}
where
$\Ii_{\rm skew} (\rho, H) = - \onehalf \tr ([ \sqrt{\rho} , H]^2 )$ is the skew information~\cite{Wigner1963}
 describing the amount of information on the values of
observables not commuting with $H$ inferred from measurements
on a system in state $\rho$. 
\end{itemize}
\end{proposition}

This proposition shows that  $D_A^{\rm SR}$ has operational interpretations related 
to parameter estimation and to quantum uncertainty in local measurements for the Bures and Hellinger distances, respectively.

If $n_A=2$ and $d = d_\Hel$ is the Hellinger distance, one finds that
$D_A^{\rm R}$ and $D_A^{\rm SR}$ (\ie, the LQU) are proportional to each other,
\begin{equation}  \label{eq-LQU_qubit} 
LQU_{ \{ 0,\pi\} } (\rho) 
= \frac{\pi^2}{4} D_A^{\rm R} (\rho)
= \frac{\pi^2}{4} \inf_{\| \uv \|=1} \Ii_{\rm skew} (\rho,  \sigma_\uv \otimes 1)
\end{equation}
(this follows from the identity $ 2 \Ii_{\rm skew} ( \rho, U) = d_\Hel ( \rho, U \rho U^\dagger )^2$ 
for $U = U^\dagger = U^{-1}$ and from the aforementioned
property of $D_A^{\rm SR}$ with respect to changes in the spectrum $\Lambda$).

\subsection{Different quantum correlation measures lead to different orderings on $\states ( \Hh)$} \label{sec-deferent_orderings}

It may appear as an unpleasant fact that the
 orderings on the states of $\states ( \Hh_\AB)$ defined by the different measures of \QCs  are in general different,
in particular they depend on the choice of the distance $d$. This means that, for instance, 
it is possible to find two states $\rho$ and $\sigma$ which satisfy 
$D^{\rm R}_A (\rho ) < D^{\rm R}_A (\sigma )$ for the Bures distance and 
the reverse inequality $D^{\rm R}_A (\rho ) > D^{\rm R}_A(\sigma )$ for the Hellinger distance.
This is illustrated in  Fig.~\ref{hstr}. This figure displays
 the distributions in the planes defined
by pairs of discords of response based on different
distances for randomly generated two-qubit states $\rho$ with a fixed purity
$P= \tr ( \rho^2)$
(a similar figure would be obtained if the purity was not fixed).
The different orderings translate into the  non-vanishing area of the plane covered by
the distribution,
which in turn reflects the absence of a functional relation between the two discords.
It is quite analogous to the
different orderings on quantum states  established by different
entanglement measures.
More strikingly,  the states
with a fixed purity $P < 1$  which are maximally quantum correlated
for one discord are not 
maximally quantum correlated for another discord, as is also illustrated in Fig.~\ref{hstr}.
The distance-dependent families of such maximally quantum correlated  two-qubit states 
have been determined  in Refs.~\cite{Roga2014,Roga_Spehner_Illuminatti2016} as a function of the purity $P$ 
 for the Bures, Hellinger, and trace distances. Note that if the purity is not fixed, axiom (v) precisely makes sure
that the family of maximally quantum correlated states is universal and is composed of the 
maximally entangled states.

\begin{figure}
\includegraphics[width=17.0cm]{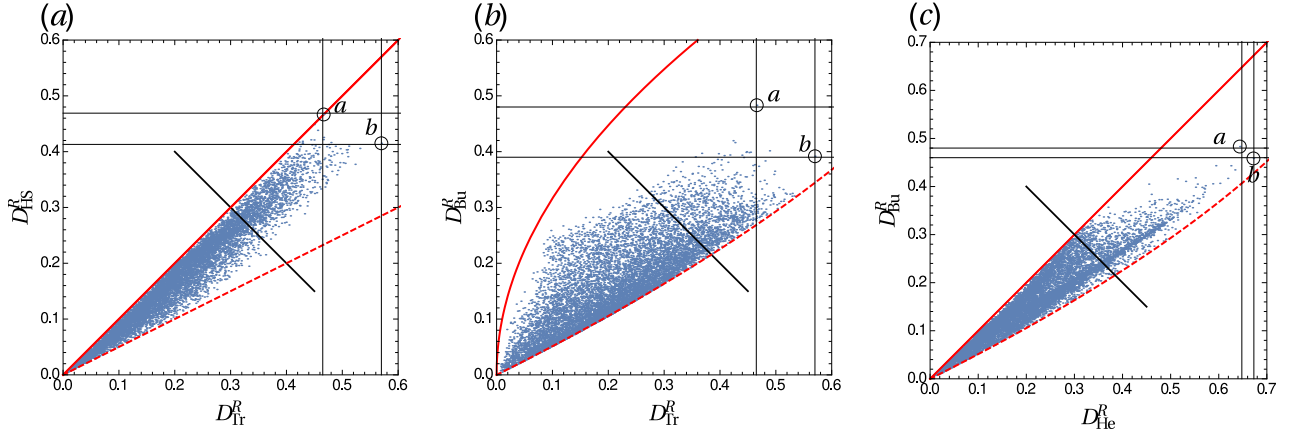}
\caption{(a) Values of the Hilbert-Schmidt and trace  discords of response
  $D^{\rm R}_\HS$ and $D^{\rm R}_{\tr}$ for
  $10^4$ random two-qubit states with the same fixed purity $P=0.6$. These states are generated from random spectra and
eigenvectors given by the column vectors of a unitary
matrix distributed according to the Haar measure.
 States on the thick black line have a hierarchy with respect to
  $D^{\rm R}_{\HS}$ that is reversed compared to the hierarchy with respect
  to $D^{\rm R}_{\tr}$. The points $a$ and $b$ represent,
  respectively, some  states  with purity
  $P$ maximizing $D^{\rm R}_\HS$ and $D^{\rm R}_{\tr}$. Note that $a$ has not
  maximal trace discord of response  $D^{\rm R}_{\tr}$, and
  similarly for $b$ and $D^{\rm R}_\HS$. 
  (b) Same for the Bures and trace discords of response $D^{\rm R}_\Bu$
  and $D^{\rm R}_{\tr}$.  (c) Same for  the Bures and Hellinger discords of response $D^{\rm R}_\Bu$
  and $D^{\rm R}_{\Hel}$. The solid and dashed lines are the borders of the regions  delimited
 by bounds on $D^{\rm R}$ derived from Table~\ref{tab3} below.
This figure is taken from Ref.~\cite{Roga_Spehner_Illuminatti2016}.
}
\label{hstr}
\end{figure}
\newpage

\section{Bures and Hellinger distances} \label{sec-Bures_and_Hellinger_dist}
\label{sec-Bures_distance}

In this section we review the properties of the Bures and Hellinger distances
between quantum states.

\subsection{Bures distance} \label{sec-def-Bures_distance}

The Bures distance was first introduced by Bures in
the context of infinite products of von Neumann
algebras~\cite{Bures69} (see also~\cite{Araki70}) and  was later
studied in a series of papers by
Uhlmann~\cite{Uhlmann76,Uhlmann86}. Uhlmann used it to define
parallel transport and related it to the fidelity generalizing the usual fidelity $| \braket{\psi}{\phi} |^2$
between pure states.  Indeed, the Bures distance is an extension to mixed states
of the Fubini-Study distance for  pure
states.
Recall that the pure states $\rho_\psi= \ketbra{\psi}{\psi}$ of a quantum system with Hilbert space $\Hh$
can be identified with elements of the projective space $P \Hh$, that is,  the set
of equivalence classes  of normalized vectors in $\Hh$ modulo a phase factor. The vectors 
$\ket{\psi_\theta}=\E^{\I \theta} \ket{\psi} \in \Hh$ with $0 \leq \theta < 2 \pi$ are called the representatives of 
$\rho_{\psi}$. The  Fubini-Study distance on $P \Hh$ is
defined by
\begin{equation} \label{eq-def_Fubini_study}
d_{\rm FS} \bigl(  \rho_{\psi}  , \sigma_{\phi}  \bigr) =
\inf_{\| \psi_\theta \|=\|\phi_\delta \|=1} \bigl\|\ket{\psi_\theta} -
\ket{\phi_\delta} \bigr\| = \bigl( 2 - 2 | \braket{\psi}{\phi} |
\bigr)^{\frac{1}{2}} \;,
\end{equation}
where the infimum in the second member is over all representatives $\ket{\psi_\theta}$ of $\rho_{\psi}$ and 
 $\ket{\phi_\delta}$ of $\sigma_{\phi}$. 
Observe that the 
third member depends on the equivalent classes $\rho_\psi$ and $\sigma_\phi$ only.  

For two mixed states $\rho$ and $\sigma \in \states ( \Hh)$, one can define analogously~\cite{Uhlmann86,Hubner92}
\begin{equation} \label{eq-def_Bures_dist_1}
d_\Bu ( \rho,\sigma) = \inf_{R,S} d_2 ( R , S)\;,
\end{equation}
where  $d_2$ is the Hilbert-Schmidt distance and
the infimum is over all Hilbert-Schmidt matrices $R$ and $S \in \Bb ( \Hh)$ 
satisfying $R R^\dagger =\rho$ and $S S^\dagger =\sigma$. Such matrices are
given by $R= \sqrt{\rho} V$ and $S=\sqrt{\sigma} W$ for some unitaries $V$ and $W$ on $\Hh$
(polar decompositions).

\begin{proposition} \label{prop-Bures_distance_is_a_distance}
 $d_\Bu$ defines a distance on the set 
of quantum states $\states ( \Hh)$, which
coincides with the Fubini-Study distance  for pure states.
\end{proposition}

{\small 

\Proof
It is clear on (\ref{eq-def_Bures_dist_1}) that $d_\Bu (\rho,\sigma)$ is  symmetric,
non-negative, and vanishes if and only if
$\rho=\sigma$. To prove the triangle inequality, let us first observe that
by the polar decomposition and the invariance property 
$d_2 ( R V , S V ) = d_2 ( R , S )$ of the Hilbert-Schmidt distance for any unitary $V$, one has
 $d_\Bu ( \rho,\sigma) = \inf_U d_2 ( \sqrt{\rho}, \sqrt{\sigma} U )$ with an infimum over all unitaries $U$.
Let $\rho$, $\sigma$, and $\tau$ be three states in $\states ( \Hh)$. The triangle inequality for 
$d_2$ and the aforementioned invariance property yield
\begin{eqnarray}
\nonumber
d_\Bu ( \rho , \tau) 
& \leq & 
\inf_{U,V} \big\{ d_2 ( \sqrt{\rho}, \sqrt{\sigma} V ) +d_2 ( \sqrt{\sigma} V , \sqrt{\tau} U ) \big\}
= \inf_{V} d_2 ( \sqrt{\rho}, \sqrt{\sigma} V ) + \inf_{W} d_2 ( \sqrt{\sigma}, \sqrt{\tau} W )
\\
& = & 
   d_\Bu ( \rho, \sigma) + d_\Bu ( \sigma, \tau)\; .
\end{eqnarray}
Hence $d_\Bu$ defines a distance on $\states ( \Hh)$. 
For pure states $\rho_{\psi} = \ketbra{\psi}{\psi}$ and $\sigma_{\phi}= \ketbra{\phi}{\phi}$,
the Hilbert-Schmidt  operators 
are of the form
$R = \ketbra{\psi}{\mu}$ and $S =\ketbra{\phi}{\nu}$ with $\| \mu \|=\|\nu \|=1$. A simple calculation then shows that 
the \RHS of  (\ref{eq-def_Fubini_study}) and (\ref{eq-def_Bures_dist_1}) coincide.
\finpro
}

\vspace{2mm}

By using the polar decompositions and the formula
$\| O \|_1=\sup_U \re \tr ( U O)$ for the trace
norm $\| \cdot \|_1$  (the supremum is over all unitaries $U$), one finds
\begin{equation} \label{eq-def_Bures_dist_3}
d_\Bu ( \rho, \sigma) =
\bigl( 2 - 2 \sup_{U} \re \tr ( U
\sqrt{\rho} \sqrt{\sigma} ) \bigr)^{\frac{1}{2}}
= 
\bigl( 2 - 2  \sqrt{F ( \rho , \sigma)}  \bigr)^{\frac{1}{2}}\;,
\end{equation}
where $F ( \rho , \sigma)= \| \sqrt{\rho} \sqrt{\sigma}\|_1^2$ is the Uhlmann fidelity.
Furthermore, the infimum in
(\ref{eq-def_Bures_dist_1}) is attained \ifif the parallel transport
condition $R^\dagger S \geq 0$ holds.  

Since the fidelity $F(\rho,\sigma)$  belongs to $[0,1]$, 
$d_\Bu (\rho,\sigma)$ takes values in the interval  $[0,\sqrt{2}]$. 
Two states $\rho$ and $\sigma$ have a
maximal distance $d_\Bu ( \rho,\sigma)=\sqrt{2}$ (\ie, a vanishing fidelity $F(\rho,\sigma)$) 
\ifif they have orthogonal supports, $\range \rho \,{\bot}\, \range \sigma$. Such orthogonal states 
are thus perfectly distinguishable.

Comparing (\ref{eq-def_Fubini_study}) and  (\ref{eq-def_Bures_dist_3}), one sees that 
the Uhlmann fidelity $F$ is a generalization of
the usual pure state fidelity $F (\ket{\psi}, \ket{\phi})= |\braket{\psi}{\phi} |^2$. More generally,
if $\sigma_\phi$ is pure, then it follows from (\ref{eq-fidelity}) that
\begin{equation} \label{eq-fidelity_pure_state}
F(\rho , \sigma_\phi) = \bra{\phi} \rho \ket{\phi} 
\end{equation}
for any $\rho \in \states ( \Hh)$.
A very useful result due to Uhlmann shows that for any states $\rho$ and $\sigma$,
$F(\rho,\sigma)$ is equal to the fidelity  between two pure states 
$\ket{\Psi}$ and $\ket{\Phi}$ belonging to an enlarged space $\Hh \otimes \Kk$ and having marginals 
$\rho = \tr_\Kk (\ketbra{\Psi}{\Psi})$ and  $\sigma = \tr_\Kk (\ketbra{\Phi}{\Phi})$. Such states 
$\ket{\Psi}$ and $\ket{\Phi}$ are called
{\it purifications} of $\rho$ and $\sigma$ on $\Hh \otimes \Kk$. More precisely, one has

\begin{theorem} { \rm \cite{Uhlmann76}}  \label{theo-Uhlmann}
Let $\rho$, $\sigma \in \states ( \Hh)$ and $\ket{\Psi}$ be a purification of $\rho$ on the Hilbert space $\Hh \otimes
\Kk$, with $\dim \Kk \geq \dim \Hh$. Then
\begin{equation} \label{eq-theo_Uhlmann}
F(\rho,\sigma) = \max_{\ket{\Phi}} | \braket{\Psi}{\Phi} |^2\;,
\end{equation}
where the maximum is over all purifications $\ket{\Phi}$ of $\sigma$ on $\Hh \otimes \Kk$.
\end{theorem}

{\small \Proof
Let us first assume  $\Kk = \Hh$. Then (\ref{eq-theo_Uhlmann}) 
follows from the definition (\ref{eq-def_Bures_dist_1}) of the Bures distance and the
fact that the map $R \mapsto \ket{\Psi_R} = \sum_{i,j} \bra{i} R \ket{j} \ket{i} \ket{j}$ is an isometry between
$\Bb ( \Hh)$ (endowed with the Hilbert-Schmidt norm $\|\cdot \|_2$) and $\Hh \otimes \Hh$
(here $\{ \ket{i}\}$ is some fixed \ONB of $\Hh$). Indeed, one easily checks
that $\rho = R R^\dagger$ if and only if $\ket{\Psi_R}$ is a purification of $\rho$ on $\Hh \otimes \Hh$.
Hence, using  (\ref{eq-def_Fubini_study}), (\ref{eq-def_Bures_dist_1}), and the invariance property of $d_2$ mentioned 
in the proof of  Proposition~\ref{prop-Bures_distance_is_a_distance}, one has
\begin{equation*}
d_\Bu ( \rho, \sigma)^2 = \inf_{S} \big\| R - S \big\|_2^2 = 
\inf_{\ket{\Phi_S}} \big\| \ket{\Psi_R} - \ket{\Phi_S} \big\|^2 = 
\inf_{\ket{\Phi_S}} d_{\rm FS} \big(\ket{\Psi_R}, \ket{\Phi_S} \big)^2 = 2 - 2 \sup_{\ket{\Phi_S}} |\braket{\Psi_R}{\Phi_S}|\;,
\end{equation*}
where the infimum and supremum are over all purifications $\ket{\Phi_S}$ of $\sigma$ on $\Hh \otimes \Hh$, and are actually
minimum and maximum. 

If $\dim \Kk > \dim \Hh$, we extend $\rho$ and $\sigma$ to a larger space ${\Hh}' \simeq \Kk$ 
by adding to them new orthonormal eigenvectors with zero eigenvalues.
As is clear from (\ref{eq-def_Bures_dist_3}), this does not change the distance, 
hence $d_{\rm B}(\rho,\sigma)= \inf_{{R}',{S}'} d_2 ( {R}', {S}')$ with an 
infimum over all ${R}' , {S}' \in  \Bb ({\Hh}')$  such that 
${R}' ({R}')^\dagger$
and ${S}' ({S}')^\dagger$ are equal to the extensions of $\rho$ and $\sigma$.
But ${R}'$ and ${S}'$ can be viewed as operators from $\Kk$ to $\Hh$ since they
have ranges $\range {R}' = \ker ({R}')^\dagger$ and 
$\range {S}' = \ker ( {S}')^\dagger$
included in $\Hh$.
Thus, one can take the infimum in (\ref{eq-def_Bures_dist_1}) over all operators $R,S :\Kk \rightarrow \Hh$ such that
$R R^\dagger= \rho$ and $S S^\dagger = \sigma$, without changing the result. 
The formula (\ref{eq-theo_Uhlmann}) then follows from the same argument as above, using the 
fact that $R \mapsto \ket{\Psi_R}$ is an isometry between the Hilbert space of all operators $\Kk \rightarrow\Hh$ and
$\Hh \otimes \Kk$.
\finpro
}

\vspace{2mm}

A direct proof of (\ref{eq-theo_Uhlmann}) from the definition  $F ( \rho , \sigma)= \| \sqrt{\rho} \sqrt{\sigma}\|_1^2$
of the fidelity has been given Ref.~\cite{Jozsa94} (see also~\cite{Nielsen}).
As the fidelity satisfies $F ( \rho_\AAA \otimes \rho_\BB , \sigma_\AAA \otimes \sigma_\BB ) 
= F( \rho_\AAA , \sigma_\AAA ) F ( \rho_\BB , \sigma_\BB )$, the Bures distance
increases by taking tensor products, \ie, 
\begin{equation} \label{eq-additivity_prop_d_B} 
d_\Bu ( \rho_\AAA \otimes \rho_\BB , \sigma_\AAA \otimes \sigma_\BB )  \geq d_\Bu ( \rho_\AAA , \sigma_\AAA )
\end{equation}
with equality \ifif $\rho_\BB=\sigma_\BB$. 
Note that the trace distance does not enjoy this property.

\subsection{Classical and quantum Hellinger distances} \label{sec-def-Hellinger_distance}

Let  $\states_{\rm clas} = \{ \pv \in \real_{+}^n ; \sum_{k} p_k = 1\}$ be  the simplex 
of classical probability distributions on the finite sample space  $\{ 1, 2,\ldots, n\}$. 
The restriction of a distance $d$ on $\states ( \Hh)$ to all  density matrices commuting with a given 
state $\rho_0$ defines a distance on $\states_{\rm clas}$.
 In particular, if $\rho$ and $\sigma$ are two commuting states with spectral decompositions
  $\rho = \sum_k p_k \ketbra{k}{k}$ and $\sigma = \sum_k q_k \ketbra{k}{k}$, then 
\begin{equation} \label{eq-classical_distance}
d_\Bu ( \rho, \sigma) = d_{\rm clas} ( \pv ,\qv) 
 \equiv \Bigl( 2 - 2 \sum_{k=1}^n \sqrt{p_k q_k} \Bigr)^\onehalf = 
\biggl( \sum_{k=1}^n ( \sqrt{p_k} - \sqrt{q_k} )^2 \biggr)^\onehalf 
\end{equation}
reduces to the classical  Hellinger distance $d_{\rm clas}$ on $\states_{\rm clas}$.
One can of course define other distances on $\states ( \Hh)$ 
which coincide with $d_{\rm clas}$
for commuting density matrices, by choosing a different ordering of the operators inside the trace
in the definition  (\ref{eq-fidelity})   of the fidelity. 
For the ``normal ordering'', one obtains the
 quantum Hellinger distance
\begin{equation} \label{eq-Q_Hellinger_distance2}
d_{\Hel}(\rho ,\sigma )
 =   
\big( 2 -2 \tr\sqrt{\rho}\sqrt{\sigma} \big)^\onehalf  
 = d_2 (\sqrt{\rho }, \sqrt{\sigma }) \;.
\end{equation}
Since  $d_2$ is a distance on $\states ( \Hh)$, this is also the case for $d_{\Hel}$. 
In the sequel, $d_{\Hel}$ will be referred to as the 
Hellinger distance when it is clear from the context
that one works with quantum states and not probability distributions.

Comparing (\ref{eq-def_Bures_dist_1}) and  (\ref{eq-Q_Hellinger_distance2}), one immediately sees that
$d_\Hel ( \rho ,\sigma) \geq d_\Bu ( \rho, \sigma)$ for any states $\rho, \sigma \in \states (\Hh)$.
Like $d_\Bu$, the Hellinger distance satisfies the monotonicity (\ref{eq-additivity_prop_d_B})
under tensor products.
A notable difference between $d_\Bu$ and $d_\Hel$ is that the latter does 
not coincide with the Fubini-Study distance for pure states 
(in fact, 
$d_\Hel ( \rho_\psi, \sigma_\phi ) = (2 - 2  | \braket{\psi}{\phi} |^2)^{\onehalf} > d_{\rm FS} ( \rho_\psi, \sigma_\phi)$ if
$\rho_\psi$ and $\sigma_\phi$ are distinct and non-orthogonal).

One can associate to two non-commuting states $\rho$ and $\sigma$ the probabilities
 $\pv=(p_1,\cdots, p_m)$ and $\qv= (q_1, \cdots,q_m)$ 
of the outcomes of a measurement performed on the system respectively in states $\rho$ and $\sigma$. A natural question is whether 
$d_\Bu( \rho, \sigma)$ or $d_\Hel ( \rho, \sigma)$ coincide
with the supremum of the classical distance $d_{\rm clas} ( \pv,\qv)$ over all such measurements.

\begin{proposition} \label{prop-link_classical_fidelity}
For any $\rho, \sigma \in \states ( \Hh)$, one has
\begin{equation} 
\label{eq-link_Bures_distance_classical}
d_\Bu ( \rho, \sigma ) = \sup_{\{ M_i\}} d_{\rm clas} ( \pv ,\qv)\;,
\end{equation}
where the supremum is over all POVMs $\{ M_i\}_{i=1}^m$ and $p_i = \tr  M_i \rho$ 
(respectively $q_i = \tr  M_i \sigma$) is the probability of
the measurement outcome $i$ in the state $\rho$ (respectively
$\sigma$). The supremum is achieved for von  Neumann measurements with rank-one projectors $M_i = \ketbra{i}{i}$. 
\end{proposition}

A proof of this result and references to the original works can be found in Nielsen and Chuang's book~\cite{Nielsen}.
Note that a similar statement also holds for the trace distance 
(with $d_{\rm clas}$ replaced by the $\ell^1$-distance). In contrast,
while $d_{\rm clas} ( \pv ,\qv) \leq d_{\rm Hel} ( \rho, \sigma )$ for 
 any POVM, the maximum over all POVMs is strictly smaller than $d_{\rm Hel} ( \rho, \sigma )$,
except when $d_{\Hel}(\rho,\sigma)=d_\Bu(\rho,\sigma)$.

\subsection{Contractivity and joint convexity} \label{sec-contractivity}

\begin{proposition} \label{prop-d_B_contractive}
The Bures and Hellinger distances $d_\Bu$ and $d_\Hel$  are contractive under quantum operations.
Moreover, $d_\Bu^2$ and $d_\Hel^2$ are jointly convex, that is, 
\begin{equation}
d_\Bu^2 \Bigl( \sum_i p_i \rho_i , \sum_i p_i \sigma_i \Bigr) \leq \sum_i
p_i d_\Bu^2 (\rho_i,\sigma_i)\;,
\end{equation}
with a similar inequality for $d_\Hel$.
\end{proposition}

The relative entropy $S(\rho || \sigma)$ is also jointly convex. This mathematical property is interpreted
as follows. Given two ensembles
$\{\rho_i, p_i \}$ and $\{\sigma_i ,  p_i \}$ of states in $\states ( \Hh)$ with the same probabilities $p_i$,
by erasing the information about which state of the ensemble is chosen, the state of the system becomes
$\rho = \sum_i p_i \rho_i$ or $\sigma = \sum_i p_i \sigma_i$. The joint convexity means that the
entropy between the two ensembles after the loss of information provoked by the state mixing is smaller 
or equal to
the average of the entropies $S ( \rho_i || \sigma_i)$.
Note that the  $L^p$-distances $d_p$ also fulfill this requirement.
According to Proposition~\ref{prop-d_B_contractive}, the same is true
for the squares of the Bures and Hellinger distances, but not for the
distances themselves.

The contractivity of $d_\Hel$ will be deduced from the following more general result, known as 
Lieb's concavity theorem~\cite{Lieb73a} (see e.g.~\cite{Nielsen} for a proof)\footnote{
The justification by Lieb and Ruskai~\cite{Lieb73} of the strong subadditivity of the von Neumann entropy 
is based on this important theorem.
}.
We denote by $\observables_{+}$ the set of all non-negative operators on $\Hh$. 

\begin{theorem} \label{lemma_Lieb_concavity_Ando_convexity}
{\rm \cite{Lieb73a}}
For any fixed operator $K \in \observables$,
$\beta \in [-1,0]$, and  $q\in [0,  1+\beta]$, the function $ ( \rho, \sigma ) \mapsto  \tr (
K^\dagger \rho^q K \sigma^{-\beta} )$ on $\observables_{+} \times
\observables_{+}$ is jointly concave in $( \rho,\sigma)$.
\end{theorem}

\vspace{2mm}

{\small 
\Proofof{Proposition~\ref{prop-d_B_contractive}} 
Let us first show that $d_\Bu^2$ is  
jointly convex. This  is a consequence of the bound
\begin{equation} \label{eq-concavity_fidelity}
\sqrt{F \Bigl( \sum_i p_i \rho_i , \sum_i q_i \sigma_i \Bigr)} \geq
\sum_i \sqrt{p_i q_i} \sqrt{F ( \rho_i, \sigma_i )}\;.
\end{equation}
To establish (\ref{eq-concavity_fidelity}), we use Theorem~\ref{theo-Uhlmann} and
introduce some
purifications $\ket{\Psi_i}$ of $\rho_i$ and $\ket{\Phi_i}$ of $\sigma_i$ on $\Hh \otimes \Hh$ such that 
$\sqrt{F ( \rho_i, \sigma_i )} = |\braket{\Psi_i}{\Phi_i}|= \braket{\Psi_i}{\Phi_i}$.  Let us define the vectors
\begin{equation}
\ket{\Psi} = \sum_i \sqrt{p_i} \ket{\Psi_i} \ket{i} \quad ,
\quad \ket{\Phi} = \sum_i \sqrt{p_i} \ket{\Phi_i} \ket{i}
\end{equation}
in $\Hh \otimes \Hh \otimes \Hh_\EE$, where $\Hh_\EE$ is an auxiliary Hilbert space with
\ONB $\{ \ket{i} \}$. Then
$\ket{\Psi}$ and $\ket{\Phi}$ are purifications of $\rho = \sum_i p_i \rho_i$ and $\sigma = \sum_i q_i \sigma_i $, 
respectively. Using Theorem~\ref{theo-Uhlmann} again, one finds
\begin{equation}
\sqrt{F ( \rho, \sigma )} \geq | \braket{\Psi}{\Phi} | = \sum_i \sqrt{p_i   q_i} \braket{\Psi_i}{\Phi_i} 
= \sum_i \sqrt{p_i q_i} \sqrt{F ( \rho_i, \sigma_i )}\;.
\end{equation}
We have thus proven that $d_\Bu^2$ is jointly convex.
The joint convexity of  $d_\Hel^2$ is a corollary of  
Theorem~\ref{lemma_Lieb_concavity_Ando_convexity},
which insures that $(\rho,\sigma) \mapsto \tr ( \sqrt{\rho} \sqrt{\sigma})$ is jointly concave. 

The following general argument shows that the
contractivity of $d_\Bu$ and $d_\Hel$ is a consequence of the joint convexity proven above 
and of Stinespring's theorem~\cite{Stinespring55} on CP maps~\cite{Uhlmann73a,Wolf,Frank13}. 
Recall that if $\mu_{\rm H}$ is the normalized Haar measure on the group $U(n)$ of $n \times n$
unitary matrices, then $\int \D \mu_{\rm H} ( U) \,U B U^\dagger = n^{-1} \tr  B$ 
for any $B \in \observables$ (in fact, all diagonal matrix
elements of  $O=\int \D \mu_{\rm H} ( U) \,U B U^\dagger$ in an arbitrary basis are equal, as a result of the
left-invariance of the Haar measure, $\D \mu_{\rm H} ( V U) = \D \mu_{\rm H} ( U)$ for any $V\in U (n)$; thus $O$ is proportional to the identity
matrix).  Let $\Mm$ be a \QO on $\observables$. One infers from the Stinespring theorem
that there exists  a pure state  $\ket{\epsilon_0}$ of an ancilla system $\EE$ and
a unitary $U$ on $\Hh \otimes \Hh_\EE$ such that 
\begin{equation}
\Mm ( \rho) \otimes ( 1 /n_\EE ) = \tr_\EE (  U \rho \otimes \ketbra{\epsilon_0}{\epsilon_0} U^\dagger ) \otimes ( 1 /n_\EE ) =
\int \D \mu_{\rm H} ( U_\EE ) \,(1 \otimes U_\EE) U \rho \otimes
\ketbra{\epsilon_0}{\epsilon_0} U^\dagger (1 \otimes U_\EE^\dagger)\;.
\end{equation}
By using the property 
$d_\Bu ( \rho \otimes \tau, \sigma \otimes \tau) = d_\Bu (\rho,\sigma)$, see (\ref{eq-additivity_prop_d_B}), 
and the joint convexity and unitary invariance  of $d_\Bu^2$, one gets
\begin{eqnarray}
\nn & & d_\Bu^2 ( \Mm(\rho) , \Mm( \sigma) ) = d_\Bu^2 \bigl( \Mm ( \rho) \otimes (1 /n_\EE ) \,,\, 
\Mm (\sigma) \otimes (1 /n_\EE ) \bigr) 
\\ \nn 
&  & \hspace*{1cm} 
\leq \int \D \mu_{\rm H} ( U_\EE ) d_\Bu^2 \bigl( (1 \otimes U_\EE) U \rho \otimes \ketbra{\epsilon_0}{\epsilon_0}  U^\dagger (1 \otimes U_\EE^\dagger) \,,\,  
(1 \otimes U_\EE) U \sigma \otimes \ketbra{\epsilon_0}{\epsilon_0} U^\dagger (1 \otimes U_\EE^\dagger) \bigr) 
\\ & & \hspace*{1cm} =
\int \D \mu_{\rm H} ( U_\EE ) d_\Bu^2 ( \rho , \sigma ) = d_\Bu^2 ( \rho , \sigma ) \;.
\end{eqnarray}
A similar reasoning applies to $d_\Hel$.
\finpro
}

\subsection{Riemannian metrics} \label{sec-Riemannian_metrics}
 
\begin{figure}
\begin{center}
\includegraphics[width=8.5cm]{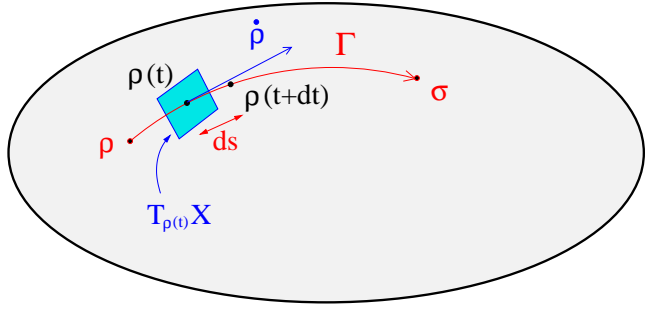}
\end{center}
\caption{
Curve $\Gamma$ joining two states $\rho$ and $\sigma$ in the 
set of quantum states $X=\states (\Hh )$.
}
\label{fig2}
\end{figure}

In Riemannian geometry, a metric on a smooth manifold $X$ is a (smooth) map $g$ 
associating to each point $x$ in $X$ a scalar product $g_x$ on the
tangent space $T_x  X$ at $x$. 
A metric $g$ induces a Riemannian distance $d$, which is such that
the square distance $\D s^2 = d ( x, x + \D x )^2$ between two infinitesimally close points $x$ and $x+ \D x$ is 
equal to $g_x ( \D x, \D x)$. 
For the manifold  $X= \states (\Hh)$ of quantum states,  the tangent spaces
$T_\rho \states (\Hh)$ can be identified with the (real)  vector space
$\saobservables^0$ of self-adjoint operators on $\Hh$ with zero trace.  
A curve $\Gamma$ in $\states ( \Hh)$ joining two states $\rho_0$ and $\rho_1$ is a 
(continuously differentiable) map 
$\Gamma : t \in [0,1] \mapsto \rho(t) \in \states(\Hh)$ with $\Gamma(0)= \rho_0$ and $\Gamma (1) = \rho_1$
(see Fig.~\ref{fig2}). 
Its length $\ell (\Gamma) $ is 
\begin{equation}
\ell (\Gamma) = \int_\Gamma \D s = \int_0^1 \D t \, \sqrt{ g_{\rho(t)} ( \dot{\rho}(t), \dot{\rho}(t) )}\;,
\end{equation} 
where $\dot{\rho}(t)$ stands for the time derivative $\D \rho/\D t$.
A curve $\Gamma_{\rm g}( \rho, \sigma)$  joining $\rho$ and $\sigma$ with the shortest length, or more generally
a curve  $\Gamma_{\rm g}( \rho, \sigma) \in {\cal C}(\rho,\sigma)=  \{ \Gamma \in C^1 ( [0,1], \states ( \Hh)) ; \Gamma(0)=\rho,\Gamma(1)=\sigma\}$ at which the map
 $\Gamma \in  {\cal C}(\rho,\sigma) \mapsto \ell (\Gamma)$ has a 
stationary point, is called
a geodesic.
The distance between two states  $\rho$ and $\sigma$ is the length of the shortest geodesic joining $\rho$ and $\sigma$,
$d ( \rho, \sigma) = \min \{ \ell (\Gamma_{\rm g}( \rho, \sigma) )\} = \min_{\Gamma \in {\cal C}(\rho,\sigma)} \ell (\Gamma)$.
Thanks to this formula, a distance $d$ on $\states ( \Hh)$  can be associated to any metric $g$. 
Conversely, one can associate a  metric $g$ to a distance $d$ 
if the following condition is satisfied (we ignore here the regularity assumptions):  
for any $\rho \in \states ( \Hh)$ and $\dot{\rho} \in \saobservables^0$,
the square distance between $\rho$ and $\rho + t \dot{\rho}$  has 
a small time  Taylor expansion of the form
\begin{equation} \label{eq-definition_metric}
\D s^2 =  d ( \rho, \rho + t \dot{\rho} )^2 =  g_\rho ( \dot{\rho} , \dot{\rho}) t^2 + \Oo ( t^3) \;.
\end{equation}
Needless to say, determining the metric induced by a given distance $d$ 
is much simpler  than finding an explicit formula for  $d(\rho,\sigma)$ 
for arbitrary states  $\rho, \sigma \in \states (\Hh)$
 from the expression of the metric $g$.

A trivial example of metric on $\states ( \Hh)$ is
\begin{equation}
g_\rho( O, O') = \langle O, O' \rangle = \tr (O O') \quad, \quad O,O' \in \saobservables^0\;,
\end{equation}
\ie,  $g_\rho$ is independent of $\rho$ and given by the Hilbert-Schmidt scalar product for matrices.
Introducing an \ONB $\{ \ket{i} \}_{i=1}^n$ of
$\Hh$, one finds that $g_\rho( O, O') =\sum_{i,j=1}^n \overline{O_{ij}} O_{ij}'$ is nothing but
 the Euclidean scalar product. Thus
the geodesics  are straight lines, $\Gamma_{\rm g}( \rho, \sigma) : t \in [0,1] \mapsto (1-t) \rho + t \sigma$,
and the distance 
between two arbitrary states $\rho$ and $\sigma$ is the Hilbert-Schmidt distance
$d_2 ( \rho,\sigma) = \langle - \rho + \sigma , - \rho + \sigma \rangle^{\onehalf} 
= ( \tr [( \rho -\sigma )^2] )^{\onehalf}$.

It is not difficult to show (see~\cite{my_review_JMP}) that the Bures and Hellinger distances 
are Riemannian and have metrics given by
\begin{equation} \label{eq-Bures_metric}
\begin{array}{ccc}
(g_\Bu )_\rho (O,O) & = & \displaystyle \onehalf \sum_{k,l=1}^n \frac{|\bra{k} O \ket{l} |^2}{p_k +
  p_l}
\\
(g_\Hel )_\rho (O,O) & = &  \displaystyle \sum_{k,l=1}^n \frac{ |\bra{k} O \ket{l} |^2}{( \sqrt{p_k} + \sqrt{p_l} )^2}
\end{array}
\quad , \quad O \in \saobservables^0\;,\;\rho >0 \;,
\end{equation}
where  $\{ \ket{k}\}$ is an \ONB of eigenvectors of $\rho$ with eigenvalues $p_k$.
 In contrast, the trace distance $d_1$ is not Riemannian. 
One deduces from (\ref{eq-Bures_metric}) that
\begin{equation} \label{eq-bounds_between_g_Bu_g_Hel}
(g_\Bu )_\rho (O,O) \leq (g_\Hel )_\rho (O,O) \leq 2 (g_\Bu )_\rho (O,O)\;.
\end{equation}
The volume of $\states ( \Hh)$ and the area of its boundary
for the Bures metric have been  determined in Ref.~\cite{Sommers03}.  

\subsection{Physical interpretations of the Bures and Hellinger metrics} \label{sec-interpretation_metrics}

The metrics $g_\Bu$ and $g_\Hel$ have interpretations in quantum metrology and
quantum hypothesis testing. 
Let us first discuss the link with quantum metrology. 
Consider the curve in $\states (\Hh)$ given by 
the unitary evolution of the state $\rho(0)= \rho$ under the Hamiltonian $H \in \saobservables$,
\begin{equation}
\rho (t) = \E^{-\I t H} \rho \,\E^{\I t H}\;.
\end{equation}
Then $\dot{\rho}(t) = - \I [ H , \rho(t) ]$. Assuming that $\rho$ is invertible, the speed  of the state evolution, 
$v (t_0) = \lim_{t \rightarrow 0} t^{-1} d_\Bu ( \rho (t_0), \rho ( t_0 + t))$, 
is given by
$\sqrt{\Ff_Q ( \rho(t_0) , H )}/2 =  \sqrt{\Ff_Q ( \rho , H )}/2$, where
\begin{equation} \label{eq-Fisher_info}
 \Ff_Q ( \rho , H )
 = 4 ( g_\Bu)_{\rho} \bigl( -\I [ H, \rho ], - \I [ H, \rho ] \bigr)
  = 2  \sum_{k,l, p_k+p_l >0} \frac{(p_k-p_l)^2}{p_k + p_l} |\bra{k} H \ket{l} |^2
\end{equation}
is the {\it quantum Fisher information}.
This quantity is related to the smallest error $\Delta t$
that can be achieved when estimating the unknown parameter $t$
by  performing  measurements on the output states
$\rho (t)$. 
Indeed, optimizing over all measurements and all unbiased statistical estimators
(that is, all functions $t_{\rm est} (i_1,\cdots, i_N)$ depending on the measurement outcomes $i_1,\cdots, i_N$ and
such that $\langle t_{\rm est} \rangle = t$), 
the best precision is given by~\cite{Braunstein94}
\begin{equation}
\label{eq-quantum_Cramer_Rao}
(\Delta t)_{\rm{best}}
= \frac{1}{\sqrt{N} \sqrt{\Ff_Q ( \rho , H ) }}\;,
\end{equation}
where $N$ is the number of measurements\footnote{
More precisely, the error $\Delta t = \langle (t_{\rm est}  - t )^2\rangle^{1/2}$ in the parameter estimation is always
larger or equal to $(\Delta t)_{\rm{best}}$ and equality  
is reached asymptotically as $N \rightarrow \infty$ by using  the maximum-likelihood estimator and an optimal measurement.
}.
Note that for pure states $\Ff_Q ( \ket{\psi} , H )= 4 \langle (\Delta H )^2 \rangle_\psi$ 
reduces to the square quantum fluctuation
$\langle (\Delta H )^2 \rangle_\psi  = \bra{\psi} H^2 \ket{\psi} - \bra{\psi} H \ket{\psi}^2$
up to a factor of four.
Hence (\ref{eq-quantum_Cramer_Rao}) takes the form of a generalized uncertainty relation
$(\Delta t)^2 \langle ( \Delta H )^2  \rangle_\psi   \geq 1/4$ (here we take $N=1$), in which
$H$ plays the role of the variable conjugated to the parameter $t$. 
We remark that the second equality in (\ref{eq-Fisher_info}) is only valid when $\rho>0$. 
The quantum Fisher information is, however, given by the last expression in (\ref{eq-Fisher_info}) for any 
state $\rho$. 

The analog of (\ref{eq-Fisher_info}) for the Hellinger metric is the {\it skew information}~\cite{Wigner1963}
\begin{equation} \label{eq-skew_info}
\Ii_{\rm skew} (\rho , H ) 
=\onehalf  ( g_\Hel )_\rho \bigl( -\I [ H, \rho ], - \I [ H, \rho  ] \bigr)
  = - \onehalf \tr \big( [ \sqrt{\rho}\,,\, H ]^2 \bigr) \;.
\end{equation}
It describes the amount of information on the values of
observables not commuting with $H$ in a system in state $\rho$.
The Fisher and skew informations
have the following properties~\cite{Wigner1963,Luo03}:
\begin{itemize}
\item[(a)] 
they are non-negative and vanish \ifif 
$[ \rho, H ] =0$ (this  follows from  the fact that  $(g_\Bu)_\rho$ and $(g_\Hel)_\rho$ are scalar products);
\item[(b)] 
they are convex in $\rho$ (this follows from the joint convexity of $d_\Bu^2$ and $d_\Hel^2$)\footnote{
Actually, let $d$ be a Riemannian distance with metric $g$ such that $d^2(\rho,\sigma)$ is jointly convex. Then 
$g_\rho ( \sum_i p_i O_i, \sum_i p_i O_i ) \leq \sum_i p_i g_{\rho_i} ( O_i, O_i)$ for any
$O_i \in \saobservables^0$ and any $\rho = \sum p_i \rho_i$. In view of their expressions
 (\ref{eq-Fisher_info}) and (\ref{eq-skew_info})  in terms of 
$g_\Bu$ and $g_\Hel$, this implies that the Fisher and skew informations are convex in $\rho$.
}.
\item[(c)] they are additive, \ie, $\Ff_Q ( \rho_\AAA \otimes \rho_\BB, H_\AAA \otimes 1 + 1 \otimes H_\BB )
= \Ff_Q ( \rho_\AAA, H_\AAA) + \Ff_Q ( \rho_\BB , H_\BB)$, with a similar identity for $\Ii_{\rm skew}$;
\item[(d)] the Fisher information is given by~\cite{Davidovich11,Toth13}
\begin{equation} \label{eq-F_Q_as_inf_over_purifications}
\frac{1}{4} \Ff_Q ( \rho, H ) = \inf_{\{ \ket{\psi_i}, \eta_i\}} \bigg\{ \sum_i \eta_i \langle ( \Delta H )^2 \rangle_{\psi_i} \bigg\}
\;,
\end{equation}
where the infimum is over all pure state decompositions $\rho = \sum_i \eta_i \ketbra{\psi_i}{\psi_i}$ of $\rho$;
\item[(e)] they obey the bounds\footnote{
This follows from (\ref{eq-bounds_between_g_Bu_g_Hel}) and, for the last bound, from
(\ref{eq-F_Q_as_inf_over_purifications}) and the concavity 
of $\rho \mapsto \langle ( \Delta H )^2 \rangle_\rho$.
}
 %
\begin{equation} \label{eq-bounds_Fisher_skew_info}
\frac{1}{8} \Ff_Q ( \rho, H ) \leq  \Ii_{\rm skew} (\rho , H ) \leq \frac{1}{4} \Ff_Q ( \rho, H ) \leq 
\langle ( \Delta H )^2 \rangle_\rho  \;,
\end{equation}
where $\langle ( \Delta H )^2 \rangle_\rho = \tr ( \rho H^2) - ( \tr \rho H )^2 $ is the variance of $H$.
The second and third inequalities are equalities for pure states.
\end{itemize}
It can be shown that if the system is composed of $N_{\rm p}$ particles, $H$ is the sum of 
the same  single particle  Hamiltonian $H_{{\rm 1p}}$ acting on each particle,  
and $\Delta h$ is the half difference between the maximal and minimal eigenvalues of $H_{{\rm 1p}}$, then
$\Ff_Q ( \rho, H) > 4 ( \Delta h)^2 N_{\rm p}$ is a sufficient (but not necessary) condition for 
particle entanglement~\cite{Pezze09,my_review_JMP}. Furthermore, 
 high values of  $\Ff_Q  ( {\rho} , H )$ imply multipartite entanglement between  
a large number of particles~\cite{Hyllus12,Toth12}.  

Let us now discuss the link with the hypothesis testing problem. 
This  problem consists in discriminating
two probability measures $\mu_1$ and $\mu_2$ given the outcomes of $N$ independent identically 
distributed random variables with laws given by 
either $\mu_1$ or $\mu_2$. In the quantum setting, this is rephrased as a
discrimination of two states $\rho$ and $\sigma$ given $N$ independent copies of $\rho$ and $\sigma$,
 by means of measurements  on the $N$ copies either in state
$\rho^{\otimes N}$ or $\sigma^{\otimes N}$. One decides among the two alternatives according to the two
possible \meas outcomes. 
According to  the quantum Chernoff bound~\cite{Audenaert07,Nussbaum09},
the  probability of error decays exponentially in the limit $N \rightarrow \infty$, 
with a rate given by a contractive function $\xi ( \rho, \sigma)$, which 
is equal to $g_\Hel ( \D \rho, \D \rho)/2$ for two infinitesimally close states $\rho$ and 
$\sigma = \rho + \D \rho$.

\subsection{Characterization of all Riemannian contractive distances} \label{sec-charac_contractive_dist}

In Ref.~\cite{Petz96}, Petz has determined the general form of all Riemannian contractive distances on 
$\states ( \Hh)$ for finite-dimensional Hilbert spaces $\Hh$.  
Such distances are induced by metrics $g$ satisfying 
\begin{equation} \label{eq-metric_contractive}
g_{\Mm ( \rho)} \bigl( \Mm (O), \Mm ( O) \bigr) \leq g_\rho ( O , O)
\quad , \quad O \in \saobservables^0\;,
\end{equation}
for any $\rho \in \states ( \Hh)$ and any \QO $\Mm : \Bb ( \Hh) \rightarrow
\Bb ( \Hh')$. We recall that  a real function $f: \real_+ \rightarrow \real$
is {\it operator monotone-increasing} if for any space dimension $n=\dim \Hh  < \infty$ and 
any $A, B \in \Bb (\Hh)_+$, one has
$A \leq B \Rightarrow f(A) \leq f ( B)$ (see e.g.~\cite{Bhatia}).

\begin{theorem} {\rm \cite{Petz96}} \label{theo-characterization_cont-metrics}
Any continuous contractive metric $g$ on $\states (\Hh)$ has the form
\begin{equation} \label{eq-characterization_monotone_metrics}
g_\rho ( O , O ) = \sum_{k, l=1}^n c(p_k,p_l) | \bra{k} O \ket{l} |^2 \quad , \quad O \in \saobservables^0\;,
\end{equation}
where $\rho = \sum_k p_k \ketbra{k}{k}$ is a spectral decomposition of $\rho$,
\begin{equation} \label{eq-def_c}
c ( p, q)= \frac{p f (q/p) + q f( p/q)}{2 pq f (p/q) f (q/p)}\;,
\end{equation}
and $f : \real_+ \rightarrow \real_+$ is an operator monotone-increasing function satisfying 
$f(x)=x  f(x^{-1})$. Conversely, the metric defined by (\ref{eq-characterization_monotone_metrics}) are
contractive for any function $f$ with these properties. 
The Bures distance is the smallest of all contractive Riemannian distances with metrics 
satisfying the normalization condition $g_\rho ( 1, 1) = \tr ( \rho^{-1})/4$.
\end{theorem}  

This theorem is of fundamental importance in geometrical approaches to
quantum information. It relies on the fact that from the classical side, 
there exists only one  (up to a normalization factor) contractive metric\footnote{
Here, the contractivity of the classical metrics refers to  Markov mappings 
$\pv \mapsto \Mm^\clas \pv $ on $\states_{\rm clas}$, with
stochastic matrices $\Mm^\clas$ having non-negative elements $\Mm^\clas_{ij}$ such that
$\sum_i \Mm^\clas_{ij} =1$ for any $j=1,\ldots , n$.
}
on  the probability simplex  $\states_{\rm clas}$, namely  
the Fisher metric $\D s^2_{\rm Fisher} =  \sum_{k=1}^n \D p_k^2/p_k$~\cite{Cencov}.
The metric  $\D s^2_{\rm Fisher}$  plays a crucial role in statistics.
It induces the Hellinger distance (\ref{eq-classical_distance}) 
up to a factor of one fourth. 
Therefore, all contractive Riemannian distances on $\states ( \Hh)$ 
satisfying the normalization condition $g_\rho ( 1, 1) = \tr ( \rho^{-1})/4$ coincide
with the classical Hellinger distance  for commuting density matrices.

It can be shown that the following functions are  operator
monotone-increasing:
\begin{equation} \label{eq-monotone_functions_for_contractive_dist}
f_{\rm KM} (x) = 4 \frac{x-1}{\ln x}  \;\;\leq \;\; 
f_\Hel (x)  = ( 1 + \sqrt{x})^2  \;\;\leq\;\;
f_\Bu (x) = 2 (x+1 )\;.
\end{equation}
Substituting them into the formula (\ref{eq-def_c}), we get 
\begin{equation}
c_{\rm KM} (p,q) = \frac{\ln p - \ln q}{4(p-q)} \;\;\geq \;\; 
c_\Hel (p,q) = \frac{1}{(\sqrt{p} + \sqrt{q} )^2} \;\;\geq \;\; 
c_\Bu (p,q) = \frac{1}{2(p+q)}\;.
\end{equation}
In view of  (\ref{eq-Bures_metric}), the last
choice $f_\Bu$ gives the Bures metrics and $f_\Hel$ gives the Hellinger metric.

The first choice in (\ref{eq-monotone_functions_for_contractive_dist}) 
corresponds to the so-called Kubo-Mori (or Bogoliubov) metric, which is associated to the relative entropy.
In fact, an explicit calculation gives~\cite{my_review_JMP}
\begin{equation} \label{eq-Kubo_Mori}
S ( \rho + t \dot{\rho} || \rho  ) = \frac{t^2}{2} (\widetilde{g}_{\rm KM})_\rho (\dot{\rho} , \dot{\rho}) + \Oo ( t^3) = 
S ( \rho||\rho  + t \dot{\rho} ) + \Oo ( t^3)\;,
\end{equation}
where we defined for convenience the metric $\widetilde{g}_{\rm KM} = 4 {g}_{\rm KM}$ satisfying the normalization
condition $(\widetilde{g}_{\rm KM})_\rho (1,1)= \tr (\rho^{-1})$.
As noted in~\cite{Balian86,Balian14}, the Kubo-Mori metric is quite natural from a physical viewpoint because
$\D \widetilde{s}^2_{\rm KM} = - \D^2 S$, 
where $S$ is the von Neumann entropy (since $S$ is concave, its second derivative is non-positive and defines a scalar product
on $\observables$). Actually, one easily deduces from (\ref{eq-Kubo_Mori}) that\footnote{
The first equality is a consequence of (\ref{eq-Kubo_Mori}) and 
the identity $S(\rho + t \dot{\rho} ) = S ( \rho)  - S ( \rho + t \dot{\rho} ||  \rho)- t \tr ( \dot{\rho} \ln \rho )$, and 
the second expression follows from 
$\ln ( \rho + t \dot{\rho}) = \ln \rho + t \int_0^\infty \D u \, ( \rho + u)^{-1} \dot{\rho} 
 ( \rho + u)^{-1} + \Oo ( t^2)$.
}
%
\begin{equation} \label{eq-Kubo_Mori_bis}
(\widetilde{g}_{\rm KM})_\rho (\dot{\rho} , \dot{\rho} ) 
= - \frac{\D^2 S( \rho + t \dot{\rho})}{\D t^2} \biggr|_{t=0}
= \tr \dot{\rho}\, \frac{\D \ln ( \rho + t \dot{\rho} )}{\D t}  \biggr|_{t=0}
\;.
\end{equation}
Let us consider the exponential mapping $\rho \in \states ( \Hh) \mapsto O \in \saobservables$ defined by
\begin{equation}
\rho = \frac{\E^O}{\tr (\E^O) }
\quad \Leftrightarrow \quad 
O - F ( O ) = \ln \rho
\quad \text{ with } \quad 
F(O) = \ln ( \tr \E^O )\;.
\end{equation}
Note that $F(O) - \tr \rho \,O =\tr \rho ( F(O) - O)= S ( \rho)$, hence
$F$ is the Legendre transform of the von Neumann entropy. As a result, 
$\D^2 F = \D^2 S + 2 \tr \D \rho\, \D O = \D \widetilde{s}^2_{\rm KM}$ (the last equality follows from 
$\D \widetilde{s}^2_{\rm KM} = - \D^2 S$, the
last expression of $(\widetilde{g}_{\rm KM})_\rho (\dot{\rho} , \dot{\rho} ) $ in (\ref{eq-Kubo_Mori_bis}), and 
$\tr ( \D {\rho}) =0$).
Hence 
the metric $\widetilde{g}_{\rm KM}$ can also be viewed as the Hessian of the free energy $F$~\cite{Balian14}.
A physical interpretation of the Kubo-Mori metric in terms of information losses in state mixing is as follows: 
the loss of information
when mixing the two
states $\rho_{t}= \rho_0 + t \dot{\rho}$ and $\rho_{-t} =\rho_0 - t \dot{\rho}$ with the same weight $p=1/2$,
 $\Delta S =  S ( \rho_0) -  S ( \rho_{t})/2+ S ( \rho_{-t})/2$,
equals $(t^2/2) (\widetilde{g}_{\rm KM})_{\rho_0} (\dot{\rho} , \dot{\rho})$ in the small $t$ limit. 
We point out that the explicit expression of the Kubo-Mori distance between two arbitrary states
$\rho$ and $\sigma$ is unknown,  except in the case of a single qubit~\cite{Balian14}.

\subsection{Comparison of the Bures, Hellinger, and trace distances} \label{sec_bounds_Bures_dist_and_trace}

One can find explicit bounds between the Bures, trace, and Hellinger distances showing that these distances
define equivalent topologies.

\begin{proposition} \label{prop_bounds_between_d_B_and_d_1}
For any $\rho, \sigma \in \states ( \Hh)$, one has
\begin{eqnarray} \label{eq-bounds_on_d_Hel}
d_\Bu ( \rho,\sigma) &  \leq &  d_\Hel ( \rho,\sigma) \;\;\leq \;\;\sqrt{2}\, d_\Bu ( \rho,\sigma)
\\ \label{eq-bounds_on_d_1} 
 d_\Hel ( \rho,\sigma)^2 & \leq  & 
d_1 ( \rho, \sigma ) \;\;\;\; \leq \;\; 2 \Bigl\{ 1 - \Bigl( 1 - \onehalf d_\Bu ( \rho, \sigma)^2 \Bigr)^2\Bigr\}^\onehalf \;.
\end{eqnarray}
The last inequality in (\ref{eq-bounds_on_d_1})
is saturated for pure states.
\end{proposition}

The bounds $d_\Bu ( \rho,\sigma)^2 \leq d_1 ( \rho, \sigma )$ and  $d_\Hel ( \rho,\sigma)^2 \leq  
d_1 ( \rho, \sigma )$, which are consequences of (\ref{eq-bounds_on_d_Hel}) and (\ref {eq-bounds_on_d_1}),
have been first proven in the
$C^\ast$-algebra setting  by Araki~\cite{Araki70} and Holevo~\cite{Holevo72}, respectively. 
An upper bound  on $d_1 ( \rho, \sigma )$ similar to the one in  (\ref {eq-bounds_on_d_1}) 
but with  $d_\Bu$ replaced by $d_\Hel$
(which is weaker than the bound in  
(\ref {eq-bounds_on_d_1}) because of (\ref{eq-bounds_on_d_Hel}))
 has been also derived by Holevo. 
Lower and upper bounds on the fidelity $F(\rho,\sigma)$ in terms of
traces of polynomials in $\rho$ and $\sigma$, which are easier to compute
than the trace distance and the fidelity itself, have been derived in~\cite{Zyczkowski09}.

\vspace{2mm}

{\small
\Proof 
The inequalities in (\ref{eq-bounds_on_d_Hel}) are consequences 
of the bounds  (\ref{eq-bounds_between_g_Bu_g_Hel}) 
on the Bures and Hellinger metrics. 
The first bound in (\ref{eq-bounds_on_d_1}) can be 
obtained as follows~\cite{Holevo72}.
We set $A=\sqrt{\rho}-\sqrt{\sigma}$ and $B= \sqrt{\rho}+\sqrt{\sigma}$ and consider the polar decomposition $A = U |A|$ with
the unitary $U=P_{+} - P_{-}$, where $P_+$ and $P_{-}=1-P_{+}$ are the spectral projectors of $A$ on $[0,\infty)$ 
and $(-\infty,0)$,  respectively.
Noting that $\rho-\sigma = (AB+BA)/2$, $U A = A U = |A|$, and $|A| P_\pm  = P_\pm |A|$, we obtain by using  
$|\tr U O  |\leq \| O\|_1$ that
\begin{equation} \label{eq-proof_bound_d_H_in_term_of_d1}
\| \rho - \sigma \|_1 
\geq 
  \tr U ( \rho-\sigma) =  
\tr  |A| B  = 
\tr  | A |^\onehalf ( P_{+} B P_{+} + P_{-} B P_{-} ) | A |^\onehalf \;.
\end{equation}
Now $-B \leq A \leq B$, so that  
\begin{equation} \label{eq-proof_bound_d_H_in_term_of_d1_bis}
- A P_{-} = -  P_{-} A P_{-} \leq   P_{-} B P_{-}
\quad , \quad A P_{+} =   P_{+} A P_{+} \leq P_{+} B P_{+}\;.
\end{equation}
 Hence the \RHS of (\ref{eq-proof_bound_d_H_in_term_of_d1}) is bounded from below by
$\tr  |A|^\onehalf A ( P_{+} - P_{-} ) |A|^\onehalf = \tr  A^2 $. This yields $\| \rho - \sigma \|_1 \geq \| \sqrt{\rho} - \sqrt{\sigma} \|_2^2$, that is,
$d_1 ( \rho,\sigma ) \geq d_\Hel ( \rho,\sigma)^2$. 

To prove the last bound in (\ref{eq-bounds_on_d_1}),
we first argue that if $\rho_\psi = \ketbra{\psi}{\psi}$ and $\sigma_\phi = \ketbra{\phi}{\phi}$ are pure states, then 
$d_1 ( \rho_\psi, \sigma_\phi ) = 2 \sqrt{ 1 - F ( \rho_\psi, \sigma_\phi ) }$, showing that this bound holds with equality.  Actually, let 
$\ket{\phi} = \cos \theta \ket{\psi} + \E^{\I   \delta} \sin \theta \ket{\psi^\bot}$, where 
$\theta , \delta \in [0,2\pi)$ and $\ket{\psi^\bot}$ is a unit vector orthogonal to $\ket{\psi}$.  Since $\rho_\psi - \sigma_\phi$ has
non-vanishing eigenvalues $\pm \sin \theta$, one has $d_1 ( \rho_\psi, \sigma_\phi ) = 2 |\sin \theta|$.  
But $F ( \rho_\psi, \sigma_\phi )= \cos^2 \theta$, hence the aforementioned statement is true.  It then follows from
Theorem~\ref{theo-Uhlmann} and from the contractivity of the trace distance under partial traces
that for arbitrary $\rho$ and $\sigma \in \states ( \Hh)$,
\begin{equation}
d_1 ( \rho, \sigma) \leq 2 \sqrt{ 1 - F(\rho,\sigma)}\;.
\end{equation}
This concludes the proof. 
\finpro 
}

\subsection{Relations with the quantum relative R\'enyi entropies} \label{sec-rel_Renyi_entropies}

The R\'enyi entropies $S_\alpha (\rho) = (1-\alpha)^{-1} \ln \tr ( \rho^\alpha )$ 
depending on a parameter
$\alpha >0$ are generalizations of the  von Neumann
entropy $S(\rho)$. For indeed, 
$S_\alpha (\rho)$ converges to $S(\rho)$ when $\alpha \rightarrow 1$. Moreover, 
$S_\alpha (\rho)$ is a non-increasing function of $\alpha$.
Similarly, the relative R\'enyi entropies generalize the relative entropy 
$S(\rho ||\sigma)= \tr [\rho ( \ln \rho -\ln \sigma)]$.
Different definitions have been proposed in
the literature. The ``sandwiched'' relative entropies
studied in~\cite{Mueller-Lennert13,Wilde13} seem to
have the nicer properties.
A family of relative R\'enyi entropies depending on two parameters $(\alpha,z)$, which
 includes the sandwiched entropies (obtained for $z=\alpha$) as special cases,
has been introduced in the context of 
fluctuation relations in quantum
statistical physics~\cite{Jaksic13,Benois16} and was later on studied
from a quantum information perspective~\cite{Audenaert14}.
These entropies are defined when $\ker \sigma \subset \ker \rho$ by
\begin{equation} \label{eq-relative_Renyi_entropy}
S_{\alpha,z} ( \rho || \sigma ) = - \frac{1}{2(1-\alpha)} \ln F_{\alpha,z} ( \rho || \sigma )
\quad , \quad 
 F_{\alpha,z} ( \rho || \sigma )
 = \Bigl( \tr \bigl[ \bigl(
  \sigma^{\frac{1-\alpha}{2 z}} \rho^{\frac{\alpha}{z}} \,\sigma^{\frac{1-\alpha}{2 z}} \bigr)^z \bigr] \Bigr)^2\;.
\end{equation}
Taking $\alpha=z \rightarrow 1$, one recovers the von Neumann relative entropy 
$S( \rho || \sigma )$~\cite{Mueller-Lennert13}. The max-entropy is obtained
in the limit  $\alpha=z \rightarrow \infty$~\cite{Mueller-Lennert13}. 
For commuting matrices $\rho$ and $\sigma$ with eigenvalues $\pv$ and $\qv$,
$S_{\alpha,z} ( \rho || \sigma )$  reduces to the classical R\'enyi divergence
$S_\alpha^{\rm clas} ( \pv || \qv ) = (\alpha-1)^{-1} \ln ( \sum_k p_k^\alpha q_k^{1-\alpha} )$.

It is known that 
$S_{\alpha,z} ( \rho || \sigma )$ is contractive and jointly convex 
when $\alpha \in ( 0,1]$ and $z \geq \max \{ \alpha, 1 - \alpha\}$
(see~\cite{Audenaert14} and references therein) and is contractive
when $\alpha =z\geq 1/2$ (see~\cite{my_review_JMP} and references
therein).  
For those values of $(\alpha,z)$, it is easy to show\footnote{
This follows from the contractivity of  $S_{\alpha,z} ( \rho || \sigma )$
applied to a \meas with rank-one projectors $\{ \ketbra{k}{k} \}$
and the fact that $S_\alpha^{\rm clas} ( \pv || \qv ) \geq 0$ with equality \ifif $\pv = \qv$.
The property is actually true for any $\alpha=z >0$ (see e.g.~\cite{my_review_JMP}) and,
probably, for other values of $(\alpha,z)$.
}
that $S_{\alpha,z} ( \rho || \sigma )\geq 0$ with equality
\ifif $\rho=\sigma$. 
Furthermore, the following monotonicity properties hold:
for any $\rho,\sigma \in \states ( \Hh)$,
$S_{\alpha,\alpha} ( \rho || \sigma )$ is non-decreasing in $\alpha$ on 
$(0,\infty)$~\cite{Mueller-Lennert13} and for any fixed $\alpha \in (0,1)$, 
$S_{\alpha,z} ( \rho || \sigma )$ is non-decreasing in $z$ on $(0,\infty)$
(this follows from the Lieb-Thirring-Araki trace inequality).

We observe that the Bures and Hellinger distances are functions of
the generalized R\'enyi relative entropies $S_{\alpha,z}$
for $(\alpha,z)=( 1/2 , 1/2)$ and $(1/2 , 1)$, respectively. In fact,
\begin{equation}
d_\Bu ( \rho, \sigma)^2 = 2 - 2  \exp \Big\{ - \onehalf S_{1/2,1/2} ( \rho|| \sigma) \Big\}
\quad , \quad 
d_\Hel ( \rho, \sigma)^2 = 2 - 2  \exp \Big\{ -\onehalf S_{1/2,1} ( \rho|| \sigma) \Big\}
\;.
\end{equation}
Thus, $S_{\alpha,z}$ 
connects monotonously and continuously to each other 
 the von Neumann relative entropy $S$, the Bures distance $d_\Bu$, and the Hellinger
distance $d_\Hel$. 

\section{Bures geometric discord} \label{sec-Bures_GD}

In this section we study the Bures geometric discord, obtained by choosing the Bures distance $d=d_\Bu$ in
(\ref{eq-def_geo_discord}),
\begin{equation} \label{eq-max_fidelity} 
D_\Bu^{\rm G} ( \rho) = d_\Bu ( \rho, \Cc_\AAA)^2 =  2  ( 1 - \sqrt{ F (\rho ,\Cc_\AAA)} ) 
\quad , \quad  F (\rho, \Cc_\AAA )= \max_{\sigma_\Aclass \in \Cc_\AAA} F (\rho, \sigma_\Aclass)  \;,
\end{equation}
where $F$ is the fidelity (\ref{eq-fidelity}).
Hereafter, we omit the lower subscript $A$  on all discords,  
as we will always take $A$ as the reference
subsystem. Instead, the chosen distance is
indicated as a lower subscript. The main result of this section is Theorem~\ref{prop_link_geo_discord_QSD}
below, which shows that 
the determination of $D^{\rm G}_\Bu (\rho)$ and of the closest $\AAA$-classical state(s) to $\rho$
are related to a minimal-error quantum state discrimination problem.

\subsection{The case of pure states} \label{sec_pure_states}

 Let us first restrict our attention to pure 
states $\rho_\Psi= \ketbra{\Psi}{\Psi}$, for which a simple formula for
the \GD in terms of the Schmidt coefficients $\mu_i$ of $\ket{\Psi}$  can be obtained. We recall that any pure state 
$\ket{\Psi} \in \Hh_A \otimes \Hh_B$ admits a Schmidt decomposition
\be \label{eq-Schmidt_decomposition}
\ket{\Psi} = \sum_{i=1}^{n} \sqrt{\mu_i} \ket{\varphi_i} \otimes \ket{\chi_i} \;,
\ee
where  $\{ \ket{\varphi_i} \}_{i=1}^{n_A}$ 
(respectively $\{ \ket{\chi_j} \}_{j=1}^{n_B}$) is
an orthonormal basis of $\Hh_A$ ($\Hh_B$) and $n =\min \{ n_A,n_B\}$.
The basis $\{ \ket{\varphi_i} \}$ (respectively $\{ \ket{\chi_j} \}$) and Schmidt coefficients $\mu_i\geq 0$ 
 are the eigenbasis and eigenvalues of the reduced
state $[\rho_\Psi]_A$ (respectively $[\rho_\Psi]_B$).

Let us  show that $D_\Bu^{\rm G} ( \ket{\Psi} )$ is equal to 
the geometric entanglement $E_\Bu^{\rm G} ( \ket{\Psi})$.
In order to calculate the latter, we write the decomposition  of separable states into pure product states, 
$\sigma_\sep = \sum_m q_m \ketbra{\phi_{\AAA}^{m} \otimes \phi_{\BB}^{m} }{\phi_{\AAA}^{m} \otimes \phi_{\BB}^{m} }$
and use the expression (\ref{eq-fidelity_pure_state}) of the fidelity and  $\sum_{m} q_{m} =1$ to get
\begin{eqnarray} \label{eq-max_fidelity_pure_state}
\nonumber
F ( \rho_{\Psi} , \Ss_\AB ) 
 \equiv \max_{\sigma_\sep \in \Ss_\AB} F (\rho_{\Psi} , \sigma_\sep) 
&  = &   \max_{ \{ \ket{\phi^m_\AAA} , \ket{\phi^m_\BB} , q_m \} }
 \Bigl\{ \sum_{m} q_{m} | \braket{\phi_{\AAA}^{m} \otimes \phi_{\BB}^{m}}{\Psi} |^2 \Bigr\}
\\
& = &  
\max_{\| \phi_\AAA \| = \| \phi_\BB \|=1}  \bigl\{ | \braket{\phi_\AAA \otimes \phi_\BB}{\Psi} |^2 \bigr\}  \;.
\end{eqnarray}
For any fixed normalized vectors
 $\ket{\phi_\AAA}\in \Hh_A$ and $\ket{\phi_\BB}\in \Hh_B$, one deduces from (\ref{eq-Schmidt_decomposition}) and the 
Cauchy-Schwarz inequality that 
\begin{eqnarray} \label{eq-bound_by_mu_max}
\nn
|\braket{\phi_\AAA \otimes \phi_\BB}{\Psi} | 
& \leq &  
\sqrt{\mu_{\rm max}}  \sum_{i=1}^n  \bigl| \braket{\phi_\AAA}{\varphi_i} \braket{\phi_\BB}{\chi_i} \bigr| 
\leq \sqrt{\mu_{\rm max}} \biggl( \sum_{i=1}^n | \braket{\phi_\AAA}{\varphi_i} |^2 \biggr)^{1/2} 
   \biggl( \sum_{j=1}^n | \braket{\phi_\BB}{\chi_j} |^2 \biggr)^{1/2}
 \\ \label{eq-bound_by_mu_max2}
& \leq & 
    \sqrt{\mu_{\rm max}} \;,
\end{eqnarray}
where $\mu_{\rm max} = \max _i \mu_i$ is the largest Schmidt 
eigenvalue.
All bounds are saturated by taking $\ket{\phi_\AAA}$ and $\ket{\phi_\BB}$
equal respectively to the eigenvectors $\ket{\varphi_{\mmax} }$  and $\ket{\chi_{\mmax}}$  
of $[ \rho_\Psi ]_\AAA$ and $[ \rho_\Psi ]_\BB$ with maximal eigenvalue $\mu_{\rm max}$.
Thus $F( \rho_\Psi , \Ss_\AB ) = \mu_\mmax$.
Furthermore, 
the pure product state $\ket{\varphi_{\mmax}} \ket{\chi_{\mmax} } $ is a closest separable state
to  $\ket{\Psi}$. Now, a product state is also an $\AAA$-classical state. Since
$d_\Bu ( \ket{\Psi}, \Cc_\AAA) \geq d_\Bu ( \ket{\Psi},  \Ss_\AB )$ (because
$\Cc_\AAA \subset \Ss_\AB  $, see Fig.~\ref{fig1}),
$\ket{\varphi_{\mmax}} \ket{\chi_{\mmax} } $ is also a closest $\AAA$-classical state to $\ket{\Psi}$ and
$D_\Bu^{\rm G} (\rho_\Psi )  = E_\Bu^{\rm G} (\rho_\Psi )$, as claimed above.

\begin{proposition} {\rm~\cite{moi_NJP}} \label{prop_closest_classical_state_pure}
The Bures \GD is given for pure states $\ket{\Psi} \in \Hh_\AB$ by  
\be \label{eq-equality_distances}
D_\Bu^{\rm G} (\ket{\Psi} )  = E_\Bu^{\rm G} (\ket{\Psi} )  
 =  2 ( 1 - \sqrt{\mu_{\rm max}} )\;.
\ee
\begin{itemize}
\item[{\rm (1)}] 
If the maximal Schmidt eigenvalue $\mu_{\rm max}$ is non-degenerate, then
the closest $\AAA$-classical (respectively classical, separable) state to $\rho_{\Psi}$ for 
the Bures distance is unique and given by the pure product 
state $\ket{\varphi_{{\rm max}}}\ket{\chi_{{\rm max}}}$.
\item[{\rm (2)}]
If $\mu_{\rm max}$ is $r$-fold degenerate, say 
$\mu_{\rm max}=\mu_1=\ldots = \mu_r> \mu_{r+1}, \ldots , \mu_{n}$, then $\rho$ has infinitely many
closest $\AAA$-classical (respectively classical, separable) states.
These closest states  are convex combinations of the pure product states 
$\ket{\widehat{\varphi}_{l}}\ket{\widehat{\chi}_{l}}$,
with 
\begin{equation} \label{eq-Schmidt_vectors}
\ket{\widehat{\varphi}_{l}} = \sum_{i=1}^r u_{il} \ket{\varphi_i}
\quad , \quad 
\ket{\widehat{\chi}_{l}}= \sum_{i=1}^r \overline{u_{il}} \ket{\chi_i} 
\quad , \quad l=1,\ldots, r\;,
\end{equation}
where  $\{ \ket{\varphi_i}\}_{i=1}^r$ and $\{ \ket{\chi_i}\}_{i=1}^r $ are some fixed orthonormal  families 
of eigenvectors of $[\rho_\Psi ]_\AAA$ and $[\rho_\Psi ]_\BB$ with eigenvalue $\mu_{\rm max}$ and
$(u_{il})_{i,l=1}^r$ is an arbitrary $r\times r$ unitary matrix.
\end{itemize}
\end{proposition}

The relation (\ref{eq-equality_distances}) is analogous to the equality between 
the entropic discord and the entanglement 
of formation for pure states (Sec.~\ref{sec-def_entropic_discord}).
It comes here from the existence of a pure product state 
which is closer or at the same distance from the pure state $\ket{\Psi}$ than any other separable state. This 
property is a special feature of the Bures distance.

We refer the reader to Refs.~\cite{moi_NJP,my_review_JMP} for a proof 
of statements (1) and (2).
It should be noticed that when $\mu_{\rm max}$ is  degenerate,  the vectors (\ref{eq-Schmidt_vectors}) provide together with
$\ket{\varphi_i}$, $\ket{\chi_i}$, $i=r+1,\ldots, n$,  a Schmidt decomposition of $\ket{\Psi}$ (in that case this decomposition is not unique). 
Conversely, disregarding degeneracies among the other eigenvalues $\mu_i < \mu_\mmax$,
all Schmidt decompositions of $\ket{\Psi}$ are of this form for some unitary matrix $(u_{il})_{i,l=1}^r$.
Thus, the existence of an infinite family of closest $\AAA$-classical states to $\ket{\Psi}$ is related to
the non-uniqueness of the Schmidt vectors associated to $\mu_\mmax$.
This  shows in particular that the maximally entangled pure states (for which $\mu_\mmax$ is $n$-fold degenerate)
are the pure states with the largest  family of closest states\footnote{
This family forms a  $(n^2+n-2)$ real-parameter submanifold of $\states ( \Hh_\AB)$.
}.

The properties of the Bures geometric entanglement $E_\Bu^{\rm G}$ 
have been investigated in~\cite{Vedral98,Wei03,Streltsov10}. 
We have already argued above
that  $E_\Bu^{\rm G}$ 
is an entanglement monotone (Sec.~\ref{sec-geo_discord}). Hence, in view of
(\ref{eq-equality_distances}), the \GD $D_\Bu^{\rm G}$  fulfills axiom (iv) of Definition~\ref{def-measure_of_QC}
and is thus a {\it bona fide} measure of \QCs
(recall that axioms (i-iii) hold for any contractive distance).
One can deduce from the Uhlmann theorem (Theorem~\ref{theo-Uhlmann}) and 
the one-to-one correspondence between
purifications and pure state decompositions of a state $\rho$ that
$F ( \rho, \Ss_\AB)$ is equal to  $\max \sum_i \eta_i F ( \ket{\Psi_i} ,\Ss_\AB )$, 
the maximum being over all  pure state decompositions  
$\rho=\sum_i \eta_i \ketbra{\Psi_i}{\Psi_i}$ of $\rho$ (convex roof)~\cite{Streltsov10}.

\subsection{Link with quantum state discrimination} \label{sec-link_QSD}

As for all other  measures of quantum correlations, determining $D_\Bu^{\rm G} (\rho)$ is harder for mixed states  
$\rho$ than for pure states. Interestingly, this problem is related  to an ambiguous \QSD task.

The objective of \QSD is to distinguish
 states taken randomly from a known ensemble of states~\cite{Helstrom,Bergou_review,my_review_JMP}.
If these states are  not orthogonal, any measurement devised to distinguish them cannot succeed
to identify exactly which state from the ensemble has been chosen. 
The \QSD problem is to find the optimal 
measurement leading to the smallest probability of equivocation.
More precisely, a receiver is given a state $\rho_i \in \states ( \Hh)$  drawn from a known ensemble
$\{ \rho_i , \eta_i \}_{i=1}^{n_A}$ with a prior probability $\eta_i$.
In order  to determine which state he has received, he performs a measurement given by a POVM $\{ M_i\}$
and concludes that the state is $\rho_j$ when he gets the measurement outcome $j$.
The probability of this outcome given that the state is
$\rho_i$ is $p_{j|i}=\tr  M_j \rho_i$. 
In the ambiguous (or minimal-error) strategy, the number of 
measurement outcomes is chosen to be equal to the number of states in the ensemble $\{ \rho_i,\eta_i\}$.  
The maximal success probability of the receiver reads 
\begin{equation} \label{eq-max_success_proba_POVM}
P_{\rm S}^{\,\rm{opt}} ( \{ \rho_i,\eta_i\}) =\max_{{\rm POVM}\; \{ M_i \} } \sum_{i=1}^{n_A} \eta_i \tr  M_i \rho_i\;.
\end{equation}
If the $\rho_i$ span $\Hh$ 
 and are linearly independent, in the sense that their eigenvectors $\ket{\xi_{ij}}$
with nonzero eigenvalues 
form a linearly independent family $\{ \ket{\xi_{ij}} \}_{i=1, \ldots , n_A}^{j=1,\ldots, n_B}$
of vectors in $\Hh$, it is known that the optimal POVM 
 is a von Neumann measurement with projectors of rank $r_i = \rank (\rho_i)$~\cite{Eldar03}. In that case,
the maximal success probability $P_{\rm S}^{\,\rm{opt}} ( \{ \rho_i,\eta_i\})$ is equal to
\begin{equation} \label{eq-max_success_proba_von_Neumann}
P_{\rm S}^{\,\rm{opt\,v.N.}} ( \{ \rho_i,\eta_i \})
= \max_{ \{ \Pi_i \} } \sum_{i=1}^{n_A} \eta_i \tr  \Pi_i \rho_i \;,
\end{equation}
the maximum being over all projective measurements with projectors $\Pi_i$ of rank $r_i$.

\begin{theorem} {\rm ~\cite{moi_NJP}} \label{prop_link_geo_discord_QSD}
For any state $\rho$ of the bipartite system $\AB$,
the largest fidelity between $\rho$ and an $\AAA$-classical state reads 
\begin{equation} \label{eq-variationnal_formula_bis} 
F (\rho, \Cc_\AAA ) = \max_{\{ \ket{\alpha_i} \} } P_{\rm S}^{\,\rm{opt\,v.N.}} ( \{ \rho_i,\eta_i \} )\;,
\end{equation}
where the maximum is over all orthonormal bases $\{ \ket{\alpha_i} \}_{i=1}^{n_\AAA}$ of $\Hh_\AAA$ 
and $P_{\rm S}^{\,\rm{opt\,v.N.}} ( \{ \rho_i,\eta_i \})$ is the maximal success probability in discriminating the 
states $\rho_i$ by 
von Neumann measurements on $\AB$ with projectors of rank $n_B$. Here, the 
states $\rho_i$ and probabilities $\eta_i$ depend on $\{ \ket{\alpha_i} \}_{i=1}^{n_\AAA}$ and are given by 
\begin{equation} \label{eq-state_Q_discrimination}
\eta_i = \bra{\alpha_i}  \rho_\AAA \ket{\alpha_i} \quad , \quad 
\rho_i = \eta_i^{-1} \sqrt{\rho} \ketbra{\alpha_i}{\alpha_i} \otimes 1 \sqrt{\rho}
\quad , \quad i=1 , \ldots , n_\AAA\;.
\end{equation}
Furthermore, the closest $\AAA$-classical states to $\rho$ are given by
\begin{equation} \label{eq-again_I_was_stupid}
\sigma_{\Bu , \rho} = \frac{1}{F(\rho, \Cc_\AAA)} 
\sum_{i=1}^{n_\AAA} \ketbra{\alpha_i^{\opt}}{\alpha_i^{\opt}} 
\otimes \bra{\alpha_i^{\opt}} \sqrt{\rho}\, \Pi_i^{\opt} \sqrt{\rho} \ket{\alpha_i^{\opt}}  \;,
\end{equation}
where $\{ \ket{\alpha_i^{\opt}} \}$ is an \ONB of $\Hh_\AAA$ maximizing  
the \RHS of (\ref{eq-variationnal_formula_bis}) and  $\{ \Pi_i^{\rm{opt}} \}$ is an optimal measurement
with projectors of rank $n_\BB$ maximizing the success probability in (\ref{eq-max_success_proba_von_Neumann}).
\end{theorem}

We postpone the proof of this theorem to Sec.~\ref{sec-proof_link_GD_QSD} and proceed with
a few comments and consequences of the theorem.
Firstly, the $\rho_i$ are quantum states because $\rho_i \geq 0$  and $\eta_i$ is chosen such that $\tr \rho_i=1$
(if $\eta_i = 0$ then $\rho_i$ is not defined but does not contribute to the sum in
(\ref{eq-max_success_proba_von_Neumann})).  
Secondly, the $\eta_i$ are the  outcome probabilities  of a \meas on $\AAA$ with rank-one projectors 
$\Pi_i^\AAA = \ketbra{\alpha_i}{\alpha_i}$, see (\ref{eq-post_meas_cond_states_B}). 
Denoting by $\rho_{\AB|i} = \eta_i^{-1} \Pi_i^\AAA \otimes 1 \rho \,\Pi_i^\AAA \otimes 1$
the corresponding  conditional states of 
$\AB$ and by  $\Mm_\AAA^\Pi$ the associated quantum operation on $\AAA$, see (\ref{eq-QO_meas}), 
we remark that $\rho_i = \Rr_{\Mm_\AAA^\Pi , \rho } ( \rho_{\AB|i})$  
is the image of $\rho_{\AB|i}$ under the Petz transpose operation $\Rr_{\Mm_\AAA^\Pi , \rho }$, 
that is, the approximate reversal operation of  $\Mm_\AAA^\Pi \otimes 1$ with respect to $\rho$
(see~\cite{my_review_JMP} for more detail).
Now, $\Mm_\AAA^\Pi \otimes 1 ( \rho)= \sum_i \eta_i \rho_{\AB|i}$ and, by definition of the transpose operation,
$\Rr_{\Mm_\AAA^\Pi , \rho } \circ \Mm_\AAA^\Pi \otimes 1 ( \rho)= \rho$. Thus
$\rho = \sum_i \eta_i \rho_i$, so that
the ensemble $\{ \rho_i,\eta_i\}_{i=1}^{n_\AAA}$ gives a convex decomposition of $\rho$
(this can also be checked directly on (\ref{eq-state_Q_discrimination})).
Another notable property of this ensemble is that the  least square measurement\footnote{
This \meas   bears several other names: it is referred to as the
``pretty good measurement'' in~\cite{Hausladen94} and is sometimes also called ``square-root measurement''~\cite{Eldar01}. 
For a pure state ensemble $\{ \ket{\psi_i} , \eta_i\}$, 
it is given by 
$\{ M_i^{\rm lsm} = \ketbra{\widetilde{\mu}_i}{\widetilde{\mu}_i} \}$ and the vectors
$\ket{\widetilde{\mu}_i} = \sqrt{\eta_i} (\sum_j \eta_j \ketbra{\psi_j}{\psi_j})^{-\onehalf} \ket{\psi_i}$  
are such that they minimize the sum of the square norms 
$\| \ket{\widetilde{\mu}_i} - \sqrt{\eta_i} \ket{\psi_i} \|^2$ under the constraint that $\{ M_i^{\rm lsm}\}$ is a POVM,
\ie, $\sum_i \ketbra{\widetilde{\mu}_i}{\widetilde{\mu}_i} = 1$~\cite{Holevo78}.  
}
associated to it, defined by the POVM $\{ M_i^{\rm lsm}\}$ with
\begin{equation}
M_i^{\rm lsm}= \eta_i \rho^{-1/2}  \rho_i \rho^{-1/2} \quad , \quad i=1, \cdots, n_\AAA \;,
\end{equation}
coincides with
$\{ \ketbra{\alpha_i}{\alpha_i} \otimes 1\}$.

\begin{corollary} \label{cor-case_rho_invertible}
If $\rho$ is invertible then
  one can substitute $P_{\rm S}^{\,\rm{opt\,v.N.}} ( \{ \rho_i,\eta_i \})$ in (\ref{eq-variationnal_formula_bis}) by
the maximal success probability $P_{\rm S}^{\opt} ( \{ \rho_i,\eta_i \})$  over all POVMs, given by (\ref{eq-max_success_proba_POVM}).
\end{corollary}

{\small 

\Proof If  $\rho>0$ then the states $\rho_i$ defined in (\ref{eq-state_Q_discrimination}) are linearly independent, thus
the optimal \meas to discriminate them is a von Neumann \meas with projectors of rank 
$r_i=\rank ( \rho_i )$ (see above). 
The linear independence can be justified as follows. 
Let us first notice that  $\rho_i$ has
rank $r_i = n_\BB$ (for indeed, it  has the same rank as 
$\eta_i \rho^{-1/2} \rho_i = \ketbra{\alpha_i}{\alpha_i} \otimes 1 \sqrt{\rho}$). A necessary and sufficient condition for 
$\ket{\xi_{ij}}$ to be an eigenvector of $\rho_i $  with eigenvalue $\lambda_{ij}>0$ is 
$\ket{\xi_{ij}}= (\lambda_{ij} \eta_i )^{-1} \sqrt{\rho} \ket{\alpha_i}\otimes \ket{\zeta_{ij}}$, where
$\ket{\zeta_{ij}} \in \Hh_\BB$ is an eigenvector of $R_i= \bra{\alpha_i} \rho \ket{\alpha_i}$
with eigenvalue $\lambda_{ij} \eta_i >0$. 
For any $i$, the  Hermitian invertible matrix
 $R_i$ admits an orthonormal 
eigenbasis $\{ \ket{\zeta_{ij}} \}_{j=1}^{n_\BB}$. Thanks to the invertibility of $\sqrt{\rho}$,
$\{ \ket{\xi_{ij}} \}_{i=1, \ldots , n_\AAA}^{j=1,\ldots, n_\BB}$ is a basis of $\Hh_\AB$ and thus
the states $\rho_i$ are linearly independent and span $\Hh_\AB$.
\finpro

}

\subsection{Quantum correlations and distinguishability of quantum states} \label{eq-interpretation}

We give in this subsection a physical interpretation of Theorem~\ref{prop_link_geo_discord_QSD}. We start by 
discussing the state discrimination problem in the special cases where
$\rho$  is either pure or $\AAA$-classical.
Of course, the values of $D_\Bu^{\rm G} ( \rho)$ are already known in these cases
(they are given by (\ref{eq-equality_distances}) and by
$D_\Bu^{\rm G} ( \rho)=0$, respectively), but it is instructive to recover that from  Theorem~\ref{prop_link_geo_discord_QSD}.

\vspace{1mm}

(a) If $\rho=\rho_\Psi$ is pure then all states $\rho_i$ with $\eta_i >0$ are identical and equal 
to $\rho_\Psi$, so that 
$P_{\rm S}^{\,\rm{opt\,v.N.}} = \max_{\{ \Pi_i\}} \{ \sum_i \eta_i \bra{\Psi} \Pi_i \ket{\Psi} \} = \eta_{\rm max}$.
One gets $F (\rho_\Psi, \Cc_\AAA )=\mu_{\rm max}$ by optimization over the basis $\{ \ket{\alpha_i} \}$.

\vspace{1mm}

(b)
If $\rho$ is an  $\AAA$-classical state, \ie, if it can be decomposed as in 
(\ref{eq-A-classical_states}), then the optimal basis $\{ \ket{\alpha_i^\opt}\}$ coincides
with the basis appearing in this decomposition. With this choice one obtains $\eta_i = q_i$ and 
$\rho_i = \ketbra{\alpha_i}{\alpha_i} \otimes \rho_{\BB|i}$ for all $i$ such that $q_i >0$. 
The states $\rho_i$ are orthogonal and can thus be perfectly discriminated by von Neumann measurements. 
This yields
$F(\rho,\Cc_\AAA )=1$ and $D_\Bu^{\rm G} ( \rho)=0$ as it should be.
Reciprocally, if $F (\rho, \Cc_\AAA )=1$ then $P_{\rm S}^{\,\rm{opt\,\vN}}( \{ \rho_i,\eta_i \})=1$ for some
basis $\{ \ket{\alpha_i} \}$ and  the corresponding $\rho_i$ must be orthogonal. Hence one can find  an 
orthonormal family $\{ \Pi_i\}$ of   projectors with rank  $n_\BB$ such that
$\rho_i = \Pi_i \rho_i \Pi_i$ for any $i$ with $\eta_i >0$. It is an easy exercise to show that this implies that
$\Pi_i = \ketbra{\alpha_i}{\alpha_i} \otimes 1$ if $\rho|_{\Pi_i \Hh_\AB}$ is invertible. Thus $\rho= \sum_i \eta_i \rho_i $ is 
$\AAA$-classical, in agreement with  
axiom (i). 

\vspace{1mm}

These special cases help us to  interpret Theorem~\ref{prop_link_geo_discord_QSD}
in the following way.
The discordant states $\rho$ are characterized by ensembles    $\{\rho_i, \eta_i\}$ 
of non-orthogonal states, which are thereby not perfectly distinguishable for any orthonormal basis
$\{ \ket{\alpha_i} \}$ of the reference system\footnote{
Note  that the entropic discord can also
be interpreted in terms of
 state distinguishability, but for states of subsystem $\BB$. 
Actually, 
the measure of classical correlations $J_{B|A}(\rho)$ 
is the maximum over all orthonormal bases $\{ \ket{\alpha_i} \}$ 
of the Holevo quantity $\chi (\{ \rho_{\BB | i}, \eta_i\} )$ (see
(\ref{def:discord}) and the footnote after this equation). The latter 
is related to the problem of decoding a message 
encoded in the post-\meas states $\rho_{\AB | i}$ when one has access to subsystem $\BB$ only. 
}.
This means that the transpose operation $\Rr_{\Mm_\AAA^\Pi , \rho }$
transforms the ensemble of orthogonal states $\{  \rho_{\AB|i} , \eta_i \}$ into a non-orthogonal ensemble
 $\{\rho_i, \eta_i \}$. Furthermore,
{\it the less distinguishable are the $\rho_i$ for the optimal basis $\{ \ket{\alpha_i^\opt} \}$, 
the most distant is $\rho$ from the
set of $\AAA$-classical states, \ie,  the most quantum-correlated is the state $\rho$.}

The states $\rho$ for which the discrimination of the ensemble $\{ \rho_i^\opt , \eta_i^\opt \}$ 
is the most difficult are the maximally entangled states.  
Actually, with the help of Theorem~\ref{prop_link_geo_discord_QSD} one can show (see~\cite{moi_NJP,my_review_JMP})
that  $D_\Bu^{\rm G}$ satisfies axiom (v)  of Sec.~\ref{sec-axioms_measure_QC},  as already anticipated in
Proposition~\ref{prop_geometric_discord_bona_fide_meas_QCs}. 
More precisely, one has
 
\begin{corollary}
If $n_A \leq n_B$ then the maximal value 
of $D_\Bu^{\rm G}(\rho)$ is equal to
$D_{\rm max}^{\rm G} = 2 (  1 - 1/\sqrt{n_\AAA} )$ 
and $D_\Bu^{\rm G}(\rho) = D_{\rm max}^{\rm G}$ \ifif $\rho$ is a maximally entangled state.
\end{corollary}

{\small

\Proofof{the value of $D_{\rm max}^{\rm G}$}
One deduces from (\ref{eq-equality_distances}) and 
the bound $\mu_{\rm max} \geq 1/n$ (which follows from $\sum_{i=1}^n \mu_i = 1$) that
for any pure state $\ket{\Psi} \in \Hh_\AB$,
\begin{equation}
D_\Bu^{\rm G}  (\ket{\Psi} ) \leq 2\Bigl(  1 - \frac{1}{\sqrt{n}} \Bigr)
\quad , \quad n = \min \{ n_\AAA, n_\BB\} \;.
\;
\end{equation}
The inequality is saturated when 
$\mu_i = 1/n$ for any $i$, \ie, for the maximally entangled states. 
Assuming that $n_\AAA \leq n_\BB$,
since a measure of \QCs is maximal for pure maximally entangled states
(Sec.~\ref{sec-axioms_measure_QC}), one has
$D_\Bu^{\rm G}( \rho) \leq D_{\rm max}^{\rm G}$ for any state $\rho \in \states ( \Hh_\AB)$.
\finpro

}

\vspace{2mm}

It is worth mentioning that finding the optimal \meas and success probability for discriminating
an ensemble of $n_\AAA > 2$ states is highly non-trivial and is still
 an open problem, even though it has been solved for particular ensembles\footnote{
In particular, if the states  $\rho_i = U^{i-1} \rho_1 (U^{i-1})^\dagger$ are related between themselves  
through conjugations by powers of a single unitary operator $U$ satisfying $U^m=\pm 1$, 
one can show that the least square \meas is optimal~\cite{Ban97,Barnett01,Chou03,Eldar01}.
}.
However, the Helstrom formula~\cite{Helstrom} provides a celebrated solution for any ensemble with
$n_\AAA=2$ states. Thus, as we shall see in the next subsection, Theorem~\ref{prop_link_geo_discord_QSD}
can be used to compute $D_\Bu^{\rm G} ( \rho)$ when the reference subsystem $\AAA$ is a qubit.
Despite our belief that this should  not be hopeless, we have not succeeded so far to solve the discrimination problem  for
the ensemble given  in (\ref{eq-state_Q_discrimination}) when $n_\AAA > 2$.  

\subsection{Computability for qubit-qudit systems} \label{sec-computability_Bu_qubit}

If subsystem $\AAA$ is a qubit then 
the ensemble $\{ \rho_i, \eta_i \}$ in Theorem~\ref{prop_link_geo_discord_QSD}
contains only $n_\AAA=2$ states and the optimal probability and measurement to discriminate the $\rho_i$ are easy to determine.
One starts by writing the projector $\Pi_1$ as $1-\Pi_0$ in the expression of the success probability,
\begin{equation}  \label{eq-success_proba}
P_{\rm S}^{\{ \Pi_i \} } ( \{ \rho_i,\eta_i\}) 
 =  \eta_0 \tr  \Pi_0 \rho_0  + \eta_1 \tr  \Pi_1 \rho_1 
 =   
\frac{1}{2} \bigl( 1 - \tr  \Lambda   \bigr)+ \tr \Pi_0 \Lambda   
\end{equation}
with $\Lambda = \eta_0 \rho_0 - \eta_1 \rho_1$.
 The maximum of $\tr  \Pi_0 \Lambda$ 
 over all projectors $\Pi_0$ of rank $n_\BB$ is achieved when $\Pi_0$ projects onto (the 
direct sum of) the eigenspaces associated to the $n_\BB$ highest  
eigenvalues $\lambda_1  \geq \cdots \geq \lambda_{n_\BB}$
of the Hermitian matrix $\Lambda$.
The maximal success probability is thus given by a variant of Helstrom's formula~\cite{Helstrom},
\begin{equation}
\label{eq-opt_success_proba_vN}
P_{\rm S}^{\,\rm opt\,\vN}( \{ \rho_i,\eta_i\})
= \frac{1}{2}\bigl(  1- \tr  \Lambda \bigr) + 
\sum_{l=1}^{n_\BB} \lambda_l \;.
\end{equation}
For the states $\rho_i$ associated to the \ONB $\{ \ket{\alpha_i}\}_{i=0}^1$ of $\complex^2$ 
via formula (\ref{eq-state_Q_discrimination}),  one has 
$\Lambda=\sqrt{\rho} \,( \ketbra{\alpha_0}{\alpha_0} - \ketbra{\alpha_1}{\alpha_1} ) \otimes 1\, \sqrt{\rho}$. 
The operator inside the parenthesis
is equal to  $\sigma_{\uv} = \uv \cdot \sigmav$ for some unit vector $\uv \in \real^3$ depending on $\{ \ket{\alpha_i}\}$
(here $\sigmav$ is the vector formed by the three Pauli matrices).
Conversely,  one can associate to  any unit vector $\uv \in \real^3$ the 
eigenbasis $\{ \ket{\alpha_i}\}_{i=0}^1$ of ${\sigma}_{\uv}$. Thus, according to Theorem~\ref{prop_link_geo_discord_QSD}, $F (\rho, \Cc_\AAA)$ is  obtained
by maximizing the \RHS of (\ref{eq-opt_success_proba_vN}) over all Hermitian matrices 
\begin{equation} \label{eq-Lambda}
 \Lambda (\uv )  =  \sqrt{\rho} \, {\sigma}_{\uv} \otimes 1 \, \sqrt{\rho} 
\end{equation}
with $\uv \in \real^3$, $\| \uv \| = 1$.
One can show~\cite{my_review_JMP}  that 
$\Lambda (\uv)$ has at most $n_\BB$ positive eigenvalues $\lambda_l ( \uv) >0$ and at most $n_\BB$ negative eigenvalues $\lambda_l ( \uv) <0$, 
counting multiplicities. 
This yields to the following formula, which shows that the computation of 
$D_\Bu^{\rm G} (\rho)$ for qubit-qudit states reduces to an optimization problem of a trace norm.

\begin{corollary} {\rm \cite{my_review_JMP}} \label{prop-geo_discord_qubit}
If $\AAA$ is a qubit ($n_\AAA=2$) and $B$ is an arbitrary
system with a $n_B$-dimensional Hilbert space (qudit), the fidelity between $\rho$ and the set of $\AAA$-classical states is given by 
\begin{equation} \label{eq-fidelity_as_min_success_discrimination_qubit}
F (\rho, \Cc_\AAA) =  \frac{1}{2} \max_{\| \uv \|=1 } \bigl\{  1  +  \| \Lambda (\uv) \|_1  \bigr\}\;,
\end{equation}
where
$\Lambda (\uv)$ is the $2n_B \times 2 n_B$ matrix defined in (\ref{eq-Lambda}).
\end{corollary}

One can also conclude from the arguments above that the closest $\AAA$-classical state(s) to $\rho$ 
is (are) given by (\ref{eq-again_I_was_stupid}) where  $\Pi_0^\opt$ is a spectral projector 
associated to the $n_\BB$ largest eigenvalues of $\Lambda ( \uv^{\,\opt} )$ and $\uv^{\,\opt} \in \real^{3}$ is a unit vector 
achieving the maximum in (\ref{eq-fidelity_as_min_success_discrimination_qubit}).
Using Corollary~\ref{prop-geo_discord_qubit}, an analytical expression for $D_\Bu^{\rm G} ( \rho)$
 can be derived  for Bell-diagonal two-qubit states $\rho$, and  the closest $\AAA$-classical states to such
Bell-diagonal states can be determined explicitly~\cite{moi_JPA}. 
The same result for $D_\Bu^{\rm G} ( \rho)$
has been found independently in Ref.~\cite{Aaronson13} 
by another method.
Analytical expressions for the geometric total and classical correlations 
$I_\AB^{\rm G} ( \rho)$ and $C_\AAA^{\rm G} ( \rho)$
for Bell-diagonal two-qubit states $\rho$ have been obtained in Ref.~\cite{Adesso2014}. 

The properties of the Bures \GD established in this section are summarized in the second column of Table~\ref{tab1}.

\begin{table}[t]
\scriptsize
\begin{center}
\begin{tabular}{|c||c|c|c|c|}

\hline
                             &   \multicolumn{4}{|c|}{Geometric discord $D^{\rm G}$}
\\[1mm]
\hline
Distance                     &        Bures   &     Hellinger     &          Trace   &          Hilbert-Schmidt \\
\hline
\hline
\begin{tabular}{c}  {\it Bona fide} measure of\\  quantum correlations \end{tabular}
                    &    \checkmark          &  \checkmark        &    proved for $n_A=2$   &    no
\\[1mm]
\hline
Satisfies  Axiom (v)
                    &  \checkmark          &  proved for $n_A=2,3$ &  proved for $n_A=2$ &
\\[1mm]
\hline
\begin{tabular}{c} Maximal value  \\ if $n_A \leq n_B$ \end{tabular}
&  $2 - 2/\sqrt{n_A}$ \hspace*{4mm} & $2 -2/\sqrt{n_A}$ \hspace*{4mm} 
& $1$ for $n_A=2$  &
\\[1mm]
\hline
Value for pure states &   $2 - 2 \sqrt{\mu_\mmax}$  &  $2 - 2 K^{-\onehalf}$  &      ?         &      $1 - K^{-1}$
\\[1mm]
\hline
\begin{tabular}{c} Relations and \\ cross inequalities \end{tabular}
                  &   \multicolumn{4}{|c|}{$2 -2 \sqrt{ 1 - D_\Hel^{\rm G}(\rho) /2} \;\leq\; D_\Bu^{\rm G} (\rho) \; \leq\;  D_\Hel^{\rm G}(\rho)
\;= \;  2 -2 \sqrt{ 1 - D_\HS^{\rm G}(\sqrt{\rho})}$}
\\[1mm]
\hline
\begin{tabular}{c} Computability  \\ for two qubits \end{tabular}
                 &   Bell-diagonal states    &    all states
                 & $\left\{ \begin{array}{l} \text{X-states} \\ \text{$B$-classical states} \end{array} \right.$
                 &   all states
\\
\hline
\end{tabular}
\end{center}
\caption{\label{tab1} Properties of the
  geometric discords with the Bures, Hellinger, trace, and
  Hilbert-Schmidt distances. Here $n_A$ is the Hilbert space
  dimension of the reference subsystem $A$, 
  $\mu_\mmax=\max \{ \mu_i \}$ is the maximal Schmidt coefficient, and $K= (\sum_i \mu_i^2 )^{-1}$ is the Schmidt number
  of a pure state. The question marks ``?'' indicate unsolved  problems.
  The results quoted in this table have been obtained in 
  Refs.~\cite{Dakic10,moi_NJP,moi_JPA,Aaronson13,Ciccarello14,Roga_Spehner_Illuminatti2016}.
The table is taken from~\cite{Roga_Spehner_Illuminatti2016}. 
}
\end{table}


\subsection{Proof of Theorem~\ref{prop_link_geo_discord_QSD}} \label{sec-proof_link_GD_QSD}

To establish Theorem~\ref{prop_link_geo_discord_QSD}, we 
rely on a slightly  more general statement summarized in the following lemma.

\begin{lemma} \label{lemma_link_geo_discord_QSD}
For a fixed family $\{ \sigma_{\AAA | i} \}_{i=1}^n$  of states  $\sigma_{\AAA |i} \in \states ( \Hh_\AAA)$ 
having orthogonal supports and spanning $\Hh_\AAA$, with $1 \leq n \leq n_\AAA$,  
let us define
\begin{equation}
\Cc_\AAA ( \{ \sigma_{\AAA | i} \} ) = \Bigl\{ \sigma = \sum_{i=1}^n q_i \sigma_{\AAA | i} \otimes \sigma_{\BB | i} \; ;\; \{ q_i, \sigma_{\BB | i} \}_{i=1}^n 
\;\text{ is a state ensemble on $\Hh_\BB$} \;\Bigr\}\;.
\end{equation}
Then
\begin{equation} \label{eq-general_link_geo_discord_QSD}
F \bigl( \rho, \Cc_\AAA ( \{ \sigma_{\AAA | i} \} ) \bigr) 
 \equiv \max_{ \sigma \in \Cc_\AAA ( \{ \sigma_{\AAA | i} \} )} \bigl\{ F ( \rho, \sigma ) \bigr\}
= \max_{U} \biggl\{  \sum_{i=1}^n \| W_i ( U) \|_2^2 \biggr\}\;,
\end{equation}
where the last maximum is over all unitaries $U$ on $\Hh_\AB$, $\| \cdot \|_2$ is the Hilbert-Schmidt norm,  and
\begin{equation} 
W_i ( U ) = \tr_\AAA \bigl( \sqrt{\sigma_{\AAA | i}} \otimes 1\,\sqrt{\rho}\, U \bigr)\;.
\end{equation}
Moreover, there exists  a unitary $U_\opt$ achieving the maximum in (\ref{eq-general_link_geo_discord_QSD}) which 
satisfies $W_i ( U_\opt ) \geq 0$.
The states $\sigma_\opt$ satisfying $F(\rho, \sigma_\opt ) =  F ( \rho, \Cc_\AAA ( \{ \sigma_{\AAA | i} \} ))$ are given in terms of this unitary  by
\begin{equation} \label{eq-general_closest_A_class_state}
\sigma_\opt =  \frac{1}{F ( \rho, \Cc_\AAA ( \{ \sigma_{\AAA | i} \} ))} \sum_{i=1}^n  \sigma_{\AAA | i} \otimes W_i ( U_\opt)^2\;.
\end{equation}
\end{lemma}

{\small 

\Proof
Using the spectral decompositions of the states $\sigma_{\BB |i}$, any $\sigma \in \Cc_\AAA ( \{ \sigma_{\AAA | i} \} )$ can be written as
\begin{equation} \label{eq-proof_geo_discord_and_QSD0}
\sigma = \sum_{i=1}^n \sum_{j=1}^{n_\BB} q_{ij} \sigma_{\AAA | i} \otimes \ketbra{\beta_{j|i}}{\beta_{j|i}} \quad \text{ with }
\quad q_{ij} \geq 0\;,\;\sum_{ij} q_{ij} = 1\;,
\end{equation}
where $\{ \ket{\beta_{j|i}} \}_{j=1}^{n_\BB}$ is an \ONB of $\Hh_\BB$ for any $i$.
By assumption,  if $i \not= i'$ then $\range \sigma_{\AAA | i}  \,\bot\,\range \sigma_{\AAA | i'}$, so that 
$\sqrt{\sigma} = \sum_{i,j} \sqrt{q_{ij}} \sqrt{\sigma_{\AAA | i}} \otimes \ketbra{\beta_{j|i}}{\beta_{j|i}} $.
We start
by evaluating the trace norm in the definition (\ref{eq-fidelity}) of the fidelity by means of the formula
$\| O \|_1 = \max_{U} | \tr  U O  |$ to obtain
\begin{eqnarray} \label{eq-proof_geo_discord_and_QSD1}
\nn
F \bigl( \rho, \Cc_\AAA ( \{ \sigma_{\AAA | i} \} ) \bigr) 
& = & 
\max_{\sigma \in  \Cc_\AAA ( \{ \sigma_{\AAA | i} \} )} \max_U \Bigl\{ \bigl| \tr  U^\dagger \sqrt{\rho} \sqrt{\sigma}  \bigr|^2 \Bigr\}
\\
& = & \max_U \biggl\{ \max_{ \{ q_{ij} \}, \{ \ket{\beta_{j|i}} \} } \biggl|  \sum_{i,j} \sqrt{q_{ij}} 
       \bra{\beta_{j|i}} W_i (U)^\dagger \ket{\beta_{j|i}} \biggr|^2 \biggr\}\;.
\end{eqnarray}
The square modulus can be bounded by invoking twice the Cauchy-Schwarz inequality and $\sum_{ij} q_{ij} =1$,
\begin{eqnarray} \label{eq-proof_geo_discord_and_QSD2}
\nn
\biggl|  \sum_{i,j}  \sqrt{q_{ij}} \bra{\beta_{j|i}} W_i (U)^\dagger \ket{\beta_{j|i}} \biggr|^2
& \leq & \sum_{i,j} \bigl|  \bra{\beta_{j|i}} W_i (U)^\dagger \ket{\beta_{j|i}} \bigr|^2 \\
&\leq & \sum_{i,j}  \bigl\| W_i (U) \ket{\beta_{j|i}} \bigr\|^2 = \sum_i \| W_i (U ) \|_2^2\;.
\end{eqnarray}
The foregoing inequalities are equalities if the following conditions are satisfied:
\begin{itemize}
\item[(1)] $W_i ( U )= W_i (U)^\dagger  \geq 0$;
\item[(2)] $q_{ij} =\bra{\beta_{j|i}}  W_i ( U) \ket{\beta_{j|i}}^2 / ( \sum_{i,j}  \bra{\beta_{j|i}}  W_i ( U) \ket{\beta_{j|i}}^2 )$;
\item[(3)] $\{ \ket{\beta_{j|i}} \}_{j=1}^{n_\BB}$ is an eigenbasis of $W_i ( U)$ for any $i$.
\end{itemize}
Therefore, (\ref{eq-general_link_geo_discord_QSD}) holds true 
provided that there is a unitary $U$ on $\Hh_\AB$ satisfying (1). For a given $U$, let us define $U_\opt = U \sum_{i} \Pi_i^\AAA \otimes V_i^\dagger$,
where $\Pi_i^\AAA$ is the projector onto $\range \sigma_{\AAA | i}$ and $V_i$ is a unitary on $\Hh_\BB$ such that
$W_i ( U ) = | W_i(U)^\dagger |  V_i$ (polar decomposition). 
Then $U_\opt$ is unitary since by hypothesis $\Pi_i^\AAA \Pi_{i'}^\AAA = \delta_{i i'} \Pi_i^\AAA$ and $\sum_i \Pi_i^\AAA = 1$.
Furthermore, 
one readily shows that $W_i ( U_\opt ) = W_i ( U_\opt )^\dagger = |W_i ( U )^\dagger | \geq 0$. 
As $\sum_i \| W_i ( U ) \|_2^2 = \sum_i \| W_i ( U_\opt ) \|_2^2$, the identity (\ref{eq-general_link_geo_discord_QSD}) follows from 
(\ref{eq-proof_geo_discord_and_QSD1}) and (\ref{eq-proof_geo_discord_and_QSD2}). 
From condition (3) one has $W_i ( U_\opt ) \ket{\beta_{j|i}^\opt} = w_{ji} \ket{\beta_{j|i}^\opt}$ with 
$\sum_{i,j} w_{ji}^2 = F ( \rho, \Cc_\AAA ( \{ \sigma_{\AAA | i} \} ) )$, see (\ref{eq-proof_geo_discord_and_QSD2}). Condition (2) entails
\begin{equation}
\sum_j q_{ij}^\opt \ketbra{\beta_{j|i}^\opt}{\beta_{j|i}^\opt}  = \frac{W_i ( U_\opt )^2 }{F ( \rho, \Cc_\AAA ( \{ \sigma_{\AAA | i} \} ) )} 
\;,
\end{equation}
which together with (\ref{eq-proof_geo_discord_and_QSD0}) leads to (\ref{eq-general_closest_A_class_state}).
\finpro

\vspace{3mm}

\Proofof{Theorem~\ref{prop_link_geo_discord_QSD}}
Let  $\{ \ket{\alpha_i} \}_{i=1}^{n_\AAA}$ be an \ONB of $\Hh_\AAA$.
Applying Lemma~\ref{lemma_link_geo_discord_QSD} with $\sigma_{\AAA | i} = \ketbra{\alpha_i}{\alpha_i}$ one gets 
\begin{eqnarray} \label{eq-proof_link_geo_discord_QSD1}
\nn
F \bigl( \rho , \Cc_\AAA ( \{ \ket{\alpha_i} \} ) \bigr)
& = & \max_U \left\{ \sum_{i=1}^{n_\AAA}  \tr  U \ketbra{\alpha_i}{\alpha_i} \otimes 1 \, U^\dagger \sqrt{\rho} 
      \,\ketbra{\alpha_i}{\alpha_i} \otimes 1\, \sqrt{\rho}  \right\}\;,
\\
& = &
\max_{ \{ \Pi_i\}} \left\{ \sum_{i=1}^{n_\AAA} 
\tr  \Pi_i \sqrt{\rho} \ketbra{\alpha_i}{\alpha_i} \otimes 1 \, \sqrt{\rho} \right\} 
= P_{\rm S}^{\,\rm{opt\,v.N.}} ( \{ \rho_i,\eta_i \} )\;.
\end{eqnarray}
The last maximum is over all orthonormal families $\{ \Pi_i\}_{i=1}^{n_\AAA}$ of projectors of rank $n_\BB$ and the success probability
$P_{\rm S}^{\,\rm{opt\,v.N.}} ( \{ \rho_i,\eta_i \} )$ is given by (\ref{eq-max_success_proba_von_Neumann}).
Since the fidelity $F ( \rho, \Cc_\AAA )$ is the maximum of $F ( \rho , \Cc_\AAA ( \{ \ket{\alpha_i} \} ) )$ over all bases $ \{ \ket{\alpha_i } \}$,
this leads to (\ref{eq-variationnal_formula_bis}) and (\ref{eq-again_I_was_stupid}).
\finpro

}

\section{Hellinger geometric discord} \label{sec-Hellinger_GD}

In this section we study the  geometric discord for the Hellinger distance, given by
(see (\ref{eq-Q_Hellinger_distance}) and~(\ref{eq-def_geo_discord}))
\begin{equation} \label{eq-Hellinger_geo_disc}
D_\Hel^{\rm G} (\rho) = 2  - 2 \max_{\sigma_{\Aclass} \in \Cc_\AAA}  \tr \sqrt{\rho} \sqrt{\sigma_{\Aclass}}  \; .
\end{equation}
%

\subsection{Values for pure states, general expression, and closest $\AAA$-classical states} \label{sec-geo_disc_Hell_mixed_states}

\begin{theorem} {\rm \cite{Roga_Spehner_Illuminatti2016}}
\label{eq-theo_geo_disc_Hell_mixed_states}
\begin{itemize}
\item[{\rm (a)}] If $\ket{\Psi} \in \Hh_\AB$ is a pure state, then
\begin{equation}  \label{eq-Hellinger_geo_discord_for_pure_states}
D_\Hel^{\rm G} ( \ket{\Psi}) = 2  - 2 K  ( \ket{\Psi} )^{-\onehalf} \; ,
\end{equation}
where $K ( \ket{\Psi})= ( \sum_i \mu_i^2 )^{-1}$ is the Schmidt number of $ \ket{\Psi}$.
Furthermore, the closest $\AAA$-classical state to $\ket{\Psi}$ for the Hellinger distance is the classical state
\begin{equation} \label{eq-Hellinger_CCQ_state}
\sigma_{\Hel, \Psi} = K ( \ket{\Psi} ) \sum_{i=1}^n \mu_i^2 \ketbra{\varphi_i}{\varphi_i} \otimes \ketbra{\chi_i}{\chi_i} \; ,
\end{equation}
where $\ket{\varphi_i}$ and $\ket{\chi_i}$ are the eigenvectors of $[\rho_\Psi]_A$ and  $[\rho_\Psi]_B$ in the Schmidt 
decomposition (\ref{eq-Schmidt_decomposition}).

\item[{\rm (b)}] If $\rho$ is a mixed state, then
\begin{equation} \label{eq-formula_Hellinger_geo_discordbis}
D_\Hel^{\rm G} ( \rho) = 2  - 2  \max_{ \{ \ket{\alpha_i} \} }  \biggl\{ \sum_{i=1}^{n_A}  {\tr}_B [ \bra{\alpha_i} \sqrt{\rho} \ket{\alpha_i}^2 ]\biggr\}^\onehalf
= 2 - 2 \max_{ \{ \ket{\alpha_i} \} } \sqrt{P_{\rm S}^{\rm lsm} ( \{ \rho_i,\eta_i \})}
\; ,
\end{equation}
where the maximum is over all \ONBs  $ \{ \ket{\alpha_i} \} $ for
$A$ and $P_{\rm S}^{\rm lsm} ( \{ \rho_i,\eta_i \})$ is 
the success probability in
discriminating the ensemble $\{ \rho_i , \eta_i\}$ defined in (\ref{eq-state_Q_discrimination}) 
by the least--square
measurement.
Let the maxima in (\ref{eq-formula_Hellinger_geo_discordbis}) be reached for the basis
$ \{ \ket{\alpha_i^\opt} \} $. Then the \CCQ state(s) to $\rho$ for the
Hellinger distance is (are)
\begin{equation} \label{eq-CCL_Hel}
\sigma_{\Hel , \rho}
  =
   \left( 1 - \frac{D_\Hel^{\rm G} ( \rho)}{2} \right)^{-2} \sum_{i=1}^{n_A} \ketbra{\alpha_i^\opt}{\alpha_i^\opt} \otimes
    \bra{\alpha_i^\opt} \sqrt{\rho} \ket{\alpha_i^\opt}^2\;.
\end{equation}
\end{itemize}
\end{theorem}

As the Schmidt number $K ( \ket{\Psi})$ is an entanglement monotone,
one infers from (a)
that $D_\Hel^{\rm G}$  satisfies Axiom  (iv) of Definition~\ref{def-measure_of_QC} and is thus a 
{\it bona fide} measure of quantum correlations, as claimed in Proposition~\ref{prop_geometric_discord_bona_fide_meas_QCs}.
Moreover,  if $n_\AAA \leq n_\BB$ then $D_\Hel^{\rm G}$ has  the same maximal value  $D_\mmax^{\rm G} = 2 -2 /\sqrt{n_A}$
as the Bures geometric discord (in fact,
$D_\Hel^{\rm G} (\rho)$ is maximum for maximally entangled pure states which have Schmidt numbers equal to $n_\AAA$).

\vspace{2mm}

{\small 

\Proof
Let us  first prove part (b) of the theorem.
By using the spectral decompositions of the states $\rho_{B|i}$ in (\ref{eq-A-classical_states}), any $\AAA$-classical state can be written as
\begin{equation} \label{eq-A_class_state}
\sigma_\Aclass
=
\sum_{i=1}^{n_A} \sum_{j=1}^{n_B} q_{ij} \ketbra{\alpha_i}{\alpha_i} \otimes \ketbra{\beta_{j|i}}{\beta_{j|i}} \; ,
\end{equation}
where  $\{ q_{ij} \}$ is a probability distribution, $\{ \ket{\alpha_i} \}_{i=1}^{n_A}$ is an orthonormal basis for $A$
and, for any $i$, $\{ \ket{\beta_{j|i}} \}_{j=1}^{n_B}$ is an orthonormal basis for $B$
(note that the $\ket{\beta_{j|i}}$ need not be orthogonal for distinct $i$'s).
The square root of $\sigma_\Aclass $ is obtained by replacing $q_{ij}$ by $\sqrt{q_{ij}}$ in the \RHS of  ~(\ref{eq-A_class_state}).
Hence, in the same way as in the proof of Sec.~\ref{sec-proof_link_GD_QSD},
\begin{equation} \label{eq-Hellinger_fidelity_to classical_state}
\tr \sqrt{\rho} \sqrt{\sigma_\Aclass}
 =  \sum_{i,j} \sqrt{q_{ij}} \bra{\alpha_i \otimes \beta_{j|i}} \sqrt{\rho} \ket{\alpha_i \otimes \beta_{j|i}}
\leq
\biggl( \sum_{i,j} \bra{\alpha_i \otimes \beta_{j|i}} \sqrt{\rho } \ket{\alpha_i \otimes \beta_{j|i}}^2 \biggr)^{\onehalf} \; .
\end{equation}
The last bound follows from the Cauchy-Schwarz inequality and the identity $\sum_{i,j} q_{ij} =1$. It is saturated when
\begin{equation} \label{eq-optimal_q_ij}
q_{ij} = \frac{\bra{\alpha_i \otimes \beta_{j|i}} \sqrt{\rho } \ket{\alpha_i \otimes \beta_{j|i}}^2}
{\sum_{i,j} \bra{\alpha_i \otimes \beta_{j|i}} \sqrt{\rho } \ket{\alpha_i \otimes \beta_{j|i}}^2}
\; .
\end{equation}
Therefore,
\begin{equation} \label{eq-Hellinger_fidelity_to classical_state2}
\max_{\{ q_{ij}\}} \tr \sqrt{\rho} \sqrt{\sigma_\Aclass }
=
\biggl( \sum_{i,j} \bra{\beta_{j|i}} B_i \ket{\beta_{j|i}}^2 \biggr)^{\onehalf}
\end{equation}
with $B_i = \bra{\alpha_i} \sqrt{\rho } \ket{\alpha_i} \in \Bb ( \Hh_\BB)_{\rm s.a.}$. 
Now, for any fixed $i$, one has
\begin{equation} \label{eq-ineq_sum_matrix_element_square}
\sum_j  \bra{\beta_{j|i}}  B_i \ket{\beta_{j|i}}^2 \leq \tr [ B_i^2 ] \; .
\end{equation}
This inequality is saturated when $\{ \ket{\beta_{j|i}} \}$ is an eigenbasis of $B_i$.
Since maximizing over all $\AAA$-classical states in (\ref{eq-Hellinger_geo_disc}) amounts to maximize over all
$\{ q_{ij}\}$, $\{ \ket{\alpha_i} \}$, and $\{ \ket{\beta_{j|i}} \}$, this gives 
\begin{equation} \label{eq-formula_Hellinger_geo_discordbis_classicalquantum}
\left( 1 - \frac{D_\Hel^{\rm G} (\rho)}{2} \right)^2
 =
\max_{ \{ \ket{\alpha_i} \} } \sum_{i=1}^{n_\AAA} \tr_B \big[ \bra{\alpha_i} \sqrt{\rho } \ket{\alpha_i}^2  \big]
\; .
\end{equation}
It has been observed in Sec.~\ref{sec-link_QSD} that the least square \meas
for the ensemble $\{ \rho_i,\eta_i \}$ defined in Theorem~\ref{prop_link_geo_discord_QSD}
is the projective \meas $\{ \ketbra{\alpha_i}{\alpha_i} \otimes 1 \}_{i=1}^{n_\AAA}$. Thus
\begin{equation} \label{eq-proba_success_lsm}
P_{\rm S}^{\rm lsm} ( \{ \rho_i,\eta_i \})= \sum_{i=1}^{n_A} \eta_i \tr \rho_i \ketbra{\alpha_i}{\alpha_i} \otimes 1 
 = \sum_{i=1}^{n_A}  \tr_B  \bra{\alpha_i}  \sqrt{\rho} \ket{\alpha_i}^2 
\;.
\end{equation}
Equation (\ref{eq-formula_Hellinger_geo_discordbis})
follows from (\ref{eq-formula_Hellinger_geo_discordbis_classicalquantum}) and (\ref{eq-proba_success_lsm}).
The \CCQ state  is given by  ~(\ref{eq-A_class_state})
in which $\ket{\alpha_i} = \ket{\alpha_i^\opt}$ are the vectors realizing the maximum
in~(\ref{eq-formula_Hellinger_geo_discordbis_classicalquantum}),
$\ket{\beta_{j|i}} = \ket{\beta_{j|i}^\opt}$ are the eigenvectors
of $B_i^\opt = \bra{\alpha_i^\opt} \sqrt{\rho} \ket{\alpha_i^\opt}$, and (see  ~(\ref{eq-optimal_q_ij})):
\begin{equation} \label{eq-optimal_q_ij_bis}
q_{ij} = \frac{\bra{\beta_{j|i}^\opt} (B_i^\opt)^2 \ket{\beta_{j|i}^\opt}}{\sum_i \tr  (B_i^\opt)^2} \; .
\end{equation}
The expression (\ref{eq-CCL_Hel}) readily follows.

We now establish part (a) of the theorem. Let $\rho = \ketbra{\Psi}{\Psi}$ be a pure state with reduced state
$\rho_A = \tr_B \ketbra{\Psi}{\Psi}$. Then $B_i =
\ketbra{\beta_i}{\beta_i}$, where $\ket{\beta_i} =
\braket{\alpha_i}{\Psi}$ has square norm $\| \beta_i\|^2 =
\bra{\alpha_i } \rho_A \ket{\alpha_i}$. Thus  ~(\ref{eq-formula_Hellinger_geo_discordbis_classicalquantum}) yields
\begin{equation} \label{eq-formula_Hellinger_geo_discord_oure_state}
\left( 1 - \frac{D_\Hel^{\rm G} ( \ket{\Psi} )}{2} \right)^2
=
\max_{ \{ \ket{\alpha_i } \} } \sum_{i=1}^{n_\AAA} \bra{\alpha_i} \rho_A \ket{\alpha_i}^2 \; .
\end{equation}
In analogy with ~(\ref{eq-ineq_sum_matrix_element_square}), the sum in the \RHS is bounded from above by
$\tr \rho_A^2 = K( \ket{\Psi})^{-1}$, the bound being saturated
when $\{ \ket{\alpha_i} \}$ is an  eigenbasis of $\rho_A$. This leads to  ~(\ref{eq-Hellinger_geo_discord_for_pure_states}). The \CCQ
state to $\ket{\Psi}$ is given by  ~(\ref{eq-CCL_Hel}) with
$\ket{\alpha_i^\opt} = \ket{\varphi_i}$, which gives~(\ref{eq-Hellinger_CCQ_state}).
\finpro
}

\subsection{Link with the Hilbert-Schmidt geometric discord} \label{sec-rel_geo_dic_HS_Hel}

In view of the definition $d_\Hel ( \rho, \sigma ) = d_2 ( \sqrt{\rho}, \sqrt{\sigma})$ of the Hellinger distance,
it should not come as a surprise that $D_\Hel^{\rm G} ( \rho )$ is related to
the Hilbert-Schmidt geometric discord $D_\HS^{\rm G} ( \sqrt{\rho} )$
of the square root of $\rho$.

\begin{proposition} {\rm \cite{Roga_Spehner_Illuminatti2016}} \label{theo-rel_geo_disc_Hel_HS}
For any $\rho \in \states ( \Hh_\AB) $, one has
\begin{equation} \label{eq-rel_geo_disc_Hel_HS}
D_\Hel^{\rm G} ( \rho )  = 2  - 2 \bigl( 1 - D_\HS^{\rm G} ( \sqrt{\rho} ) \bigr)^{\onehalf} \; .
\end{equation}
\end{proposition}

Note that the Hilbert-Schmidt geometric discord is evaluated for the
square root of $\rho$, which is not a state but is nevertheless a
non-negative operator. Thus $\sigma = \sqrt{\rho} \, /\tr  \sqrt{\rho}
$ is a density operator and $D_\HS^{\rm G} ( \sqrt{\rho})$ is defined as
$D_\HS^{\rm G} ( \sqrt{\rho}) \equiv  ( \tr \sqrt{\rho} )^2 D_{\HS}^{\rm G} ( {\sigma } )$.

\vspace{2mm}

{\small

\Proof 
The following expression of $D_\HS^{\rm G} ( \rho)$   has been found by Luo and Fu~\cite{Luo_Fu10}:
\begin{equation}  \label{eq-HS_geo_discord}
D_\HS^{\rm G}  ( \rho )
 = \tr  \rho^2 - \max_{ \{ \ket{\alpha_i} \}}  \sum_{i=1}^{n_A} \tr_B \bra{\alpha_i} \rho \ket{\alpha_i}^2
= \min_{ \{ \ket{\alpha_i} \}} \sum_{i \not= j}^{n_A} \tr_B  | \bra{\alpha_i} \rho \ket{\alpha_j} |^2 \; .
\end{equation}
For completeness, let us give a simple derivation of (\ref{eq-HS_geo_discord}).  
By definition,
\begin{equation}
D_\HS^{\rm G}  ( \rho )
= \min_{\sigma_\Aclass \in \Cc_\AAA} \| \rho - \sigma_\Aclass \|_2^2
=  \tr \rho^2 + \min_{\sigma_\Aclass \in \Cc_\AAA} \tr ( \sigma_\Aclass^2- 2 \rho \sigma_\Aclass ) \; .
\end{equation}
Thanks to (\ref{eq-A_class_state}), the last trace is equal to
\begin{equation}
\sum_{i,j}
  \Bigl\{ \bigl( q_{ij} - \bra{\alpha_{i} \otimes \beta_{j|i}} \rho \ket{\alpha_{i} \otimes \beta_{j|i}} \bigr)^2
    - \bra{\alpha_{i} \otimes \beta_{j|i}} \rho \ket{\alpha_{i} \otimes \beta_{j|i}}^2  \Bigr\} \; .
\end{equation}
The minimum over the  probability distribution $\{ q_{ij}\}$ is
obviously achieved  for $ q_{ij} = \bra{\alpha_{i} \otimes
  \beta_{j|i}} \rho \ket{\alpha_{i} \otimes \beta_{j|i}}$.
Minimizing also over the \ONBs $\{ \ket{\alpha_i} \}$ and
$\{ \ket{\beta_{j|i}}\}$ and using~(\ref{eq-ineq_sum_matrix_element_square}) again, one finds
the first equality in (\ref{eq-HS_geo_discord}).
The second equality follows from the relation
$\tr  \rho^2 = \sum_{i,j} \tr_B  | \bra{\alpha_i} \rho \ket{\alpha_j} |^2 $.
The result of Proposition~\ref{theo-rel_geo_disc_Hel_HS} is now obtained by 
comparing~(\ref{eq-formula_Hellinger_geo_discordbis}) and~(\ref{eq-HS_geo_discord}).
\finpro

}

\begin{remark} {\rm
By using similar arguments as in the proof of Theorem~\ref{eq-theo_geo_disc_Hell_mixed_states}, one finds
that the closest $\AAA$-classical
state  to $\rho$ for the Hilbert-Schmidt distance coincides
with the  post-measurement state $\Mm_\AAA^\Pi \otimes 1 ( \rho)$, where  
 $\Mm_\AAA^\Pi$ is the \QO (\ref{eq-QO_meas}) associated to a  measurement on $A$ with projectors 
$\Pi_i^\AAA = \ketbra{\alpha_i^\opt }{\alpha_i^\opt }$,
$\{ \ket{\alpha_i^\opt} \}$ being the \ONB
maximizing  the first sum in (\ref{eq-HS_geo_discord}). Therefore, as already observed in Ref.~\cite{Luo_Fu10}, 
for the Hilbert-Schmidt distance the geometric and measurement-induced geometric discords are equal, 
$D^{\rm G}_\HS =D^{\rm M}_\HS$.
Furthermore,  the known value  $D_\HS^{\rm G} ( \ket{\Psi} ) = 1 - K  ( \ket{\Psi} )^{-1}$ for pure states~\cite{Luo2013}
 is recovered by noting that  (\ref{eq-rel_geo_disc_Hel_HS}) implies
$D_\Hel^{\rm G} ( \ket{\Psi} ) = 2  - 2 (1 - D_\HS^{\rm G} ( \ket{\Psi} ))^\onehalf$ and by
comparing with  (\ref{eq-Hellinger_geo_discord_for_pure_states}).
}
\end{remark}

\subsection{Comparison between the Bures and Hellinger geometric discords}

As pointed out in Sec.~\ref{sec-deferent_orderings}, the Bures and Hellinger geometric discords are not 
functions of each other
and thereby define different orderings on $\states ( \Hh_\AB)$. A large number of inequalities enabling to compare 
$D^{\rm G}$, $D^{\rm M}$, and $D^{\rm R}$  for the Bures, Hellinger, trace, and Hilbert Schmidt distances
have been established in Ref.~\cite{Roga_Spehner_Illuminatti2016} (some of these inequalities are given in 
Tables~\ref{tab1}-\ref{tab3}).
A particular bound is as follows.

\begin{proposition} {\rm \cite{Roga_Spehner_Illuminatti2016}} \label{theo-bounds_Bures_geo}
The Bures and Hellinger geometric discords satisfy
\begin{equation} \label{eq-bounds_Bures_geo}
g^{-1} ( D_\Hel^{\rm G} ( \rho ))
\leq D_\Bu^{\rm G} ( \rho ) \leq D_\Hel^{\rm G} ( \rho ) \; ,
\end{equation}
where  the increasing function $g(d)$ and its inverse are defined by
\begin{equation} \label{eq-function_g}
g (d) = 2 d - \onehalf d^2
\quad , \quad 
g^{-1} (d) = 2 - 2 \sqrt{1-d/2}\;.
\end{equation}
If $\AAA$ is a qubit, the stronger 
bound $D_\Hel^{\rm G} ( \rho ) \leq g^{-1} \circ h ( D_\Bu^{\rm G} ( \rho ))$ holds
and is saturated for pure states, with $h(d) = 2 g(d) - g(d)^2$.
\end{proposition}

{\small 

\Proof 
The first statement
is a consequence of Theorem~\ref{prop_link_geo_discord_QSD} and of an upper
bound on the probability of success in \QSD due to  Barnum and Knill~\cite{Barnum02}. According to such bound, the maximum
probability of success $P_{\rm S}^{\,\rm{opt\,v.N.}} ( \{ \rho_i,\eta_i \})$ is at most equal to the square root of the probability of success
$P_{\rm S}^{\rm lsm} ( \{ \rho_i,\eta_i \})$ obtained by discriminating the states $\rho_i$ with the least--square
measurement. Hence 
\begin{equation} \label{eq-inequality_square_root_meas}
 \max_{\{ \ket{\alpha_i } \} }  P_{\rm S}^{\rm lsm} ( \{ \rho_i,\eta_i \})
\leq
F (\rho , \Cc_\AAA )
\leq
\max_{\{ \ket{\alpha_i} \} } P_{\rm S}^{\rm lsm} ( \{ \rho_i,\eta_i \})^{\frac{1}{2}} \; .
\end{equation}
The second inequality 
together with (\ref{eq-max_fidelity}) and~(\ref{eq-formula_Hellinger_geo_discordbis}) yields
to the first bound in~(\ref{eq-bounds_Bures_geo}).
The second
bound in (\ref{eq-bounds_Bures_geo}) is an immediate consequence of the fact that the Bures distance is always smaller or
equal to the  Hellinger distance\footnote{
We remark that by exploiting~(\ref{eq-max_fidelity})
and~(\ref{eq-formula_Hellinger_geo_discordbis}), this second bound is equivalent
precisely to  the lower bound in~(\ref{eq-inequality_square_root_meas}). 
}
 (Proposition~\ref{prop_bounds_between_d_B_and_d_1}).
The stronger bound when $\AAA$ is a qubit follows from the inequality 
$D_\Hel^{\rm R} ( \rho) \leq 1 - (1 - D_\Bu^{\rm R} ( \rho))^2$ on the discords of response\footnote{
This inequality follows from the definitions of $D_\Hel^{\rm R}$ and $D_\Bu^{\rm R}$ and from the trace inequality
$F ( \rho, U_\AAA \otimes 1 \,\rho\,U_\AAA^\dagger \otimes 1) = \| \sqrt{\rho}\,  U_\AAA \otimes 1 \,\sqrt{\rho}\|_1^2 \leq
\tr ( \sqrt{\rho} \, U_\AAA \otimes 1 \,\sqrt{\rho}\,  U_\AAA^\dagger \otimes 1 )$.
It is saturated for pure states (see~\cite{Roga_Spehner_Illuminatti2016} for more detail).
}
 and from the identities
$ D_\Bu^{\rm R} ( \rho) = g ( D_\Bu^{\rm G} ( \rho))$ and $ D_\Hel^{\rm R} ( \rho) = g ( D_\Hel^{\rm G} ( \rho))$, 
see Table~\ref{tab3} and Ref.~\cite{Roga_Spehner_Illuminatti2016}.
\finpro

}

\vspace{2mm}

Proposition~\ref{prop_bounds_between_d_B_and_d_1} also yields bounds on 
$D_\Hel^{\rm G}$ and $D_\Bu^{\rm G}$ in terms of the trace  \GD $D_{\tr}^{\rm G}$:
\begin{equation}
 [ D_\Hel^{\rm G} (\rho) ]^2  \leq   D_{\tr}^{\rm G} ( \rho)
\leq 2 g( D_{\Bu}^{\rm G} (\rho ) )
\;.
\end{equation}
Similar bounds hold for the measurement-induced geometric discord $D^{\rm M}$ and discord of response $D^{\rm R}$ 
(but one has to take care of the different normalization factors in the definition
of $D^{\rm R}$, see Sec.~\ref{sec-disc_response}).

\subsection{Computability for qubit-qudit systems} \label{sec-computability_Hellinger_GD}

We show in this subsection that the Hellinger geometric discord is an easily computable quantity, at least
 when $A$ is a qubit.
For indeed,  we will determine with the help of
(\ref{eq-formula_Hellinger_geo_discordbis}) an explicit expression for $D_\Hel^{\rm G}(\rho)$
for arbitrary qubit-qudit states $\rho$.

 Let us introduce the
vector $\vec{\gamma}$ formed by the $(n_B^2-1)$ self-adjoint operators
$\gamma_p$ on $\Hh_\BB$ satisfying  $\tr \gamma_p = 0$ and  $\tr \gamma_p
\gamma_q  = n_B \delta_{pq}$ for any  $p,q=1,\ldots , n_B^2-1$
(this means that $\{ 1/\sqrt{n_B} , \gamma_p/\sqrt{n_B} \}$ is an \ONB of the Hilbert
space of all $n_B \times n_B$ matrices). 
This vector is the analog for $\BB$ of the vector $\sigmav$ formed by the three Pauli matrices for $\AAA$.
The square root of $\rho$ can be decomposed as
\begin{equation} \label{eq-Boch_dec_square_root}
\sqrt{\rho} = \frac{1}{\sqrt{2 n_B}}
\Bigl( t_0 1 \otimes 1 + \vec{x} \cdot  \vec{\sigma} \otimes 1 + 1 \otimes \vec{y} \cdot \vec{\gamma} + \sum_{m=1}^3\sum_{p=1}^{n_B^2-1}
t_{mp} \, \sigma_m \otimes \gamma_p \Bigr)
\end{equation}
with $t_0 \in [-1,1]$, $\vec{x} \in \real^3$, and $\vec{y} \in
\real^{n_B^2-1}$. We denote by $T$ the $3 \times
(n_B^2-1)$ complex matrix with coefficients $t_{mp}$. The condition $\tr ( \sqrt{\rho})^2  = 1$ entails $t_0^2 + \| \vec{x} \|^2 + \| \vec{y} \|^2 + \tr (T T^{\rm T} ) = 1$ (here $T^{\rm T}$ stands for the transpose of $T$). For any \ONB  $\{ \ket{\alpha_i} \}_{i=0,1}$ for qubit $A$, one finds
\begin{equation}
\sum_{i=0,1}  \tr  \bra{\alpha_i} \sqrt{\rho} \ket{\alpha_i}^2  =
t_0^2 + \| \vec{y} \|^2 + \vec{u}^{\rm T} ( \vec{x} \vec{x}^{\rm T} + T T^{\rm T} ) \vec{u} \; ,
\end{equation}
where we have introduced the unit vector $\vec{u}= \bra{\alpha_0} \vec{\sigma} \ket{\alpha_0} = - \bra{\alpha_1} \vec{\sigma} \ket{\alpha_1}$.
Maximizing over all such vectors and using~(\ref{eq-formula_Hellinger_geo_discordbis}), we have~\cite{Roga_Spehner_Illuminatti2016}
\begin{equation} \label{eq-explicit_formula_geo_disc_2_qubits}
D_\Hel^{\rm G} ( \rho) = 2 - 2 \sqrt{ t_0^2 + \| \vec{y}\|^2 + k_{\rm max} } \; \; ,
\end{equation}
where  $k_{\rm max}$ is the largest eigenvalue of the $3 \times 3$ matrix $K = \vec{x} \vec{x}^{\rm T} + T T^{\rm T}$.
Therefore, the calculation of $D_\Hel^{\rm G} ( \rho) $ is straightforward
once one has determined the decomposition~(\ref{eq-Boch_dec_square_root}) of the square root of $\rho$.

A formula for the Hilbert-Schmidt geometric discord for two-qubit states
has been given in Ref.~\cite{Dakic10}.
An alternative derivation of (\ref{eq-explicit_formula_geo_disc_2_qubits}) consists in using this formula and
Proposition~\ref{theo-rel_geo_disc_Hel_HS}.
The trace geometric discord $D^{\rm G}_{\tr}$ seems harder to compute than
 $D^{\rm G}_{\Hel}$ and $D^{\rm G}_{\Bu}$, but analytical expressions have been found in Ref.~\cite{Ciccarello14} 
for two-qubit $X$-states and two-qubit $\BB$-classical  states.

The results of this section are summarized in the third column of Table~\ref{tab1}.

\section{Measurement-induced geometric discord and discord of response} \label{sec_meas_ind_geo_disc_and_disc_resp}

The properties of the measurement-induced geometric discord $D^{\rm M}$
and discord of response $D^{\rm R}$ for the Bures, Hellinger, trace, and Hilbert-Schmidt distances 
are summarized in Tables~\ref{tab2} and~\ref{tab3}.
We refer the reader to Ref.~\cite{Roga_Spehner_Illuminatti2016} for the proofs and 
references to the original works.
For any $\rho \in \states ( \Hh_\AB)$, 
the following general expressions and bounds on $D^{\rm M}$ and  $D^{\rm R}$ 
 can be derived. For the Bures distance, one has (compare with (\ref{eq-variationnal_formula_bis}))~\cite{Roga_Spehner_Illuminatti2016}
\begin{equation}
\label{eq-formula_Bures_meas_ind_discord}
\begin{array}{cclcc}
D^{\rm G}_\Bu ( \rho) & \leq & 
D_\Bu^{\rm M} ( \rho)
 =  
\dss 2 - 2 \max_{ \{ \ket{\alpha_i} \}} \tr \sqrt{\sum_{i=1}^{n_A} \eta_i^2 \rho_i^2}  
& \leq & g( D^{\rm G}_\Bu ( \rho) )
\\
1 - \sqrt{1-D_\Hel^{\rm R} (\rho)}           & \leq & 
D_\Bu^{\rm R} ( \rho)
 = 
\dss 1 -  \max_{ \{ \ket{\alpha_i} \} } \tr \bigg| \sum_{i=1}^{n_A} \eta_i e^{-\I \frac{2 \pi i}{n_A}} \rho_i \bigg| 
& \leq & D_\Hel^{\rm R} ( \rho) 
\end{array}
\; , 
\end{equation}
where  $\{ \rho_i, \eta_i\}$ is the state ensemble defined in (\ref{eq-state_Q_discrimination})
and $g$ is the function (\ref{eq-function_g}).
Similarly, one finds for the Hellinger distance
(compare with (\ref{eq-formula_Hellinger_geo_discordbis}))~\cite{Roga_Spehner_Illuminatti2016}
\begin{equation}
\label{eq-formula_Hellinger_meas_ind_discord}
\begin{array}{cclcc}
D^{\rm G}_\Hel ( \rho) 
& \leq & 
D_\Hel^{\rm M} ( \rho)
 =  
\dss 2 - 2 \max_{\{ \ket{\alpha_i}\} } \sum_{i=1}^{n_A} {\tr}_B \bra{\alpha_i} \sqrt{\rho} \ket{\alpha_i} \sqrt{\bra{\alpha_i} \rho \ket{\alpha_i}} 
& \leq & g( D^{\rm G}_\Hel ( \rho) )
\\
 \sin^2 \left( \frac{\pi}{n_A} \right) g ( D^{\rm G}_{\Hel}  (\rho) )         
& \leq & 
D_\Hel^{\rm R} ( \rho)
 = 
\dss 2 \min_{ \{ \ket{\alpha_i} \} }   \sum_{i,j=1}^{n_A} \sin^2 \Big( \frac{\pi (i-j)}{n_A} \Bigr)
 {\tr}_B  \big| \bra{\alpha_i} \sqrt{\rho} \ket{\alpha_j} \big|^2   
& \leq &
g ( D^{\rm G}_{\Hel}  (\rho) )
\end{array}
\; .
\end{equation}
The first inequality in the last line is an equality when  $n_\AAA=2$ or $3$. Thus,
for the Hellinger distance 
the  discord of response is a function of the geometric discord when $\AAA$ is a qubit or a qutrit. 
This is also true for the Bures and trace distances when $\AAA$ is a qubit (see Table~\ref{tab3}).
In that case, $D^{\rm R}_\Hel ( \rho)= g ( D^{\rm G}_\Hel (\rho))$ 
 can be evaluated analytically by relying on the formula 
(\ref{eq-explicit_formula_geo_disc_2_qubits}), showing that 
$D^{\rm R}_\Hel$ is an easily computable measure of quantum correlations.
In fact, when $n_\AAA=2$ then $D^{\rm R}_\Hel (\rho)$ is related to the LQU (see (\ref{eq-LQU_qubit})), which has been determined 
for arbitrary qubit-qudit states in Ref.~\cite{Girolami2013}.

\begin{table}
\scriptsize
\begin{center}
\begin{tabular}{|c||c|c|c|c|}

\hline
                             &   \multicolumn{4}{|c|}{Measurement-induced geometric discord $D^{\rm M}$}
\\[1mm]
\hline
Distance                     &          Bures     &     Hellinger &          Trace   &          Hilbert-Schmidt \\
\hline
\hline
\begin{tabular}{c}  {\it Bona fide} measure of\\  quantum correlations \end{tabular}
                    &    \checkmark          &  \checkmark        &     \checkmark   &    no
\\[1mm]
\hline
Satisfies  Axiom (v)
                    &  \checkmark          &   for $n_A=2$ (conjecture) &  proved for $n_A=2$ &
\\[1mm]
\hline
\begin{tabular}{c} Maximal value  \\ if $n_A \leq n_B$ \end{tabular}      &  \begin{tabular}{r} $2 - 2/\sqrt{n_A}$ \hspace*{4mm} \\  \end{tabular}
  &  \begin{tabular}{r} $2 - 2/\sqrt{n_A}$ \hspace*{4mm} \\  \end{tabular} &   $(2-2/n_A)^2$  &
\\[1mm]
\hline
Value for pure states &   $2 - 2 K^{-\onehalf}$   &   $2 - 2 \sum_i \mu_i^{\frac{3}{2}}$  &    see Theorem 3.3 in~\cite{Piani14}         &      $1 - K^{-1}$
\\[1mm]
\hline
\begin{tabular}{c} Comparison with the \\ geometric discord \end{tabular}
                  &  $D_\Bu^{\rm G} \leq D^{\rm M}_\Bu  \leq g( D^{\rm G}_\Bu )$
                  &  $D_\Hel^{\rm G} \leq D^{\rm M}_\Hel  \leq   g ( D^{\rm G}_\Hel )$
                  &  $\begin{cases} D_{\tr}^{\rm M} = D_{\tr}^{\rm G} & \text{for } n_A=2 \\  D_{\tr}^{\rm M} \geq  D_{\tr}^{\rm G} & \text{for } n_A>2 \end{cases}$
                  &  $D_\HS^{\rm M} = D_\HS^{\rm G}$
\\[1mm]
\hline
\begin{tabular}{c} Computability \\ for two qubits \end{tabular}
                 &    ?    &    ?
                 & $\left\{ \begin{array}{l} \text{X-states} \\ \text{$B$-classical states} \end{array} \right.$
                 &   all states
\\
\hline

\end{tabular}
\end{center}
\caption{\label{tab2} Properties of the measurement-induced geometric  discords with the Bures, Hellinger, trace, and
  Hilbert-Schmidt distances. The function $g$ is given by (\ref{eq-function_g}).
 The remaining notations are the same as in the caption of Table~\ref{tab1}.
The results quoted in this table have been obtained in Refs.~\cite{Ciccarello14,Dakic10,Luo_Fu10,Nakano13,Piani14,Roga_Spehner_Illuminatti2016}.
 This table is taken from~\cite{Roga_Spehner_Illuminatti2016}.
}
\end{table}

\begin{table}
\scriptsize
\begin{tabular}{|c|l||c|c|c|c|}

\hline
\multicolumn{2}{|c||}{}                 &   \multicolumn{4}{|c|}{Discord of response $D^{\rm R}$}
\\[1mm]
\hline
\multicolumn{2}{|c||}{Distance}    & Bures    &    Hellinger    &          Trace   &          Hilbert-Schmidt \\
\hline
\hline
\multicolumn{2}{|c||}{\begin{tabular}{c}  {\it Bona fide} measure of \\  quantum correlations \end{tabular}}
                              &    \checkmark          &  \checkmark        &   \checkmark   &    no
\\[1mm]
\hline
\multicolumn{2}{|c||}{Satisfies  Axiom (v)}          &  \checkmark
     &  \checkmark   &  \checkmark  & no if $n_B \geq 2 n_A$
\\[1mm]
\hline
\multicolumn{2}{|c||}{\begin{tabular}{c} Maximal value  \\ if $n_A
    \leq n_B$ \end{tabular}}                 &   $1$     &   $1$    &
$1$   & $1$
\\[1mm]
\hline
\multicolumn{2}{|c||}{Value for pure states}
                              &  $1 - \sqrt{1- E^{\rm R}}$ &   $E^{\rm R}$ &      $E^{\rm R}$        &      $E^{\rm R}$
\\[1mm]
\hline
           & $n_A=2$         &  $D^{\rm R}_{\Bu}   = g ( D^{\rm G}_{\Bu})$
                             &  $D^{\rm R}_{\Hel}   = g ( D^{\rm G}_{\Hel})$
                             &  $D^{\rm R}_{\tr}=  D^{\rm G}_{\tr} $
                             &  $D^{\rm R}_\HS = 2 D^{\rm G}_\HS$
\\[2mm] \cline{2-6}
\begin{tabular}{c} Functional  \\  relation with \\ $D^{\rm G}$ \end{tabular}             &  $n_A=3$         &
\multirow{2}[2]{*}{
no
}
                             &  $D^{\rm R}_{\Hel}   =  \frac{3}{4} g (D^{\rm G}_{\Hel})$
                             &  \multirow{2}[2]{*}{no }
                             &   $D^{\rm R}_\HS = \frac{3}{2} D^{\rm G}_\HS$
\\[2mm] \cline{2-2}\cline{4-4}\cline{6-6}
         &  $n_A >3$        &
                             &  no
                             &
                             &  no
\\[2mm]
\hline
       & $n_A = 2$       &   $D_\Bu^{\rm M} \precsim 2 - \sqrt{2} \sqrt{1  +  (1-D_\Bu^{\rm R})^2}$
                            & \multirow{3}[2]{*}{ 
         $\begin{array}{c} \\[1mm] \sin^2 \big( \frac{\pi}{n_A} \big) g ( D_{\Hel}^{\rm G} ) 
         \\[2mm] \leq D^{\rm R}_{\Hel} \leq  \\[2mm] g( D_{\Hel}^{\rm G} )  \\[1mm] \end{array}$ 
}
                            & $D_{\tr}^{\rm R} = D_{\tr}^{\rm M}=D_{\tr}^{\rm G}$
                           & $D_\HS^{\rm R}= 2 D^{\rm M}_\HS$
\\[2mm] \cline{2-3} \cline{5-6}
\begin{tabular}{c} Comparison  \\  with $D^{\rm G}$ \\ and $D^{\rm M}$ \end{tabular}
     &  $n_A=3$         &  $D_\Bu^{\rm M} \precsim 2 - \frac{2}{\sqrt{3}} \sqrt{1  + 2 (1-D_\Bu^{\rm R})^2}$
                         &
                         &     \multirow{2}[2]{*}{$\begin{array}{c}
    \frac{1}{n_An_B} \sin^2 \bigl(\frac{\pi}{n_A} \bigr) D^{\rm G}_{\tr}
    \\[2mm] \leq D^{\rm R}_{\tr} \leq \\[2mm] n_A n_B D^{\rm G}_{\tr}\end{array}$ }
                         &     $D^{\rm R}_\HS = \frac{3}{2} D^{\rm M}_\HS$
\\[2mm] \cline{2-3} \cline{6-6}
         &  $n_A >3$     & $D^{\rm M}_\Bu \leq 2 - \frac{2}{\sqrt{n_A}} (1 - D^{\rm R}_\Bu )$
                         &
                         &
                         & $\begin{array}{c} 2 \sin^2 ( \frac{\pi}{n_A}) D^{\rm G}_{\HS} \\ \leq D^{\rm R}_\HS  
                            \leq 2 D^{\rm G}_{\HS} \end{array}$
\\[2mm]
\hline
\multicolumn{2}{|c||}{\begin{tabular}{c} Cross inequalities \\ and relations \end{tabular}}
                  &   \multicolumn{4}{|c|}{
$D^{\rm R}_\Bu \leq D^{\rm R}_\Hel \precsim  1 - ( 1 - D^{\rm R}_\Bu )^2\;\; $,
$\;\;  ( D^{\rm R}_\Hel )^2 \leq D^{\rm R}_{\tr}
  \leq 1 - ( 1 - D^{\rm R}_\Bu )^2 \;\; $,   $\;\;  D^{\rm R}_\Hel ( \rho) = D_\HS^{\rm R}( \sqrt{\rho})$}
\\[1mm]
\hline
\multicolumn{2}{|c||}{\begin{tabular}{c} Computability \\ for two qubits \end{tabular}}
                 &   Bell-diagonal states
                 &   all states
                 &   $\left\{ \begin{array}{l} \text{X-states} \\ \text{$B$-classical states} \end{array} \right.$
                 &   all states
\\
\hline

\end{tabular}
\caption{\label{tab3} 
 Properties of the discords of response with the Bures, Hellinger, trace, and
 Hilbert-Schmidt distances. 
 Here  $E^{\rm R}$ is the entanglement of response~\cite{Giampaolo2007,Monras2011} and
 the function $g$ is given by (\ref{eq-function_g}). Inequalities denoted by the symbol $\precsim$ instead of $\leq$ 
 are saturated for pure states.
 The remaining notations are the same as  in the caption of Table~\ref{tab1}.
 The results quoted in this table have been obtained in Refs.~\cite{Roga2014,Roga_Spehner_Illuminatti2016}. This table is taken from~\cite{Roga_Spehner_Illuminatti2016}.
}
\end{table}

\newpage
 
\section{Conclusion} \label{sec-conclusion}

We have presented the properties of three
classes of geometric measures of quantum correlations, namely
the geometric discord $D^{\rm G}$, the measurement-induced geometric discord $D^{\rm M}$, 
and the discord of response  $D^{\rm R}$,
for two distinguished distances on the set of quantum states, the Bures and Hellinger distances.
These measures satisfy all  the axiomatic criteria for
{\it bona fide} measures of quantum correlations while being
 easier to compute than the entropic quantum discord and having operational interpretations.
Indeed, we have  found that the geometric discord may be interpreted in terms of  
a probability of success in a quantum state discrimination task.
The discords of response  for the Hellinger and Bures distances are related respectively to the
Local Quantum Uncertainty (LQU)~\cite{Girolami2013} and the interferometric power~\cite{Girolami2014}. The latter 
are in fact local geometrical versions of $D^{\rm R}$ (called here the discords of speed of response) and
 enjoy clear interpretations in local measurements and quantum metrology.
The geometric measures $D^{\rm G}$, $D^{\rm M}$, and $D^{\rm R}$ are likely to appear as figures of merit in 
other protocols of quantum information 
and quantum technologies (for instance, $D^{\rm R}$ provides
 upper and lower bounds on the probability of error in quantum reading~\cite{Roga15}).  
We have addressed the issue of the explicit evaluation of the geometric measures
when the reference subsystem $\AAA$ is a qubit. We have found in particular that
the Hellinger geometric discord and Hellinger discord of response are easily computable for any qubit-qudit states.
When $\AAA$ is a qubit or a qutrit, 
 different  measures may be linked by algebraic relations.
This is what happens for instance for the Hellinger geometric discord, Hellinger discord of response, and LQU.  
When $\AAA$ has a higher dimensional Hilbert space, however, each geometric measure  defines its own
 ordering on the set of quantum states. In this sense, the different measures are not equivalent.  
Some bounds enabling to compare them have been given.

From a broader perspective, we have tried in this chapter to show that
the study of the geometry  on the set of quantum states 
defined by contractive Riemannian distances sheds new light on  
quantum correlations in bipartite systems and, more generally, on the whole field of quantum information theory.

\vspace{5mm}

\noindent {\bf Acknowledgments.}
We acknowledge support from the French ANR project No. ANR-13-JS01-0005-01,
the EU FP7 Cooperation STREP Projects iQIT No. 270843  and EQuaM 
No. 323714, the Italian
Minister of Scientific Research (MIUR) national PRIN programme, and 
the Chilean Fondecyt project No. 1140994.


\newpage


\end{document}